\documentclass[numberedappendix]{emulateapj}
\usepackage{ifthen}
\usepackage{comment}

\newboolean{emulateapj}
\setboolean{emulateapj}{true}

\newcommand{\NUMredppmx}{\ensuremath{471,970}}
\newcommand{\NUMredppmxinhat}{\ensuremath{33,177}}
\newcommand{\NUMredppmxinhatwithlc}{\ensuremath{32,831}}
\newcommand{\NUMredppmxinhatwithlcnongiant}{\ensuremath{28,785}}
\newcommand{\NUMredppmxinhatwithlcnongiantrpmcut}{\ensuremath{1225}}
\newcommand{\NUMtotsample}{\ensuremath{27,560}}
\newcommand{\NUMtotpotentialvar}{\ensuremath{2120}}

\newcommand{\NUMrobustplusflares}{\ensuremath{1806}}
\newcommand{\NUMrobustplusflaresnonblend}{\ensuremath{1568}}
\newcommand{\NUMrobustplusflaresnonblendknownvar}{\ensuremath{38}}

\newcommand{\NUMrobustplusflaresnonblendnotknownvar}{\ensuremath{1530}}
\newcommand{\NUMEB}{\ensuremath{95}}
\newcommand{\NUMEBnonblend}{\ensuremath{78}}
\newcommand{\NUMEBnonblendknown}{\ensuremath{5}}

\newcommand{\NUMinvars}{\ensuremath{2033}}
\newcommand{\NUMinvarsorEBcatalog}{\ensuremath{2095}}
\newcommand{\NUMinvarsorEBcatalogunblended}{\ensuremath{507}}
\newcommand{\NUMinvarsorEBcatalogunlikelyblend}{\ensuremath{340}}
\newcommand{\NUMinvarsorEBcatalogpotentialblend}{\ensuremath{943}}
\newcommand{\NUMinvarsorEBcatalogprobableblend}{\ensuremath{305}}
\newcommand{\NUMrobustvarswithEPDamp}{\ensuremath{1279}}

\newcommand{\NUMrobustvarsnonblendnoperioderr}{\ensuremath{}}
\newcommand{\NUMrobustnonpotblendnoneighborswithEPD}{\ensuremath{301}}

\newcommand{\NUMcross}{\ensuremath{61}}
\newcommand{\NUMnotcross}{\ensuremath{2059}}
\newcommand{\NUMmatchgcvs}{\ensuremath{29}}
\newcommand{\NUMmatchnsv}{\ensuremath{2}}
\newcommand{\NUMmatchnsvs}{\ensuremath{6}}
\newcommand{\NUMmatchrotse}{\ensuremath{7}}
\newcommand{\NUMmatchasas}{\ensuremath{3}}
\newcommand{\NUMmatchswasp}{\ensuremath{23}}
\newcommand{\NUMmatchswaspone}{\ensuremath{2}}

\newcommand{\NUMmatchrotsegcvs}{\ensuremath{3}}
\newcommand{\NUMcrosscorrect}{\ensuremath{33}}
\newcommand{\NUMcrosswrong}{\ensuremath{28}}
\newcommand{\NUMcrosswrongprobableblend}{\ensuremath{17}}
\newcommand{\NUMcrosswrongpotentialblend}{\ensuremath{6}}
\newcommand{\NUMcrosswrongunlikelyblend}{\ensuremath{3}}
\newcommand{\NUMcrosswrongunblended}{\ensuremath{2}}
\newcommand{\NUMmatchgcvscorrect}{\ensuremath{12}}

\newcommand{\NUMflarestars}{\ensuremath{60}}
\newcommand{\NUMflareevents}{\ensuremath{64}}
\newcommand{\NUMflarelcsearch}{\ensuremath{23,589}}
\newcommand{\NUMflarelcnotsearch}{\ensuremath{3971}}
\newcommand{\NUMflareeventautoselect}{\ensuremath{320}}
\newcommand{\NUMflarestarautoselect}{\ensuremath{281}}
\newcommand{\NUMflareswithperorEB}{\ensuremath{35}}
\newcommand{\NUMflareswithrobustperorEB}{\ensuremath{35}}

\newcommand{\NUMflareswithrobustperorEBnonpererrnonblend}{\ensuremath{32}}

\newcommand{\NUMmatchrosat}{\ensuremath{237}}
\newcommand{\NUMmatchrosatvars}{\ensuremath{226}}
\newcommand{\NUMmatchrosatEB}{\ensuremath{14}}
\newcommand{\NUMmatchrosatflares}{\ensuremath{24}}
\newcommand{\NUMmatchrosatdistinct}{\ensuremath{232}}

\newcommand{\EBminimumperiod}{\ensuremath{ 0.193}}
\newcommand{\EBmaximumperiod}{\ensuremath{24.381}}
\newcommand{\NUMEBprobableblend}{\ensuremath{17}}
\newcommand{\NUMEBpotentialblend}{\ensuremath{28}}
\newcommand{\NUMEBunlikelyblend}{\ensuremath{21}}
\newcommand{\NUMEBunblended}{\ensuremath{29}}

\newcommand{\NUMaovharmepdselect}{\ensuremath{1337}}
\newcommand{\NUMaovharmepdpass}{\ensuremath{1082}}
\newcommand{\NUMaovharmepdEB}{\ensuremath{34}}
\newcommand{\NUMaovharmepdquest}{\ensuremath{185}}
\newcommand{\NUMaovharmepdrej}{\ensuremath{36}}
\newcommand{\NUMaovharmtfaselect}{\ensuremath{1717}}
\newcommand{\NUMaovharmtfapass}{\ensuremath{1443}}
\newcommand{\NUMaovharmtfaEB}{\ensuremath{43}}
\newcommand{\NUMaovharmtfaquest}{\ensuremath{217}}
\newcommand{\NUMaovharmtfarej}{\ensuremath{14}}

\newcommand{\NUMblsbothEB}{\ensuremath{52}}

\newcommand{\NUMblseitherselect}{\ensuremath{752}}

\newcommand{\NUMblseitherEB}{\ensuremath{89}}

\newcommand{\NUMblsepdEBonly}{\ensuremath{8}}
\newcommand{\NUMblstfaEBonly}{\ensuremath{29}}

\newcommand{\NUMtotstarsinVarColorHistplotLeft}{\ensuremath{8944}}
\newcommand{\NUMtotstarsinVarColorHistplotRight}{\ensuremath{26884}}
\newcommand{\NUMtotvarsinVarColorHistplotLeft}{\ensuremath{298}}
\newcommand{\NUMtotvarsinVarColorHistplotRight}{\ensuremath{1190}}

\newcommand{\VARFRACCOEFFA}{\ensuremath{0.0016}}
\newcommand{\VARFRACCOEFFB}{\ensuremath{0.929}}
\newcommand{\VARFRACERRApos}{\ensuremath{0.0011}}
\newcommand{\VARFRACERRAneg}{\ensuremath{0.00066}}
\newcommand{\VARFRACERRB}{\ensuremath{0.13}}
\newcommand{\VARFRACCOEFFAnocor}{\ensuremath{0.001}}
\newcommand{\VARFRACCOEFFBnocor}{\ensuremath{0.969}}
\newcommand{\VARFRACERRAposnocor}{\ensuremath{0.00069}}
\newcommand{\VARFRACERRAnegnocor}{\ensuremath{0.00041}}
\newcommand{\VARFRACERRBnocor}{\ensuremath{0.14}}
\newcommand{\VARFRACCOEFFBLEND}{\ensuremath{0.0162}}
\newcommand{\VARFRACERRBLEND}{\ensuremath{0.0021}}
\newcommand{\NUMfracexpectedblend}{\ensuremath{26\%}}

\shorttitle{K/M Dwarf Variability Survey}
\shortauthors{Hartman et al.}

\begin{document}

\title{A Photometric Variability Survey of Field K and M Dwarf Stars with HATNet}
\author{
    J.~D.~Hartman\altaffilmark{1}, 
    G.~\'{A}.~Bakos\altaffilmark{1}, 
    R.~W.~Noyes\altaffilmark{1}, 
    B.~Sip\H{o}cz\altaffilmark{1,2}, 
    G.~Kov\'acs\altaffilmark{3}, 
    T.~Mazeh\altaffilmark{4}, 
    A.~Shporer\altaffilmark{4}, 
    A.~P\'al\altaffilmark{1,2}
}
\altaffiltext{1}{Harvard-Smithsonian Center for Astrophysics, 60 Garden St., Cambridge, MA~02138, USA}
\altaffiltext{2}{Department of Astronomy,
  E\"otv\"os Lor\'and University, Budapest, Hungary.}
\altaffiltext{3}{Konkoly Observatory, Budapest, Hungary.}
\altaffiltext{4}{Wise Observatory, Tel Aviv University, Tel Aviv, Israel}

\begin{abstract}
Using light curves from the HATNet survey for transiting extrasolar
planets we investigate the optical broad-band photometric variability
of a sample of \NUMtotsample{} field K and M dwarfs selected by color
and proper-motion ($V - K \ga 3.0$, $\mu > 30~{\rm mas/yr}$, plus
additional cuts in $J-H$ vs. $H-K_{S}$ and on the reduced proper
motion). We search the light curves for periodic variations, and for
large-amplitude, long-duration flare events. 
A total of \NUMtotpotentialvar{} stars exhibit potential variability,
including \NUMEB{} stars with eclipses and \NUMflarestars{} stars with
flares. Based on a visual inspection of these light curves and an
automated blending classification, we select
\NUMrobustplusflaresnonblend{} stars, including \NUMEBnonblend{}
eclipsing binaries, as secure variable star detections that are not
obvious blends. We estimate that a further $\sim 26\%$ of these stars
may be blends with fainter variables, though most of these blends are
likely to be among the hotter stars in our sample.
We find that only
\NUMrobustplusflaresnonblendknownvar{} of the
\NUMrobustplusflaresnonblend{} stars, including \NUMEBnonblendknown{}
of the eclipsing binaries, have previously been identified as
variables or are blended with previously identified variables. One of
the newly identified eclipsing binaries is 1RXS~J154727.5+450803, a
known $P = 3.55~{\rm day}$, late M-dwarf SB2 system, for which we
derive preliminary estimates for the component masses and radii of
$M_{1} = M_{2} = 0.258 \pm 0.008~{\rm M_{\odot}}$ and $R_{1} = R_{2} =
0.289 \pm 0.007~{\rm R_{\odot}}$. The radii of the component stars are
larger than theoretical expectations if the system is older than $\sim
200~{\rm Myr}$. The majority of the variables are heavily spotted BY
Dra-type stars for which we determine rotation periods.
Using this sample, we investigate the relations between period, color,
age, and activity measures, including optical flaring, for K and M
dwarfs, finding that many of the well-established relations for F, G
and K dwarfs continue into the M dwarf regime. We find that the
fraction of stars that are variable with peak-to-peak amplitudes
greater than 0.01~mag increases exponentially with the $V-K_{S}$ color
such that approximately half of field dwarfs in the solar neighborhood
with $M \la 0.2~M_{\odot}$ are variable at this level. Our data hints
at a change in the rotation-activity-age connection for stars with $M
\la 0.25~M_{\odot}$.  
\end{abstract}

\keywords{
	stars: late-type, low-mass, rotation, flare, fundamental parameters (masses, radii) ---
	binaries: eclipsing ---
        surveys ---
        catalogs ---
	techniques: photometric
}

\section{Introduction}\label{sec:intro}

Main-sequence stars smaller than the Sun are known to exhibit the
following types of photometric variability: variability due to
binarity (either eclipses or proximity effects such as ellipsoidal
variability), variability due to the rotational modulation or temporal
evolution of starspots, and flares. Below we discuss each type of
variability, and what might be learned from studying it.

\subsection{Low-Mass Eclipsing Binaries}

Recent discoveries of eclipsing binaries (EBs) with K and M dwarf
components have led to the realization that the radii of early M
dwarfs and late K dwarfs are somewhat larger than predicted by theory
\citep[the number typically stated is
  10\%;][]{Torres.02,Ribas.03,Lopez-Morales.05,Ribas.06,Beatty.07,Fernandez.09}.
It has been suggested that the discrepancy between theory and
observation for these binary star components may be due to their
enhanced magnetic activity inhibiting convection.
\citep{Ribas.06,Torres.06,Lopez-Morales.07,Chabrier.07}. Support for
this hypothesis comes from interferometric measurements of the
luminosity-radius relation for inactive single K and M dwarfs which
appears to be in agreement with theoretical predictions
\citep{Demory.09}. 
Testing this hypothesis will require finding additional binaries
spanning a range of parameters (mass, rotation/orbital period,
metallicity, etc.).

As transiting planets are discovered around smaller stars, the need
for models that provide accurate masses and radii for these stars has
become acute. For example, the errors in the planetary parameters for
the transiting Super-Neptune GJ 436b appear to be dominated by the
uncertainties in the stellar parameters of the $0.45~M_{\odot}$
M-dwarf host star \citep[][]{Torres.07}. Having a large sample of low
mass stars with measured masses and radii would enable the
determination of precise empirical relations between the parameters
for these stars.

\subsection{Stellar Rotation}
The rotation period is a fundamental measurable property of a
star. For F, G, K and early M main sequence stars there is a
well-established relation between rotation, magnetic activity and age
\citep[e.g.][]{Skumanich.72,Noyes.84,Pizzolato.03}. In addition to
illuminating aspects of stellar physics, this relation in practice
provides the best method for measuring the ages of field main sequence
stars \citep[][]{Mamajek.08,Barnes.07}.

For late M-dwarfs the picture is less clear. From a theoretical
standpoint one might expect that fully convective stars should not
exhibit a rotation-activity connection if the connection is due to the
$\alpha\Omega$-dynamo process, which operates at the interface of the
radiative and convective zones in a star, and is thought to generate
the large scale magnetic fields in the Sun \citep[see for example the
  discussion by][]{Mohanty.03}. Nonetheless, several studies have
found evidence that the rotation-activity connection may continue even
to very late M-dwarfs \citep{Delfosse.98, Mohanty.03, Reiners.07}. In
these studies the rotation period is inferred from the projected
rotation velocity $v \sin i$, which is measured spectroscopically.
Rotation studies of this
sort suffer both from the inclination axis ambiguity, and from low
sensitivity to slow rotation. 
Moreover, because low-mass stars are intrinsically faint,
these studies require large telescopes to obtain high-resolution, high
S/N spectra, so that typically only a few tens of stars are studied at
a time.

There are two techniques that have been used to directly measure
stellar rotation periods. The first technique, pioneered by
\citet{Wilson.57}, is to monitor the emission from the cores of the
CaII H and K lines, searching for periodic variations. The
Mount-Wilson Observatory HK project has used this technique to measure
the rotation periods of more than 100 slowly rotating dwarfs and giant
stars
\citep[][]{Wilson.78,Duncan.91,Baliunas.95,Baliunas.96}. Alternatively,
if a star has significant spot-coverage it may be possible to measure
its rotation period by detecting quasi-periodic variations in its
broad-band photometric brightness. Studies of this sort have been
carried out in abundance for open clusters \citep[e.g.][]{Radick.87}
as well as for some field stars
\citep[e.g.][]{Strassmeier.00,Kiraga.07}. While there are rich samples
of rotation periods for K and M dwarfs in open clusters with ages $\la
600~{\rm Myr}$
\citep{Irwin.06,Irwin.07,Irwin.09a,Meibom.09,Hartman.09}, the data for
older K and M dwarfs is quite sparse. As such, there are few
observational constraints on the rotational evolution of these stars
after $\sim 0.5~{\rm Gyr}$.

Given the existing uncertainties in the
rotation-activity connection for low mass stars and the potential to
use rotation as a proxy for age, a large, homogeneously collected
sample of photometric rotation periods for field K and M dwarfs could
potentially be of high value.

\subsection{Flares}
Flaring is known to be a common phenomenon among K and M
dwarfs. Studies of open cluster and field flare stars have shown that
the frequency of flaring increases with decreasing stellar mass, and
decreases with increasing stellar age \citep[e.g.][]{Ambartsumyan.70,
  Mirzoyan.89}. Significant flaring on these low-mass dwarfs is likely
to impact the habitability of any planets they may harbor
\citep[e.g.][]{Kasting.93,Lammer.07,Guinan.09}, so determining the
frequency of flares, and its connection with other stellar properties
such as rotation, has important implications for the study of
exoplanets.

\subsection{The HATNet Survey}
To address these topics we use data from the Hungarian-made Automated
Telescope Network \citep[HATNet;][]{Bakos.04} and the Wise-HAT project
\citep[WHAT;][]{Shporer.09} to conduct a variability survey of K and M
dwarfs. The on-going HATNet project is a wide-field search for
transiting extrasolar planets (TEPs) orbiting relatively bright
stars. To date, the survey has announced the discovery of 26 TEPs,
including HAT-P-11b \citep{Bakos.09}, one of only a handful of known
transiting Super-Neptune planets. While the primary focus of the
HATNet project has been the discovery of TEPs, some results not
related to planets have also been presented. This includes studies of
low-mass stars in single-lined eclipsing binary systems
\citep{Beatty.07, Fernandez.09}, as well as searches for variable
stars \citep{Hartman.04, Shporer.07}.

\subsection{Overview of the Paper}
The structure of the paper is as follows. In \S~\ref{sec:data} we
describe the HATNet photometric data, and select the sample of field K
and M dwarfs. In \S~\ref{sec:selection} we discuss our methods for
selecting variable stars. We estimate the degree of blending for
potential variables in \S~\ref{sec:blend}. We match our catalog of
variables to other catalogs in \S~\ref{sec:match}. We discuss the
properties of the variables in \S~\ref{sec:discussion} including an
analysis of one of the EB systems found in the survey. We conclude in
\S~\ref{sec:conclusion}. 
In appendix~\ref{sec:cat} we
present the catalog of variable stars.

\section{Observational Data}\label{sec:data}

\subsection{HATNet Data}\label{sec:hatnet}

The HATNet project, which has been in operation since 2003, uses a
network of 6 small (11\,cm aperture), autonomous telescopes to obtain
time-series observations of stars. The telescopes are distributed in
longitude, with 4 in Arizona at Fred Lawrence Whipple Observatory
(FLWO), and 2 in Hawaii on the roof of the Sub-Millimeter Array at
Mauna Kea Observatory (MK). An additional seventh telescope, known as
WHAT, is located at Wise Observatory in Israel and is identical in
design to the HATNet telescopes. For details on the system design and
observing procedure see \citet{Bakos.04, Bakos.06}; here we briefly
review a few points that are relevant to the survey presented in this
paper.

The HATNet telescopes have a fast focal ratio of f/1.8 yielding a
plate-scale of $1031.315\,\arcsec\,{\rm mm}^{-1}$. Prior to 2007
September each telescope employed a 2K$\times$2K CCD and a Cousins
$I_{C}$ filter \citep{Cousins.76}. The 2K$\times$2K CCDs covered an
$8.2\degr \times 8.2\degr$ field of view (FOV) at a pixel scale of
$14\arcsec$. With these CCDs, stars with $7.5 \la I_{C} \la 14.0$ were
observed with 5 minute exposures yielding a typical per-image
photometric precision of a few mmag at the bright end, 0.01 mag at
$I_{C} \sim 11$, and 0.1 mag at $I_{C} \sim 13.5$.  After this date
the telescopes were refitted with 4K$\times$4K CCDs and Cousins
$R_{C}$ filters. The new CCDs cover a $10.6\degr \times 10.6\degr$ FOV
at a pixel scale of $9\arcsec$. With these CCDs, stars with $8.0 \la
R_{C} \la 15.0$ are observed with 5 minute exposures yielding a
typical per-image photometric precision of a few mmag at the bright
end, 0.01 mag at $R_{C} \sim 12$, and 0.1 mag at $R_{C} \sim 15$. The
exact magnitude limits and precision per exposure as a function of
magnitude vary within a field due to vignetting, and from field to
field due to differences in the reduction procedure used (which has
evolved over time) and in the degree of stellar crowding. In 2008
September the filters were changed to Sloan $r$, though we do not
include any observations taken through the new filters in the survey
presented here. For both CCD formats, the typical full width at half
maximum (FWHM) of the point spread function (PSF) is $\sim 2$ pixels
(i.e. $\sim 30\arcsec$ for the 2K fields and $\sim 20\arcsec$ for the
4K fields).

The typical observing procedure is to assign to each telescope one to
three fields to monitor, with several fields assigned concurently to a
telescope at MK and a telescope at FLWO. These fields are defined by
dividing the sky into 838 $7.5\degr \times 7.5\degr$ tiles. Each field
is continuously observed at a cadence of 5.5 minutes for as long as it
is visible (typically meaning the airmass is less than 2.0), or until
another higher priority field rises. The total time spanned by the
observations for a given field varies from 45 days to 2.5 years, with
a median time span of half a year. The total number of exposures
obtained for a given field varies from 1600 to 11,600 with a median
value of 4800.

The data for this survey comes from 72 HATNet fields with declinations
between $+15\degr$ and $+52\degr$. The survey covers approximately
4000 square degrees, or roughly 10\% of the sky.

Both aperture photometry (AP) and image subtraction photometry (ISM)
have been used for reductions. Both pipelines were developed from
scratch for HATNet. See \citet{Pal.09} for detailed
descriptions of both methods \citep[see also][]{Bakos.04}.

For both pipelines the Two Micron All-Sky Survey
\citep[2MASS;][]{Skrutskie.06} is used as the astrometric
reference. The astrometric solutions for the images are determined
using the methods described by \citet{Pal.06}~and
\citet{Pal.09}. Photometry is performed at the positions of 2MASS
sources transformed to the image coordinate system. For each resulting
light curve the median magnitude is fixed to the $I_{C}$ or $R_{C}$
magnitude of the source based on a transformation from the 2MASS
$J$,$H$ and $K_{S}$ magnitudes. We used the transformations:
\begin{eqnarray}
R_{C} & = & 0.0606 + 2.7823J + 0.8922H - 2.6713K_{S} \\
I_{C} & = & 0.0560 + 2.0812J + 0.4074H - 1.4889K_{S}
\end{eqnarray}
which were determined from 590 Landolt standard stars
\citep{Landolt.92} with 2MASS photometry. The RMS scatter of the
residuals for these relations are $0.11~{\rm mag}$ and $0.06~{\rm
  mag}$ respectively. For the ISM reduction this magnitude is also
used as the reference magnitude for each source in converting
differential flux measurements into magnitudes. For each source,
photometry is performed using three separate apertures. The set of
apertures used has changed over time; the most recent reductions use
aperture radii of 1.45, 1.95 and 2.35 pixels, which we have found to
give light curves with the lowest RMS for both the 2K and 4K era
observations, older observations which have not yet been re-reduced
use aperture radii of 2.10, 2.55 and 3.20 pixels. Following the
post-processing routines discussed below, we adopt a single ``best''
aperture for each light curve.

Both the AP and ISM pipelines produce light curves that are calibrated
against ensemble variations in the flux scale. For ISM this is an
automatic result of the method \citep[see section 2.9
  of][]{Pal.09}. To select the photometric reference for the ISM
method we manually choose an image from the set of images with the
narrowest, but not distorted, PSF, which were observed at low airmass
on photometric nights with little background light from the Moon, and
with the Sun well below the horizon. For AP the calibration is done
using the method described in section 2.7.3 of
\citet{Pal.09}. Briefly, for the AP pipeline a master reference image
is selected by hand from a photometric night, and the magnitude
transformation from each observed image to the reference image is fit
for using a function that consists of a fourth order polynomial in the
spatial coordinates and a linear dependence on the 2MASS $J-K_{S}$
color. After obtaining initial light curves for the stars the
transformation is then redetermined this time rejecting stars that are
outliers in the magnitude-RMS relation from the fit and weighting each
star by the scatter of its light curve.  Finally we reject images for
which a substantial fraction of the transformed magnitudes are
$3\sigma$ outliers in their respective light curves. Note that for
fields observed by multiple telescopes the above procedure is
performed on the combined set of observations. Following this
procedure, the typical RMS of the light curves of the brightest
non-saturated stars in a frame is $\sim 5~{\rm mmag}$.

The calibrated light curves for each aperture are then passed through
two routines that filter systematic variations from the light curves
that are not corrected in calibrating the ensemble. The first routine
(EPD) decorrelates each light curve against a set of external
parameters including parameters describing the shape of the PSF, the
sub-pixel position of the star on the image, the zenith angle, the
hour angle, the local sky background, and the variance of the
background \citep[see][]{Bakos.09, Pal.09}. For fields including
observations from multiple telescopes the decorrelation is done
independently on the data from each telescope. This procedure is
applied assuming that each star has a constant magnitude. As discussed
below, for large amplitude variable stars this will distort the signal
and may lower the S/N of the detection, however such large amplitude
variables will generally still be detectable, as confirmed by
injecting sinusoid signals into raw light curves and attempting to
recover them after applying the EPD procedure.

After applying EPD, the light curves are then processed with the
Trend-Filtering Algorithm \citep[TFA;][]{Kovacs.05, Pal.09} which
decorrelates each light curve against a representative sample of other
light curves from the field. The number of template light curves used
differs between the fields, typically the number is $\sim 8\%$ of the
total number of images for that field. In applying the TFA routine we
also perform $\sigma$-clipping on the light curves since this
generally reduces the number of false alarms when searching for
transits. For the remainder of the paper we will refer to light curves
that have been processed through EPD only, without application of TFA,
as EPD light curves, and will refer to light curves that have been
processed through both EPD and TFA as TFA light curves. We note that
for some fields the EPD light curves were not stored and only TFA
light curves are available.

Both of these algorithms tend to improve the signal to noise ratio of
transit signals or low amplitude variations in the light curves, but
they may distort the light curves of stars that show large-amplitude,
long-period, continuous variability. Additionally the decorrelation
against the zenith and hour angles in the EPD routine will tend to
filter out real variable star signals with periods very close to a
sidereal day or an integer multiple of a sidereal day. Since these
variables would be difficult to distinguish from systematic errors in
the photometry, losing them from the sample may be unavoidable. The
TFA routine in particular can distort long-period signals while
improving the signal to noise ratio of short-period signals. For this
reason we analyze both the EPD and TFA light curves, when available,
to select variable stars (\S~\ref{sec:selection}). We note that for
the analysis in this paper we do not use the signal-reconstruction
mode TFA presented by \citet{Kovacs.05}. Once a signal is detected,
TFA can be run in this mode to obtain a trend-filtered light curve
that is free of signal distortions, however for signal detection one
must use zero-signal TFA since the signal is not known a priori.

Finally, an optimal aperture is chosen for each star. For stars
fainter than a fixed limit the smallest aperture is used (to minimize
the sky noise), for brighter stars the aperture with the smallest
root-mean-square (RMS) light curve is used.

\subsection{Composite Light Curves}\label{sec:lightcurve}

Because the separation between the HATNet field centers is smaller
than the FOV of the HATNet telescopes for both the 2K and 4K CCDs,
some stars are observed in multiple fields. These stars may have more
than one light curve, which we combine into composite light curves. In
making a composite light curve we subtract the median magnitude from
each component light curve. For fields reduced with both ISM and AP we
use the light curve with the lowest RMS, and in the case of equal RMS
we use the ISM light curve. Note that the composite light curve for a
star may include a mix of $I_{C}$ and $R_{C}$ photometry. While the
amplitude of variability may be different from filter to filter, the
period and phasing for variations due to eclipses or the rotational
modulation of starspots will be independent of bandpass. 

We do not allow for independent amplitudes of different filters in
searching for variability using the methods described in
\S~\ref{sec:selection}. To verify that this does not significantly
affect the detection of periodic signals, we injected sinusoids with
periods between 0.1 and 100~days drawn from a uniform logarithmic
distribution, and amplitudes between 0.01 and 1.0 mag taken from a
uniform logarithmic distribution into the light curves of 553
nonvariable stars which have both $I_{C}$ and $R_{C}$ photometry. We
adopted independent amplitudes for the $I_{C}$ and $R_{C}$ light
curves of a given star, but adopted the same phase and period for the
variations. We then attempted to recover the injected signals in the
combined light curves using the AoVHarm technique discussed in
\S~\ref{sec:selection}. We find that in 99\% of the cases the peak
frequency found in the AoVHarm periodogram is within $1/T$ of the
injected frequency, where $T$ is the timespan of the composite light
curve. We also performed the simulations allowing the $I_{C}$ and
$R_{C}$ light curves to have independent phases as well as independent
amplitudes, and find that the recovered period agrees with the
injected period in 96\% of the cases. We note that this simple merging
of photometry from different filters may result in spurious side lobes
in the power spectrum, so for a more detailed analysis of individual
objects, including searches for multiple frequencies of variation,
this effect should be considered.

\subsection{Selection of the K and M Dwarf Sample}\label{sec:otherdat}

Combining photometric observations with proper motion measurements is
an effective method for selecting nearby dwarf stars. This technique
is routinely used in the search for cool stars in the solar
neighborhood \citep[e.g.][]{Reyle.04}, and has been suggested as an
effective method for screening giants from transit surveys
\citep{Gould.03}. 

To select our sample of stars that are probable K and M dwarfs we
apply cuts on the proper motion and on the color. Proper motion
measurements are taken from the PPM-Extended catalog
\citep[PPMX;][]{Roser.08} which provides proper motions with
precisions ranging from 2~mas/yr to 10~mas/yr for 18 million stars
over the full sky down to a limiting magnitude of $V \sim 15.2$. The
PPMX catalog provides complete coverage of the HATNet survey for stars
with $V-I_{C} \la 1.2$ for the 2K fields and for stars with $V - R \la
0.7$ for the 4K fields. For stars redder than these limits, the faint
limit of HATNet is deeper than the faint limit of PPMX. We select all
stars from this catalog with a proper motion $\mu > 30~{\rm mas/yr}$,
as these are stars for which the proper motion is detected with better
than $3\sigma$ confidence. For the color selection we use the 2MASS
$JHK_{S}$ photometry and, where available, $V$-band photometry from
the PPMX catalog, which is taken from the Tycho-2 catalog
\citep{Hog.00} and transformed to the Johnson system by
\citet{Kharchenko.01}. Only $\sim 4\%$ of the stars have $V$
photometry given in the PPMX catalog, for the majority of stars that
do not have $V$ photometry we calculate an initial approximate $V$ magnitude
using
\begin{equation}\label{eq:2MASSVtrans}
V = -0.0053 + 3.5326J + 1.3141H - 3.8331K_{S}.
\end{equation}
which is determined from 590 Landolt Standard stars \citep{Landolt.92}
with 2MASS photometry, and is used internally by the HATNet project to
estimate the $V$ magnitudes of transit candidates for follow-up
observations. We then select stars with $V-K_{S} > 3.0$ which
corresponds roughly to stars with spectral types later than K6
\citep{Bessell.88}. This selects a total of \NUMredppmx{} stars from
the PPMX catalog, of which \NUMredppmxinhat{} fall in a reduced HATNet
field; of these \NUMredppmxinhatwithlc{} have a HATNet light curve
containing more than 1000 points.

Extrapolating a $V$ magnitude from near-infrared photometry
is not generally reliable to more than a few tenths of a magnitude. We
therefore also obtained $V$ magnitudes for stars in our sample by
matching to the USNO-B1.0 catalog \citep{Monet.03}. This matching was
done after the variability search described in \S~\ref{sec:selection};
we choose not to redo the sample selection and the subsequent
variability selection. Note that low-mass sub-dwarfs, which have
anomalously blue $V-K_{S}$ values, will pass a selection on $V-K_{S} >
3.0$ computed using eq.~\ref{eq:2MASSVtrans} to extrapolate the $V$
magnitude, while they would not necessarily pass a selection using the
measured value of $V-K_{S}$.  To transform from the photographic
$B_{U}$, $R_{U}$ magnitudes in the USNO-B1.0 catalog to the $V$-band
we use a relation of the form:
\begin{equation}\label{eq:BURUVtrans}
V = aB_{U} + bB_{U}^2 + cR_{U} + dR_{U}^2 + eB_{U}R_{U} + f
\end{equation}
with coefficients given separately in table~\ref{tab:fitparams} for
the $(B_{U,1},R_{U,1})$, $(B_{U,1},R_{U,2})$, $(B_{U,2},R_{U,1})$ and
$(B_{U,2},R_{U,2})$ combinations, where $B_{U,1}$ and $B_{U,2}$ are
the first and second epoch blue photographic magnitudes respectively,
and $R_{U,1}$ and $R_{U,2}$ are the first and second epoch red
photographic magnitudes. These transformations were determined using
$\sim 1100$ stars with both $V$ photometry in the PPMX catalog and
USNO-B1.0 $(B_{U},R_{U})$ photometry. Based on the root-mean-square
(RMS) scatter of the post-transformation residuals, we used the
$(B_{U,2},R_{U,2})$, $(B_{U,1},R_{U,2})$, $(B_{U,2},R_{U,1})$ and
$(B_{U,1},R_{U,1})$ transformations in order of preference. For stars
with neither PPMX $V$ photometry nor USNO-B1.0 photometry, we used
eq.~\ref{eq:2MASSVtrans}. For the remainder of the analysis in this
paper, the $V$ magnitude is taken from PPMX (Tycho-2) for 3.1\% of the
stars in our sample, from USNO-B1.0 for 93.6\% of the stars, and is
transformed from the 2MASS magnitudes for 3.3\% of the stars.

\begin{figure}[!ht]
\ifthenelse{\boolean{emulateapj}}{\epsscale{1.2}}{\epsscale{1.0}}
\plotone{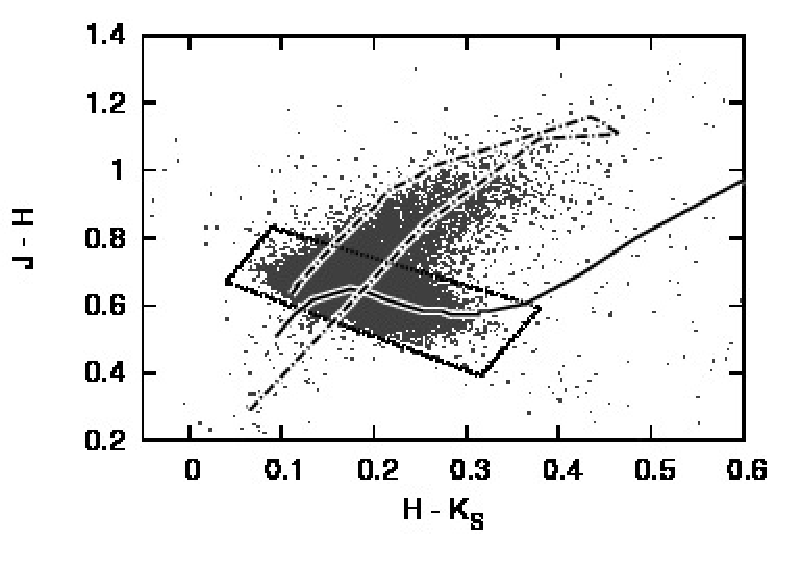}
\caption{$J-H$ vs. $H-K_{S}$ color-color diagram for
  \NUMredppmxinhatwithlc{} stars that have $V-K_{S} > 3.0$ with $V$
  taken either from the PPMX catalog (Tycho-2) or extrapolated from
  the 2MASS $J$, $H$ and $K_{S}$ magnitudes using
  eq.~\ref{eq:2MASSVtrans}, $\mu > 30~{\rm mas/yr}$ from the PPMX
  catalog, and that have a HATNet light curve containing more than
  1000 points (gray-scale points). The solid line shows the expected
  relation for cool dwarfs \citep{Baraffe.98,Chabrier.00}, while the
  dot-dashed line shows the expected relation for giants
  \citep{Marigo.08, Bonatto.04}. Stars outside the area enclosed by
  the dotted line are rejected. For display purposes we have added
  slight Gaussian noise to the observed colors in the plot.}
\label{fig:HKJH}
\end{figure}

Figure~\ref{fig:HKJH} shows the $J-H$ vs. $H-K_{S}$ color-color
diagram for the selected sample. We also show the expected relations
for dwarf stars and for giants. The relation for dwarfs is taken from
a combination of the \citet{Baraffe.98} 1.0 Gyr isochrone for solar
metallicity stars with $0.15~M_{\odot} \leq M \leq 0.7~M_{\odot}$ and
the \citet{Chabrier.00} models for objects with $M \leq
0.075~M_{\odot}$. The $JHK$ magnitudes for these isochrones were
transformed from the CIT system \citep{Elias.82,Elias.83} to the 2MASS
system using the transformations determined by
\citet{Carpenter.01}. The relation for giant stars with $\log g < 2.0$
is taken from the 1.0 Gyr, solar metallicity Padova isochrone
\citep{Marigo.08, Bonatto.04}\footnote{The isochrone was obtained from
  the CMD 2.1 web interface
  http://stev.oapd.inaf.it/cgi-bin/cmd\_2.1}. While the majority of
stars lie in the expected dwarf range, a significant number of stars
fall along the giant branch. Some of these stars may be rare carbon
dwarfs, but the majority are most likely giants with inaccurate proper
motion measurements in the PPMX catalog. Of the 2445 selected stars
with $J - H > 0.8$ that have HATNet light curves, 87\% have undetected
proper motions or proper motions less than 10 mas/yr in the USNO-B1.0
catalog, this is compared to 28\% of the sample with $J - H < 0.8$. We
note that the PPMX and USNO-B1.0 proper motions agree to within 20\%
for $\sim 62\%$ of the stars with $J - H < 0.8$ and which have proper
motions given in USNO-B1.0, and to within 50\% for $\sim 86\%$ of
these stars. A visual inspection of the POSS-I and POSS-II Digitized
Sky Survey images for a number of the sources with $J - H > 0.8$ and
$\mu > 100~{\rm mas/yr}$ revealed none with visually detectable proper
motion, and in many cases the object consists of two close, comparably
bright stars, for which misidentification of sources may be to blame
for the spurious proper motion detection. This includes several stars
where the PPMX and USNO-B1.0 proper motion values are comparable. We
therefore apply an additional cut in the $J - H$ vs $H-K_{S}$
color-color diagram as shown in figure~\ref{fig:HKJH} to reduce the
sample to \NUMredppmxinhatwithlcnongiant{} stars.

\begin{figure}[!ht]
\ifthenelse{\boolean{emulateapj}}{\epsscale{1.2}}{\epsscale{1.0}}
\plotone{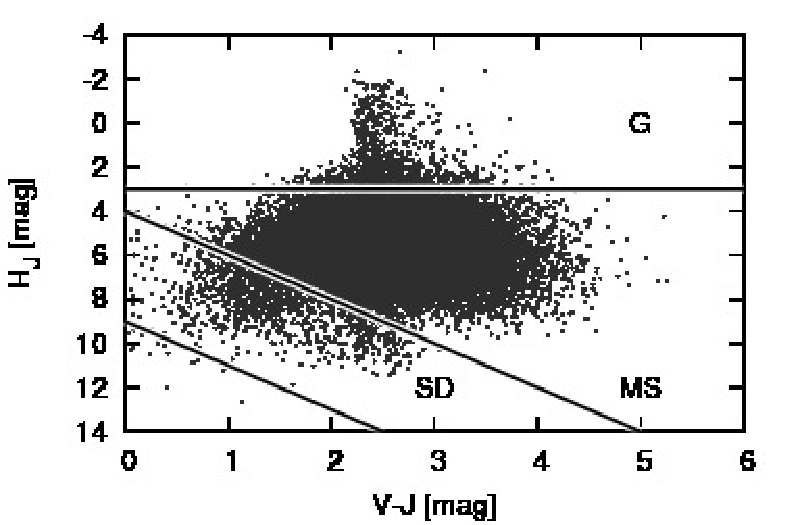}
\caption{$V-J$ vs. $H_{J}$ RPM diagram for stars with $\mu > 30~{\rm
    mas/yr}$ passing $V-K_{S}$ and $JHK$ color cuts. Note that in this
  plot we use $V$ magnitudes that are transformed from the USNO-B1.0
  photographic magnitudes for the majority stars, and not the $V$
  magnitudes transformed from $JHK$ that were used in applying the
  initial $V-K_{S}$ cut. The lines separate main sequence dwarfs from
  sub-dwarfs and giants. We reject the
  \NUMredppmxinhatwithlcnongiantrpmcut{} stars with $H_{J} < 3.0$. The
  lines separating main sequence dwarfs from sub-dwarfs are taken from
  \citet{Yong.03}.}
\label{fig:rpm}
\end{figure}

\begin{figure}[!ht]
\ifthenelse{\boolean{emulateapj}}{\epsscale{1.2}}{\epsscale{1.0}}
\plotone{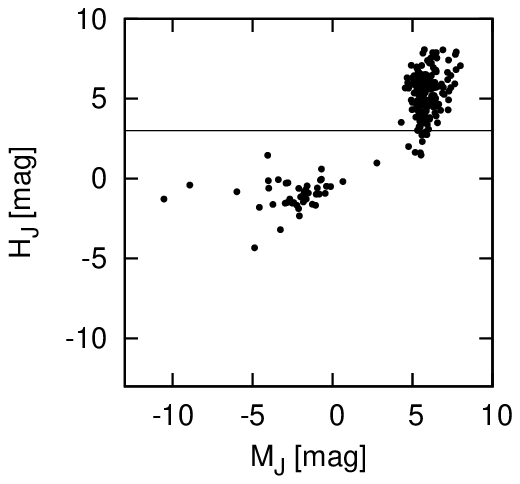}
\caption{The absolute magnitude $M_{J}$ vs. the RPM for 239 stars in
  the sample which have a Hipparcos parallax. This confirms that the
  RPM provides a rough estimate of the absolute magnitude for this
  sample of stars. We reject stars with $H_{J} < 3.0$, as these appear to be
  predominately giants.}
\label{fig:MJRPMcomp}
\end{figure}

Figure~\ref{fig:rpm} shows a $V-J$ vs. $H_{J}$ reduced proper-motion
\citep[RPM;][]{Luyten.22} diagram for the sample. Here the RPM, $H_{J}$, is calculated as
\begin{equation}
H_{J} = J + 5\log_{10}(\mu/1000)
\end{equation} 
and gives a rough measure of the absolute magnitude $M_{J}$ of a
star. We show roughly the lines separating main sequence dwarfs from
sub-dwarfs and giants. In figure~\ref{fig:MJRPMcomp} we compare the
RPM to $M_{J}$ for 239 stars in the sample which have a Hipparcos
parallax \citep{Perryman.97}. To remove additional giants from the
sample we reject \NUMredppmxinhatwithlcnongiantrpmcut{} stars with
$H_{J} < 3.0$, leaving our final sample of \NUMtotsample{} stars.

\ifthenelse{\boolean{emulateapj}}{\begin{deluxetable*}{ccrrrrrrr}}{\begin{deluxetable}{ccrrrrrrr}}
\tabletypesize{\scriptsize}
\tablewidth{0pc}
\tablecaption{Coefficients for transformations from USNO-B1.0 $B_U$ and $R_U$ magnitudes to $V$ (eq.~\ref{eq:BURUVtrans}).}
\tablehead{
\colhead{$B_{U}$} &
\colhead{$R_{U}$} &
\colhead{$a$} &
\colhead{$b$} &
\colhead{$c$} &
\colhead{$d$} &
\colhead{$e$} &
\colhead{$f$} &
\colhead{RMS [mag]}
}
\startdata
1 & 1 & $-0.76 \pm 0.19$ & $-0.02 \pm 0.01$ & $1.71 \pm 0.15$ & $-0.14 \pm 0.01$ & $0.15 \pm 0.02$ & $1.63 \pm 0.52$ & 0.27 \\
1 & 2 & $0.72 \pm 0.08$ & $-0.061 \pm 0.002$ & $0.38 \pm 0.06$ & $-0.057 \pm 0.006$ & $0.114 \pm 0.007$ & $-0.89 \pm 0.26$ & 0.19 \\
2 & 1 & $0.85 \pm 0.09$ & $-0.045 \pm 0.006$ & $0.44 \pm 0.08$ & $-0.051 \pm 0.003$ & $0.079 \pm 0.008$ & $-1.52 \pm 0.24$ & 0.22 \\
2 & 2 & $0.73 \pm 0.14$ & $-0.164 \pm 0.008$ & $0.46 \pm 0.13$ & $-0.20 \pm 0.01$ & $0.35 \pm 0.02$ & $-0.75 \pm 0.26$ & 0.19 \\
\enddata
\label{tab:fitparams}
\ifthenelse{\boolean{emulateapj}}{\end{deluxetable*}}{\end{deluxetable}}

\section{Selection of Variable Stars}\label{sec:selection}

To search for periodic variations we use the harmonic-fitting Analysis
of Variance periodogram \citep[AovHarm;][]{SchwarzenbergCzerny.96} and
the Box-Least-Squares \citep[BLS;][]{Kovacs.02} algorithm as
implemented in the VARTOOLS program\footnote{The VARTOOLS program is
  available at http://www.cfa.harvard.edu/\~{}jhartman/vartools/}
\citep{Hartman.08}. We also conduct a search for flare-like events in
the light curves. We apply the AoVHarm and BLS algorithms to both the
EPD and TFA light curves of sources, when available. For the
flare-search we use only the EPD light curves because the
$\sigma$-clipping applied to the TFA light curves may remove real
flares.

In addition to the searches mentioned above, we also tested the
phase-binning AoV periodogram \citep[AoV;][]{SchwarzenbergCzerny.89},
and the Discrete Auto-Correlation Function \citep[DACF;][]{Edelson.88}
to search for periodic and quasi-periodic signals. We found that both
of these techniques yield a substantial number of false alarm
detections that must then be culled by eye (for the DACF technique
more than half of the selected potential variables were determined to
be false alarms during a visual inspection of the light curves), and
only a relatively small number of detections that are visually judged
to be robust and are not also selected by the AoVHarm technique. We
therefore do not give further consideration to these methods. 

In the following subsections we
discuss the AoVHarm, BLS and flare searches in turn. The resulting
catalog of variable stars is presented in appendix~\ref{sec:cat}.

\subsection{Harmonic AoV}

We run the AoVHarm algorithm \citep{SchwarzenbergCzerny.96} on the
full sample of stars. We first apply an iterative $5\sigma$ clipping
to the light curve before searching for periods between $0.1$ and
$100~{\rm days}$. We run the algorithm using a sinusoid model with no
higher harmonics (it is thus comparable to DFT methods, or to the
popular Lomb-Scargle technique; \citealp{Lomb.76,Scargle.82}), and
generate the periodogram at a frequency resolution of $0.1 / T$ where
$T$ is the total time-span covered by a given light curve. We then
determine the peak at 10 times higher resolution. As our figure of
merit we use the signal-to-noise ratio (S/N), with an iterative
$5\sigma$ clipping applied to the periodogram in calculating the noise
(the RMS of the periodogram). 

\begin{figure}[!ht]
\ifthenelse{\boolean{emulateapj}}{\epsscale{1.2}}{\epsscale{0.7}}
\plotone{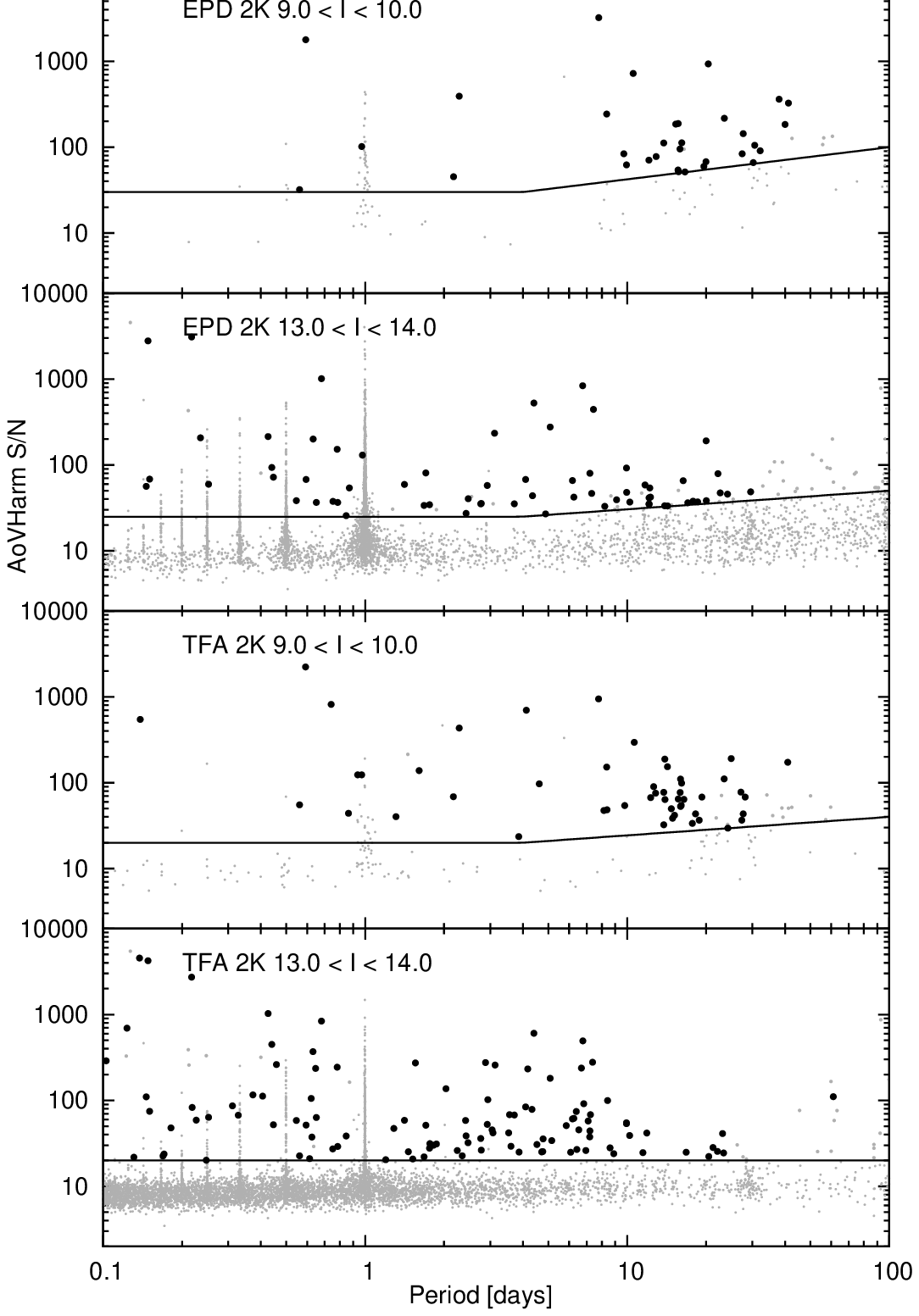}
\caption{Period vs. S/N from AoVHarm for 4 representative light curve
  groups.
  We show the light curves that pass the automatic selection and are flagged
  as reliable detections during the by-eye inspection (dark filled
  points) and the light curves that do not pass the selection or are
  flagged as unreliable detections (grey filled points) separately. 
  The lines show the adopted S/N
  cut-off as a function of period. Note that in addition to the
  cut-off shown with the line, we also automatically reject light
  curves for which the peak period is close to one sidereal day or a
  harmonic of one sidereal day.}
\label{fig:AOVHarm_SNvsP}
\end{figure}

To select the variables we group the light curves by the
post-processing applied (EPD or TFA), the CCD format (2K or 4K) and
the magnitude of the source. We adopt a separate selection threshold
on S/N for each group.
The thresholds have the form
\begin{equation}
{\rm S/N}_{\rm min} = \left\{ \begin{array}{ll}
{\rm S/N}_{0} & \mbox{if $P < P_{0}$} \\
{\rm S/N}_{0}\left( P/P_{0} \right) ^{\alpha} & \mbox{if $P \geq P_{0}$}
\end{array}
\right..
\end{equation}
The adopted values of S/N$_{0}$ range from 20-40, while the adopted
values of $\alpha$ range from 0 to $0.5$. We use $P_{0} = 4~{\rm
  days}$ for all groups. To establish the false alarm probability
associated with these S/N thresholds, we simulated white noise light
curves with time sampling drawn randomly from the HATNet light curves,
calculate the AoVHarm periodogram for each simulation, and determine
the S/N of the highest peak in each periodogram using the same
procedure as for the real light curves. We find that the false alarm
probability (FAP) as a function of S/N is well fitted by a function of
the form:
\begin{equation}
{\rm FAP} = 1 - \left[ 1 - e^{-0.90({\rm S/N}+1)} \right] ^{3350}
\end{equation}
A cut-off at S/N$ > 20$ thus corresponds to a FAP of $\sim 2 \times
10^{-5}$ while a cut-off at S/N$ > 40$ corresponds to a FAP of $\sim 4
\times 10^{-13}$. The expected total number of false alarms for our
sample of stars is thus fewer than 1. Figure~\ref{fig:AOVHarm_SNvsP}
shows the AoVHarm S/N as a function of the peak period for various
light curve groups. The adopted thresholds increase as a function of
period to account for temporally correlated systematic errors in the
photometry. In addition to this selection we also reject detections
with periods near 1 sidereal day, one of its harmonics, or other
periods which appear as spikes in the histogram of detected periods
(the latter includes periods between 5.71 and 5.80 days). For
composite light curves that contain both 2K and 4K observations we
take
\begin{equation}
{\rm S/N}_{\rm min} = f_{\rm 2K}{\rm S/N}_{\rm min,2K} + f_{\rm 4K}{\rm S/N}_{\rm min,4K}
\label{eqn:aovsnmin}
\end{equation}
where $f_{\rm 2K}$ and $f_{\rm 4K}$ are the fraction of points in the light
curve that come from 2K and 4K observations, and S/N$_{\rm min,2K}$
and S/N$_{\rm min,4K}$ are the 2K and 4K thresholds at the period of
the composite light curve.

Our selection
threshold passes a total of \NUMaovharmepdselect{} EPD light curves, and
\NUMaovharmtfaselect{} TFA light curves. These are inspected by eye to
reject obvious false alarms and to identify EBs. There are
\NUMaovharmepdpass{} EPD light curves that we judge to show clear,
continuous, periodic variability, \NUMaovharmepdEB{} that show eclipses,
\NUMaovharmepdquest{} that we consider to be questionable (these are
included in the catalog, but flagged as questionable), and
\NUMaovharmepdrej{} that we reject. For the TFA light curves the
respective numbers are \NUMaovharmtfapass{}, \NUMaovharmtfaEB{},
\NUMaovharmtfaquest{}, and \NUMaovharmtfarej{}. Note that the distinction
between ``clear'' variability and ``questionable'' cases is fairly
subjective. Generally we require the variations to be obvious to the
eye for periods $\ga$ 30 days, for shorter periods we consider the
selection to be questionable if it appears that the variability
selection may be due to enhanced scatter on a few nights.

\subsection{BLS - Search for Eclipses}
\begin{figure}[!ht]
\ifthenelse{\boolean{emulateapj}}{\epsscale{1.2}}{\epsscale{0.7}}
\plotone{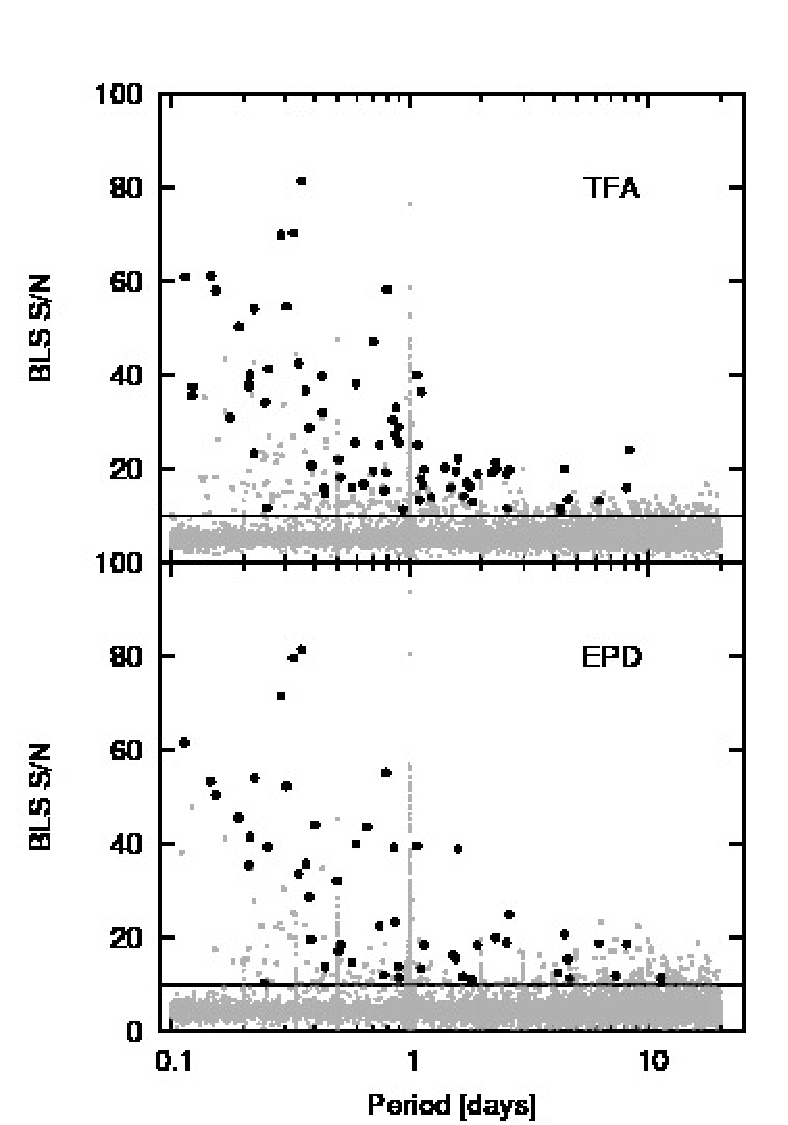}
\caption{Period vs. S/N from BLS for EPD and TFA light curves. The
  dark points show light curves that pass the ${\rm S/N} > 10.$ cut, do not
  have a period near a sidereal day or one of its harmonics, and are
  selected as eclipsing binaries during the by eye inspection. The
  grey points show all other light curves.}
\label{fig:BLS_SNvsP}
\end{figure}

The BLS algorithm, primarily used in searches for transiting planets,
detects periodic box-like dips in a light curve. This algorithm may be
more sensitive to detached binaries with sharp-featured light curves
than the other methods used. For our implementation of the BLS
algorithm we search 9,000 evenly spaced frequency points covering a
period range of 0.1 to 1.0 days and 100,000 evenly spaced frequency
points covering a period range of 1.0 to 20.0 days. At each trial
frequency we bin the phased light curve into 200 bins, and search over
fractional eclipse durations ranging from 0.01 to 0.1 in
phase. Figure~\ref{fig:BLS_SNvsP} shows the S/N vs the period for the
EPD and TFA light curves. We select \NUMblseitherselect{} stars with
${\rm S/N} > 10.0$ and with a period not close to 1 sidereal day or a
harmonic of a sidereal day as potentially eclipsing systems.
The selected light curves are inspected by eye to identify eclipsing
systems. A total of \NUMblseitherEB{} candidate EB systems are found
in this manner, \NUMblsepdEBonly{} are found in the EPD light curves
only, \NUMblstfaEBonly{} are found in the TFA light curves only, and
\NUMblsbothEB{} are found in both the EPD and TFA light curves.

\subsection{Search for Flares}\label{sec:findflares}

As noted in the introduction, flaring is a common phenomenon among K
and M dwarfs. While we were inspecting the light curves of candidate
variable stars we noticed a number of stars showing significant
flares. We therefore decided to conduct a systematic search for flare
events in the light curves. Most optical stellar flares show a very
steep rise typically lasting from a few seconds to several
minutes. \citet{Krautter.96} notes that flares can be divided into two
classes based on their decay times: ``impulsive'' flares have decay
times of a few minutes, to a few tens of minutes, while ``long-decay''
flares have decay times of up to a few hours. Due to the 5-minute
sampling of the HATNet light curves, flares of the former type will
only affect one or two observations in a light curve, unless the
peak-height is exceptionally large, while flares of the latter type
might affect tens of observations. In general it is very difficult to
determine whether a given outlier in a light curve is due to a flare
or bad photometry without inspecting the images from which an
observation was obtained. This is impractical to do for tens of
thousands of light curves when each light curve may contain tens to
hundreds of outliers. While observations that are potentially
corrupted are flagged, in practice the automated routines that
generate these flags do not catch all cases of bad photometry. We
therefore do not attempt to identify individual ``impulsive'' flares
affecting only a few points in the light curves. Long-decay flares, or
very high-amplitude impulsive flares, on the other hand, may be
searched for in an automated fashion if a functional form for the
decay is assumed \citep[this is similar to searching for microlensing
  events, see for example][]{Nataf.09}. To search for long-decay
flares we used the following algorithm:
\begin{enumerate}
\item Compute $m_{0}$, the median magnitude of the light curve, and
  $\delta_{0}$ the median deviation from the median.
\item Identify all sets of consecutive points with $m - m_{0} <
  -3\delta_{0}$. Let $t_{0}$ be the time of the brightest observation
  in a given set, and let $N$ be the number of consecutive points
  following and including $t_{0}$ with $m - m_{0} < -2\delta_{0}$. We
  proceed with the set if $N > 3$.
\item\label{step:fitflare} Use the Levenberg-Marquardt algorithm
  \citep{Marquardt.63} to fit to the $N$ points a function of the
  form:
\begin{equation}
m(t) = -2.5\log_{10}\left( A e^{-(t-t_{0})/\tau} + 1\right) + m_{1}
\label{eqn:flare}
\end{equation}
where $A$, $\tau$ and $m_{1}$ are the free parameters. Here $A$ is the
peak intensity of the flare relative to the non-flaring intensity,
$\tau$ is the decay timescale, and $m_{1}$ is the magnitude of the
star before the flare. For the initial values we take $m_{1} = m_{0}$,
$\tau = 0.02~{\rm days}$, and $A = 10^{-0.4(m_{p} - m_{0})} - 1$,
where $m_{p}$ is the magnitude at the peak.
\item\label{step:Ftestflare} Perform an F-test
  \citep[see][]{Lupton.93} on the statistic
\begin{equation}
f=\frac{(\chi^{2}_{N-1} - \chi^{2}_{N-3})/2}{\chi^{2}_{N-3}/(N-3)}
\label{eqn:ftest}
\end{equation}
  where $\chi^{2}_{N-1}$ is the $\chi^{2}$ value about the mean and
  $\chi^{2}_{N-3}$ is the $\chi^{2}$ value from fitting
  eq.~\ref{eqn:flare}. If the false alarm probability is greater than
  1\%, reject the candidate. If not, increase the number of points by
  one and repeat step~\ref{step:fitflare}. Continue as long as the
  false alarm probability decreases.
\item We reject any flare candidate for which there are at least two
  other candidate flares from light curves in the same field that
  occur within 0.1 days of the flare candidate.
\item Let the number of points with $t< t_{0}$ and $t_{0} - t <
  0.05~{\rm days}$ be $N_{\rm before}$ and the number of points with
  $t > t_{1}$ and $t - t_{1} < 0.05~{\rm days}$ be $N_{\rm
    after}$. Here $t_{1}$ is the time of the last observation included
  in the fit. Reject the candidate flare if $N_{\rm before} < 2$ or
  $N_{\rm after} < 2$. Also reject the candidate flare if $A < 0.$, $A
  > 10.0$, $\tau < 0.001~{\rm days}$, $\tau > 0.5~{\rm days}$, $A <
  \sigma_{A}$, $\tau < \sigma_{\tau}$, or if the false alarm
  probability from step~\ref{step:Ftestflare} is greater than
  0.1\%. Here $\sigma_{A}$ and $\sigma_{\tau}$ are the formal
  uncertainties on $A$ and $\tau$ respectively. The selection on $A$
  is used to reject numerous light curves with significant outliers
  which appear to be due to artifacts in the data rather than flares.
\end{enumerate}

\begin{figure}[]
\ifthenelse{\boolean{emulateapj}}{\epsscale{1.2}}{\epsscale{1.0}}
\plotone{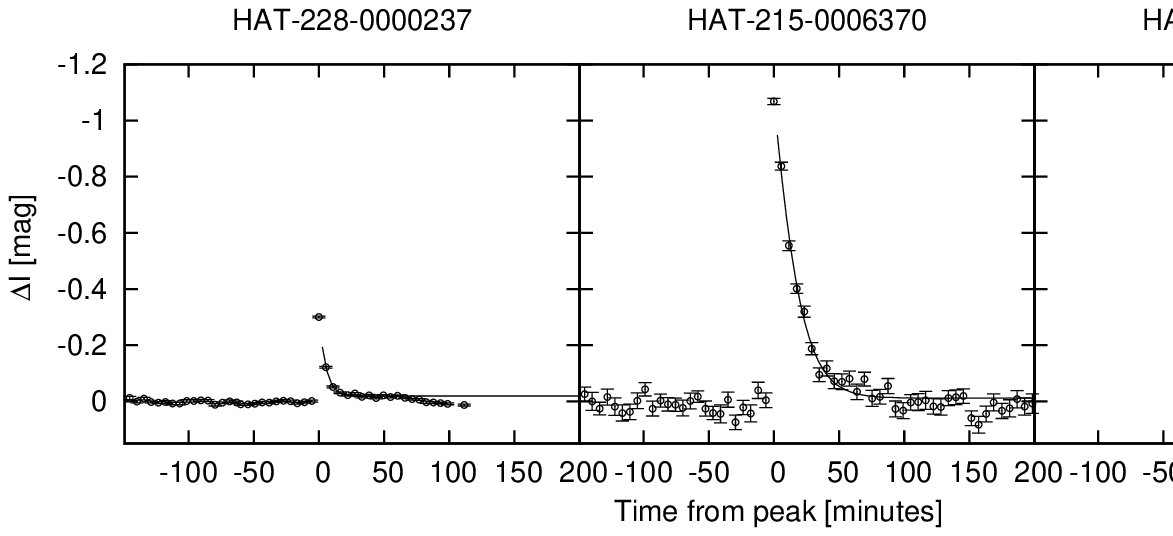}
\caption{Three examples of flares seen in the HATNet light curves. In
  each case the solid line shows the fit of eq.~\ref{eqn:flare} to the
  light curve.}
\label{fig:exampleflares}
\end{figure}

We apply the above algorithm to the non-composite EPD light curves
(i.e.~light curves from each field are processed independently for
stars with light curves from multiple fields). The algorithm is
applied both on the raw EPD light curves, and on EPD light curves that
are high-pass filtered by subtracting from each point the median of
all points that are within 0.1 day of that point. There are a total of
\NUMflarelcsearch{} stars with EPD light curves that are analyzed, we
exclude from the analysis \NUMflarelcnotsearch{} stars from the full
sample for which only $\sigma$-clipped TFA light curves are
available. A total of \NUMflareeventautoselect{} candidate flare
events from \NUMflarestarautoselect{} stars are selected. These are
inspected by eye to yield the final sample of \NUMflareevents{} flare
events from \NUMflarestars{} stars. During the visual inspection we
improved the automated fit for some of the flares by manually
adjusting the range of points used in the
fit. Figure~\ref{fig:exampleflares} shows two examples of these
large-amplitude, long-decay flares. The identified flares have peak
intensities that range from $A = 0.07$ to $A = 4.21$ and decay
time-scales that range from $\tau \sim 5$~minutes to $\tau \sim 1$~hour.

\subsection{Independent Search for Periodic Variables in a Subset of the K/M Dwarf Sample}\label{sec:gksearch}

As a check on the reliability of our variable star detections, an
independent selection of variables was conducted on a subset of the
K/M dwarf sample. A total of 3187 light curves were analyzed from 7
different HATNet fields using the DFT-based method described by
\citet{Szulagyi.09}. The light curves were cleaned of trends using 700
TFA template light curves for each field, and a polynomial is fit and
subtracted from the DFT power spectrum for each light curve to further
correct for red noise. Light curves with S/N$> 8$ were selected as
variables. An iterative search for additional frequencies of
variability was conducted on each of these light curves by
successively fitting a Fourier series to the light curve and
calculating the DFT power spectrum of the residual. Two examples of
multiperiodic variables identified in this way are discussed in
Section~\ref{sec:multiperiod}. A total of 4.6\% of the analyzed light
curves were selected by this method, 80\% of these were also selected
as variables by the AoVHarm method. On the other hand, 40\% of the
variables selected by AoVHarm applied to the TFA light curves from
these fields were not selected by DFT. There are several factors which
contribute to these differences, especially important are differences
in the number of TFA templates used, in the adopted selection
thresholds, in filtering applied to the power spectrum, and in the use
of composite light curves (the DFT analysis did not use composite
light curves for stars observed in more than one field). In general,
TFA light curves with AoVHarm S/N$ \ga 100$ are selected by both the
DFT and AoVHarm methods.

\section{Variability Blending}\label{sec:blend}

While the wide FOV of the HATNet telescopes allows a significant
number of bright stars to be simultaneously observed, the downside to
this design is that the pixel scale is necessarily large, so a given
light curve often includes flux contributions from many
stars. Blending is a particularly significant issue in high stellar
density fields near the Galactic plane. A star blended with a nearby
variable star may be incorrectly identified as a variable based on its
light curve. If the stars are separated by more than a pixel or two,
it may be possible to distinguish the real variable from the blend by
comparing the amplitudes of their light curves. However, because
photometry is only obtained for stars down to a limiting magnitude
(the value used varies from field to field), in many cases we do not
have light curves for all the faint neighbors near a given candidate
variable star, so we cannot easily determine which star is the true
variable. In these cases we can still give an indication of whether or
not a candidate is likely to be a blend by determining the expected
flux contribution from all neighboring stars to the candidate's light
curve.

\subsection{Blending From Other Stars With Light Curves}\label{sec:blendwithlc}

To determine whether or not a candidate is blended with a nearby
variable star that has a light curve, we measure the peak-to-peak
light curve amplitude (in flux) of all stars within 2$\arcmin$ of the
candidate. If any neighbor has an amplitude that is greater than twice
the flux amplitude of the candidate, the candidate is flagged as a
probable blend. If any neighbor has an amplitude that is between half
and twice the flux amplitude of the candidate, the candidate is
flagged as a potential blend. If any neighbor has an amplitude that is
between 10\% and half the flux amplitude of the candidate, the
candidate is flagged as an unlikely blend. And finally we flag the
candidate as a non-blend if all neighbors have amplitudes that are
less than 10\% that of the candidate. We determine the amplitude of a
light curve by fitting to it 10 different Fourier series of the form:
\begin{equation}
m(t) = m_{0} + \sum_{i=1}^{N}a_{i}\sin \left( 2\pi it/P + \phi_{i}\right)
\end{equation}
with $N$ ranging from 1 to 10. Here $P$ is the period of the light
curve. We perform an F-test to determine the significance of each fit
relative to fitting a constant function to the light curve, and choose
the amplitude of the Fourier series with the lowest false alarm
probability. If the lowest false alarm probability is greater than
10\% we set the amplitude to zero. We try all periods identified for
each candidate by the variability searches described in
\S~\ref{sec:selection}, and adopt the largest amplitude found. For
candidates that have light curves from multiple fields, or that have
both ISM and AP reductions, we do the amplitude test on each separate
field/reduction and adopt the most significant blending flag found for
the candidate. If the amplitude of the candidate variable star is set
to zero for a given field/reduction we do not use that field/reduction
in determining the blending flag. We use the EPD
light curves in doing this test.

\subsection{Blending From Faint Stars Without Light Curves}\label{sec:blendnolc}

To determine whether or not a candidate is potentially blended with a
nearby faint variable star that does not have a light curve, we
compare the observed amplitude of the candidate to its expected
amplitude if a neighboring star were variable with an intrinsic
amplitude of $1.0~{\rm mag}$. We assume that an amplitude of $1.0~{\rm
  mag}$ is roughly the maximum value that one might expect for a short
period variable star. If the measured amplitude is less than the
expected amplitude then we flag the candidate as a potential blend, if
it is greater than the expected amplitude and less than twice the
expected amplitude we flag the candidate as an unlikely blend, and if
it is greater than twice the expected amplitude, we flag it as a
non-blend. The test is done for all stars in the catalog (see the
following subsection) within 2$\arcmin$ of the candidate that do not
have a light curve, and we adopt the most significant blending flag
found for the candidate. To determine the expected amplitude of the
candidate star induced by the neighbor, we note that a star with
magnitude $m_{1}$ located near a variable star with magnitude $m_{2}$
and amplitude $\Delta m_{2} > 0$, has an expected light curve
amplitude that is given by
\begin{eqnarray}
\lefteqn{\Delta m_{1,AP} = 2.5\log_{10} \left[ f_{1}10^{-0.4m_{1}} + f_{2}10^{-0.4(m_{2} - \Delta m_{2})} \right]} \nonumber\\
&&\mbox{} - 2.5\log_{10} \left[ f_{1}10^{-0.4m_{1}} + f_{2}10^{-0.4m_{2}} \right] \hspace{0.9in}
\label{eqn:blendAP}
\end{eqnarray}
for the case of aperture photometry, and by
\begin{eqnarray}
\lefteqn{\Delta m_{1,ISM} = } \nonumber \\
&&2.5\log_{10} \left[ f_{1}10^{-0.4m_{1}} + f_{2}10^{-0.4m_{2}}\left( 10^{0.4\Delta m_{2}} - 1\right) \right] \nonumber \\
&&\mbox{} - 2.5\log_{10} \left[ f_{1}10^{-0.4m_{1}} \right] \hspace{2.0in}
\label{eqn:blendISM}
\end{eqnarray}
for the case of image subtraction photometry. The two expressions
differ because in the ISM pipeline photometry is done on difference
images (only differential flux is summed in the aperture), whereas in
the AP pipeline photometry is done directly on the science images (all
stellar flux is summed in the aperture). Here $f_{1,2}$ is the
fraction of the flux from star 1(2) that falls within the aperture,
and we use the catalog values (transformed in the case of 2MASS) for
$m_{1}$ and $m_{2}$. To determine $f_{1}$ and $f_{2}$ we integrate the
intersection between the circular aperture and a Gaussian PSF which we
assume to have a FWHM of 2 pixels (this is a typical effective
``seeing'' for both the 2K and 4K images), we do not consider
pixelation effects in making this estimate.

\subsection{Blending Results}

We compare the results from the two blending tests for each candidate,
and adopt the most significant blending flag from among the two tests
for the catalog. We do not run the test on the candidate flare stars,
unless the star was selected as a variable by another method as
well. 

We carried out this blending analysis twice: once using 2MASS for
estimating the magnitudes of stars and identifying potentially
contaminating neighbors, and a second time using the Sloan Digital Sky
Survey Data Release 7 \citep[SDSS DR7;][]{Abazajian.09}. The latter
survey is deeper than 2MASS and also provides magnitudes in filters
that are closer to those used by HATNet, however unlike 2MASS, SDSS
DR7 does not provide complete coverage of the HATNet fields (43\% of
our stars have SDSS DR7 observations). We find that in the majority of
cases (53\%) the blending flags are the same when using SDSS or 2MASS
to estimate the magnitudes of stars in the HATNet filters and to
identify faint neighbors for which we have not obtained light
curves. However, for a signficant fraction (42\%) the blending flag is
higher when using SDSS than when using 2MASS. Typically the blending
flag from SDSS is greater by one than the blending flag from 2MASS. We
find that 6\% of the objects are flagged as probable blends when using
SDSS but unlikely blends when using 2MASS. To combine the results from
2MASS and SDSS for the analysis below we adopt the greater of the two
blending flags, except for when there is no SDSS coverage, in which
case we adopt the 2MASS blending flag. We report the 2MASS and SDSS
blending flags separately in our catalogs.

Out of the \NUMinvarsorEBcatalog{} stars that are selected as
potential periodic variables, \NUMinvarsorEBcatalogunblended{} are
flagged as unblended, \NUMinvarsorEBcatalogunlikelyblend{} are flagged
as unlikely blends, \NUMinvarsorEBcatalogpotentialblend{} are flagged
as potential blends, and \NUMinvarsorEBcatalogprobableblend{} are
flagged as probable blends.

\begin{figure}[!ht]
\ifthenelse{\boolean{emulateapj}}{\epsscale{1.2}}{\epsscale{0.6}}
\plotone{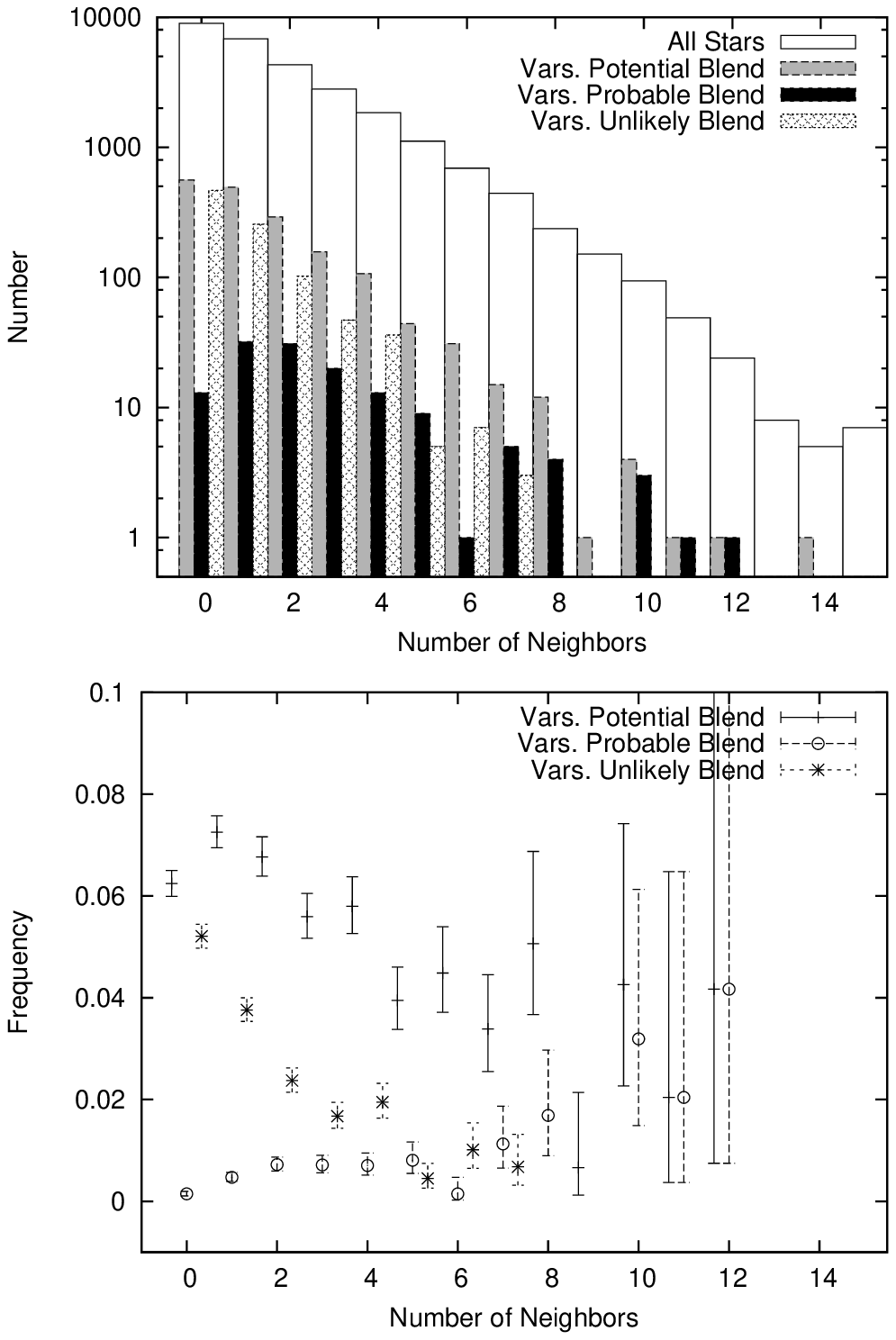}
\caption{Top: The distribution of the number of stars within
  $30\arcsec$ of a given source shown for all stars in the sample, and
  for variable stars with different blending classifications. For the
  potential and unlikely blend groups we include all stars with that
  blending classification or a lesser blending classification. For the
  probable blend group we include only stars classified as probable
  blends. We only include variables classified by eye as reliable
  detections. Bottom: The fraction of all stars in the sample selected
  as variables using the same groups as in the top panel. The
  errorbars show the standard $1\sigma$ uncertainties from the
  binomial sampling distribution. The relative frequency of
  variability for stars not classified as probable blends does not
  increase for more crowded regions (it decreases due to increased
  photometric noise in more crowded regions), while the relative
  frequency of probable blends does increase for more crowded
  regions. This indicates that the blending classification scheme is
  consistent.}
\label{fig:VarFreqvsNumNeighbors}
\end{figure}

As a consistency check on the blending classification scheme, we show
in figure~\ref{fig:VarFreqvsNumNeighbors} the fraction of stars
detected as variable as a function of the number of 2MASS sources
within $30\arcsec$ of the star. The results are shown separately for
stars with different blending classifications. The frequency of
variability for stars not classified as probable blends does not
increase with the number of neighbors, which shows the consistency of
this method.

\section{Match to Other Catalogs}\label{sec:match}

\subsection{Match to Other Variable Star Surveys}\label{sec:varmatch}

We match all \NUMtotpotentialvar{} stars selected as potential
variables to the combined General Catalogue of Variable Stars
\citep[GCVS;][]{Samus.06}, the New Catalogue of Suspected Variable
Stars \citep[NSV;][]{Kholopov.82} and its supplement
\citep[NSVS;][]{Kazarovets.98}\footnote{The GCVS, NSV and NSVS were
  obtained from http://www.sai.msu.su/groups/cluster/gcvs/gcvs/ on
  2009 April 7}. We also match to the ROTSE catalog of variable stars
\citep{Akerlof.00}, to the ALL Sky Automated Survey Catalogue of
Variable Stars \citep[ACVS;][]{Pojmanski.02}\footnote{Version 1.1
  obtained from http://www.astrouw.edu.pl/asas/?page=catalogues}, and
to the Super-WASP catalogue of periodic variables coincident with
ROSAT X-ray sources \citep{Norton.07}. In all cases we use a
$2\arcmin$ matching radius. We use a large matching radius to include
matches to known variables that may be blended with stars in our
sample. In total \NUMcross{} of our candidate variables lie within
$2\arcmin$ of a source in one of these catalogs, meaning that
\NUMnotcross{} are new identifications. This includes \NUMmatchgcvs{}
that match to a source in the GCVS, \NUMmatchnsv{} that match to a
source in the NSV, \NUMmatchnsvs{} that match to a source in the NSVS,
\NUMmatchasas{} that match to a source in the ACVS, \NUMmatchrotse{}
that match to a source in the ROTSE catalog (\NUMmatchrotsegcvs{} of
which are in the GCVS as well), and \NUMmatchswasp{} that match to a
Super-WASP source (\NUMmatchswaspone{} of these are in their catalogue
of previously identified variables). Two of the \NUMmatchgcvs{}
candidate variables that match to a source in the GCVS
(HAT-215-0001451 and HAT-215-0001491) actually match to the same
source, V1097~Tau, a weak emission-line T~Tauri star. Both stars are
flagged as probable blends in our catalog, in this case
HAT-215-0001491 is the correct variable while HAT-215-0001451 is the
blend.

We inspect each of the \NUMcross{} candidates with a potential match
and find that the match is incorrect for \NUMcrosswrong{} of them and
correct for \NUMcrosscorrect{}. For \NUMcrosswrongprobableblend{} of
the \NUMcrosswrong{} incorrect matches the candidate variable is
flagged as a probable blend in our catalog. In
\NUMcrosswrongpotentialblend{} cases the candidate variable is flagged
as a potential blend, in \NUMcrosswrongunlikelyblend{} cases it is
flagged as an unlikely blend, and in \NUMcrosswrongunblended{} cases it is
flagged as unblended. The match appears to be correct
for three of the candidate variables flagged as probable blends. In
addition to HAT-215-0001491, the stars HAT-239-0000221 and
HAT-239-0000513 both match correctly to sources in the GCVS. These
stars form a common proper motion, low mass binary system. Both stars
are flagged as probable blends in our catalog. Each matches
separately, and correctly, to a flare star in the GCVS (V0647~Her and
V0639~Her respectively).

The \NUMmatchgcvscorrect{} variables that match correctly to a source
in the GCVS include 4 BY Draconis-type rotational variables, 6 UV
Ceti-type flare stares, 1 INT class Orion variable of the T Tauri
type, and 5 eclipsing systems. The EBs include the two W UMa-type
contact systems DY CVn and V1104 Her, the two Algol-type systems DK
CVn and V1001 Cas, and the M3V/white dwarf EB DE CVn
\citep[][]{VanDenBesselaar.07}.

Table~\ref{tab:cross}, at the end of the paper, lists the first ten
cross-identifications, the full table is available electronically with
the rest of the catalog.

\subsection{Match to ROSAT}\label{sec:xray}

We match all \NUMtotpotentialvar{} stars selected as potential
variables to the ROSAT All-sky survey source catalog \citep{Voges.99}
using the US National Virtual Observatory catalog matching
utilities. We use a 3.5$\sigma$ positional matching criterion. A total
of \NUMmatchrosat{} of the variables match to an X-ray source,
including \NUMmatchrosatvars{} stars in our catalog of noneclipsing
periodic variables, \NUMmatchrosatEB{} of the EBs, and
\NUMmatchrosatflares{} of the \NUMflarestars{} flare stars. A few of
the variable stars are close neighbors where one is likely to be a
blend of the other, so there are \NUMmatchrosatdistinct{} distinct
X-ray sources that are matched to. Table~\ref{tab:xray} at the end of
the paper gives the cross-matches. The full table is available
electronically with the rest of the catalog. In Section~\ref{sec:rot}
we discuss the X-ray properties of the rotational variables.

\section{Discussion}\label{sec:discussion}

\subsection{Eclipsing Binaries}\label{sec:eb}

\begin{figure}[!ht]
\ifthenelse{\boolean{emulateapj}}{\epsscale{1.2}}{\epsscale{1.0}}
\plotone{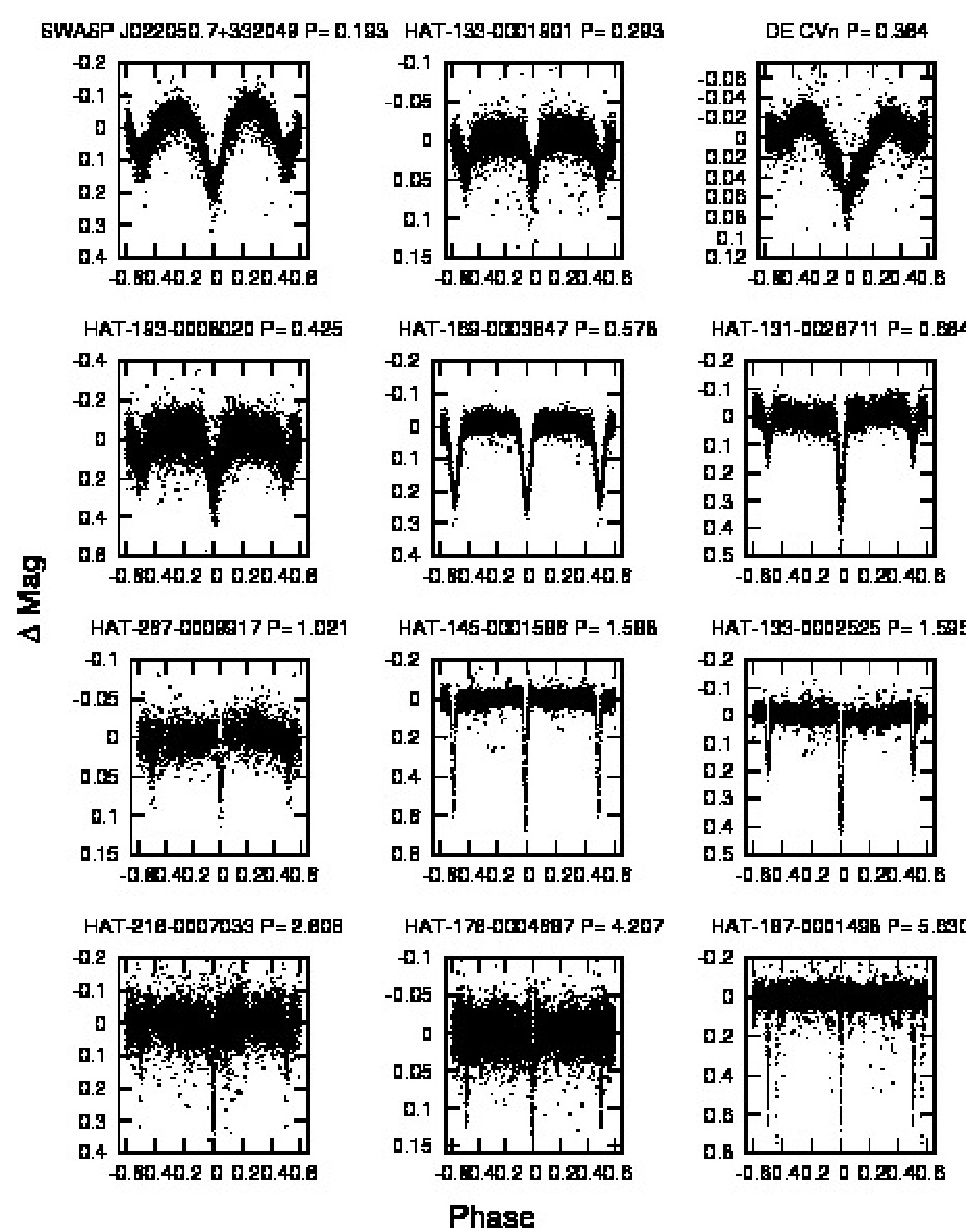}
\caption{Example phased light curves for 12 of the \NUMEB{} potential EBs found in the survey. The period listed is in days.}
\label{fig:exampleEBs}
\end{figure}

The \NUMEB{} stars that we identify as potential EBs have periods
ranging from $P = \EBminimumperiod{}~{\rm days}$ to
$P=\EBmaximumperiod{}~{\rm days}$. We flag \NUMEBprobableblend{} of
the candidate EBs as probable blends, \NUMEBpotentialblend{} as
potential blends, \NUMEBunlikelyblend{} as unlikely blends, and
\NUMEBunblended{} as unblended. Figure~\ref{fig:exampleEBs} shows
phased light curves for 12 of the EBs.

In addition to matching the candidate EBs to other variable star
catalogs (\S~\ref{sec:varmatch}) and to ROSAT (\S~\ref{sec:xray}) we
also checked for matches to previously studied objects using
SIMBAD. The following objects had noteworthy matches:
\begin{enumerate}
\item \emph{HAT-148-0000574}: matches to the X-ray source 1RXS
  J154727.5+450803, and was previously discovered to be an SB2 system
  by \citet{Mochnacki.02}, we discuss this system in detail in
  \S~\ref{sec:RXJ1547}.
\item \emph{HAT-216-0002918} and \emph{HAT-216-0003316}: match to CCDM
  J04404+3127A and CCDM J04404+3127B, respectively, which form a
  common proper-motion 15$\arcsec$ binary system. The two stars are
  both selected as candidate EBs with the same period, and are both
  flagged as potential blends in the catalog; based on a visual
  inspection of the light curves we conclude that the fainter
  component (HAT-216-0003316) is most likely the true $P=2.048~{\rm
    day}$ EB. The fainter component, which has spectral type M3 on
  SIMBAD, also matches to the X-ray source RX J0440.3+3126.
\item \emph{HAT-127-0008153}: matches to CCDM J03041+4203B, which is
  the fainter component in a common proper-motion $20\arcsec$ binary
  system. This star is flagged as a potential blend, the brighter
  component appears to match to the X-ray source 1RXS
  J030403.8+420319. Based on a visual inspection of the light curves
  we conclude that HAT-127-0008153 is likely the true variable.
\item \emph{HAT-169-0003847}: is $24\arcsec$ from the Super-WASP
  variable 1SWASP~J034433.95+395948.0, the two stars are blended in
  the HATNet images, however from a visual inspection of the light
  curves we conclude that HAT-169-0003847 is likely the true variable.
\item \emph{HAT-192-0001841}: is $46\arcsec$ from a high
  proper-motion, K0 star BD+41~2679. The two stars may be members of a
  common proper-motion binary system (the former has a proper motion
  of $65.79$, $-151.77~{\rm mas/yr}$ in RA and DEC respectively, while
  the latter has $89.76$, $-117.14~{\rm mas/yr}$).
\item \emph{HAT-169-0003847}: is flagged as an unlikely blend, and we
  confirm that BD+41~2679 is not the true variable.
\item \emph{HAT-193-0008020}: matches to GSC~03063-02208 and has a
  spectral type of M0e listed on SIMBAD \citep[see also][]{Mason.00}.
\item \emph{HAT-216-0007033}: matches to the X-ray source
  RX~J0436.1+2733 and has spectral type M4 listed on SIMBAD.
\item \emph{HAT-341-0019185}: is $43\arcsec$ from TYC~1097-291-1, the
  two stars appear to be members of a common proper motion binary
  system. HAT-341-0019185 is flagged as a probable blend in our
  catalog, though it does not appear that TYC~1097-291-1 is the real
  variable.
\end{enumerate}

\subsubsection{The Low-mass EB 1RXS~J154727.5+450803}\label{sec:RXJ1547}

The EB HAT-148-0000574 matches to 1RXS J154727.5+450803. Using RV
observations obtained with the Cassegrain spectrograph on the David
Dunlap Observatory (DDO) 1.88~m telescope\footnote{Based on data
  obtained at the David Dunlap Observatory, University of Toronto},
\citet{Mochnacki.02} found that this object is a $P=3.54997 \pm
0.00005~{\rm day}$ double-lined spectroscopic binary system with
component masses $\ga 0.26~M_{\odot}$. This system, however, was not
previously known to be eclipsing. Here we combine the published RV
curves from \citet{Mochnacki.02} with the HATNet I-band light curve to
provide preliminary estimates for the masses and radii of the
component stars.

\begin{figure}[!ht]
\ifthenelse{\boolean{emulateapj}}{\epsscale{1.2}}{\epsscale{0.8}}
\plotone{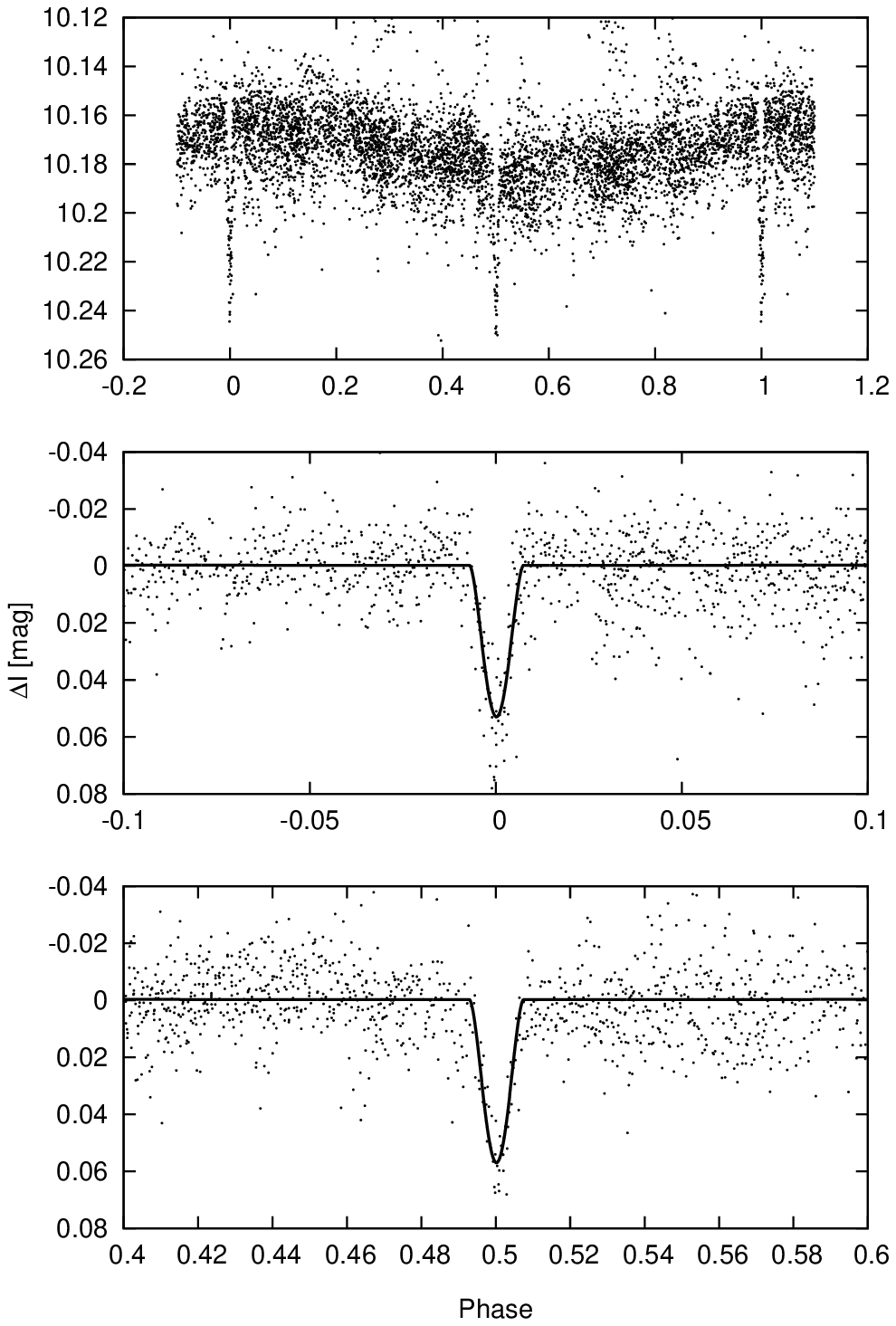}
\caption{Phased HATNet I-band light curve for the low-mass EB 1RXS J154727.5+450803. The top panel shows the full EPD light curve, the bottom two panels show a model fit to the two eclipses after subtracting a harmonic series fit to the out of eclipse portion of the light curve.}
\label{fig:RXJ1547lc}
\end{figure}

\begin{figure}[!ht]
\ifthenelse{\boolean{emulateapj}}{\epsscale{1.2}}{\epsscale{1.0}}
\plotone{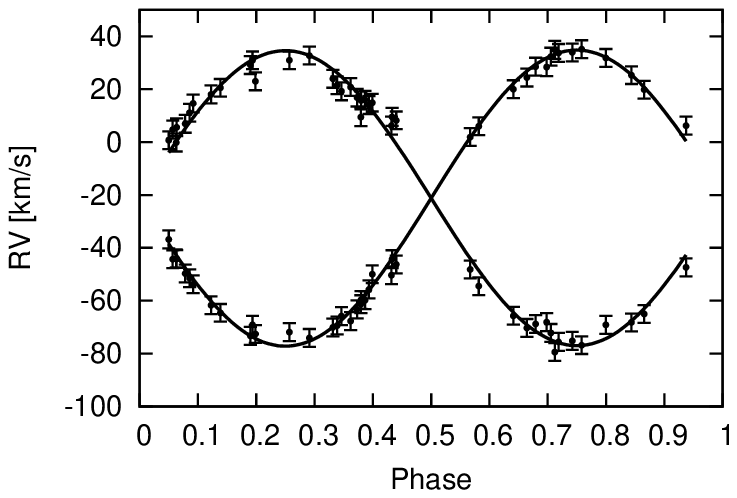}
\caption{Circular orbit fit to the RV curves for the low-mass EB 1RXS J154727.5+450803. The observations are taken from \citet{Mochnacki.02}.}
\label{fig:RXJ1547RV}
\end{figure}

Figure~\ref{fig:RXJ1547lc} shows the EPD HATNet light curve phased at
a period of $P = 3.550018$ days together with a model fit, while
figure~\ref{fig:RXJ1547RV} shows a fit to the radial velocity
observations taken from \citet{Mochnacki.02}. Note the out of eclipse
variations in the light curve, presumably due to spots on one or both
of the components, which indicates that the rotation period of one or
both of the stars is tidally locked to the orbital period. Since the
HATNet light curve is not of high enough quality to measure the radii
to better than a few percent precision, we do not fit a detailed spot
model to the light curve, and instead fit a harmonic series to the out
of eclipse observations and then subtract it from the full light
curve. We model the light curve using the JKTEBOP program
\citep{Southworth.04a,Southworth.04b} which is based on the Eclipsing
Binary Orbit Program \citep[EBOP;][]{Popper.81,Etzel.81,Nelson.72},
but includes more sophisticated minimization and error analysis
routines. We used the DEBiL program \citep{Devor.05} to measure the
eclipse minimum times from the light curve, which in turn were used
with the RV curves to determine the ephemeris. In modeling the RV
curves we fix $e = 0$ \citep[][found $e = 0.008 \pm 0.007$; with the
  new ephemeris the $2\sigma$ upper limit on the eccentricity is $e <
  0.04$]{Mochnacki.02}, and we fix $k = R_{2}/R_{1} = 1.0$ in
modelling the light curve given $q = 1.00 \pm 0.02$ from the fit to
the RV curves (the light curve is not precise enough to provide a
meaningful constraint on $k$, however the constraint on $(R_{1} +
R_{2})/a$ is robust). For completeness we note that we assumed
quadratic limb darkening coefficients of $a = 0.257$, $b = 0.586$ for
both stars \citep{Claret.00}, which are appropriate for a $T_{\rm eff}
= 3000~{\rm K}$, $\log g = 4.5$, solar metallicity star. The results
are insensitive to the adopted limb darkening coefficients; we also
performed the fit using the coefficients appropriate for a $T_{\rm
  eff} = 4000~{\rm K}$, $\log g = 4.5$ star and found negligible
differences in the resulting parameters and uncertainties. The
parameters for the system are given in
table~\ref{tab:RXJ1547param}. Note that the $1\sigma$ errors given on
the masses and radii are determined from a Monte Carlo simulation
\citep{Southworth.05}. These are likely to be overly optimistic given
the inaccurate treatment of the spots, and our assumption that the
component radii are equal.

\begin{deluxetable}{lr}
\tabletypesize{\scriptsize}
\tablewidth{0pc}
\tablecaption{Parameters for the EB 1RXS J154727.5+450803}
\ifthenelse{\boolean{emulateapj}}{}{\tablehead{
\colhead{Parameter} & \colhead{Value}
}}
\startdata
\cutinhead{Coordinates and Photometry}
RA (J2000) & 15:47:27.42\tablenotemark{a} \\
DEC (J2000) & +45:07:51.39\tablenotemark{a} \\
Proper Motions [mas/yr] & -259, 200\tablenotemark{b} \\
J & $9.082~{\rm mag}$\tablenotemark{c} \\
H & $8.463~{\rm mag}$\tablenotemark{c} \\
K & $8.215~{\rm mag}$\tablenotemark{c} \\
\cutinhead{Ephemerides}
P & $3.5500184 \pm 0.0000018~{\rm day}$ \\
HJD & $2451232.89534 \pm 0.00094$ \\
\cutinhead{Physical Parameters}
$M_{1}$ & $0.2576 \pm 0.0085~M_{\odot}$\\
$M_{2}$ & $0.2585 \pm 0.0080~M_{\odot}$\\
$R_{1}=R_{2}$ & $0.2895 \pm 0.0068~R_{\odot}$\\
\cutinhead{RV Fit Parameters}
$\gamma$ & $-21.21 \pm 0.41~{\rm km/s}$ \\
$K_{1}$ & $55.98 \pm 0.76~{\rm km/s}$ \\
$K_{2}$ & $55.78 \pm 0.83~{\rm km/s}$ \\
$e$ & 0.0\tablenotemark{d}\\
\cutinhead{LC Fit Parameters}
$J_{2}/J_{1}$ & $1.0734 \pm 0.030$ \\
$(R_{1} + R_{2})/a$ & $0.0737 \pm 0.0014$ \\
$R_{2}/R_{1}$ & 1.0\tablenotemark{e}\\
$i$ & $86.673^{\circ} \pm 0.068^{\circ}$ \\
\enddata
\tablenotetext{a}{SIMBAD}
\tablenotetext{b}{PPMX}
\tablenotetext{c}{2MASS}
\tablenotetext{d}{fixed}
\tablenotetext{e}{fixed based on $q = 1.004 \pm 0.020$ from fitting the RV curves.}
\label{tab:RXJ1547param}
\end{deluxetable}

\begin{figure}[!ht]
\ifthenelse{\boolean{emulateapj}}{\epsscale{1.2}}{\epsscale{1.0}}
\plotone{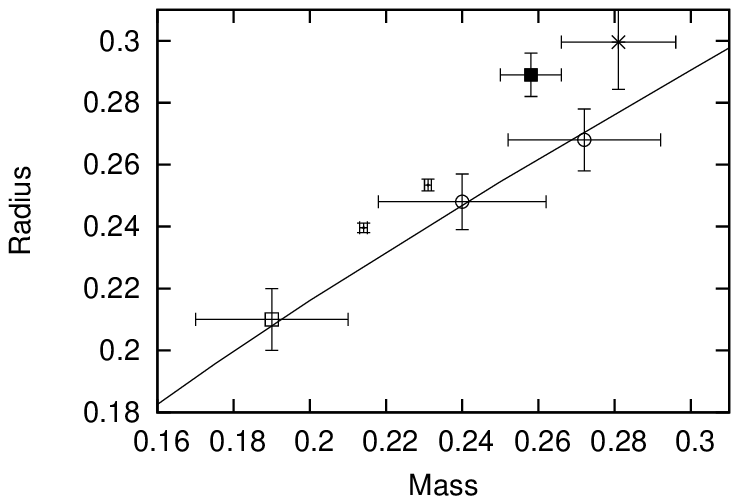}
\caption{Mass-radius relation for 7 main sequence stars in
  double-lined DEBs with $M < 0.3 M_{\odot}$. The two points with the
  smallest error bars are the components of CM Dra, the open circles
  are the components of SDSS-MEB-1, the open square is the secondary
  component of 2MASSJ04463285+190432, the X is the secondary component
  of GJ~3236, and the filled square marks the two components of 1RXS
  J154727.5+450803. The solid line shows the $1.0~{\rm Gyr}$, solar
  metallicity isochrone from \citet{Baraffe.98}. Note that the error
  bars for 1RXS~J154727.5+450803 do not incorporate systematic errors
  that may result from not properly modelling the spots or allowing
  the stars to have unequal radii.}
\label{fig:EBmodelcomp}
\end{figure}

Table~\ref{tab:otherEBS} lists the masses and radii of the 4 other
known double-lined detached EBs with at least one main sequence
component that has a mass less than $0.3 M_{\odot}$. We do not include
RR Cae which is a white-dwarf/M-dwarf EB that has presumably undergone
mass transfer \citep{Maxted.07}. In figure~\ref{fig:EBmodelcomp} we
plot the mass-radius relation for stars in the range $0.15 M_{\odot} <
M < 0.3 M_{\odot}$. Like the components of CM Dra and the secondary of
GJ~3236, and unlike the components of SDSS-MEB-1 and the secondary of
2MASSJ04463285+190432, the components of 1RXS J154727.5+450803 have
radii that are larger than predicted from the \citet{Baraffe.98}
isochrones (if the age is $\ga 200~{\rm Myr}$). The radii are $\sim
10$\% larger than the predicted radius in the $1.0~{\rm Gyr}$, solar
metallicity isochrone. High precision photometric and spectroscopic
follow-up observations, and a more sophisticated analysis of the data
are need to confirm this.

\ifthenelse{\boolean{emulateapj}}{\begin{deluxetable*}{lrrrrr}}{\begin{deluxetable}{lrrrrr}}
\tabletypesize{\scriptsize}
\tablewidth{0pc}
\tablecaption{Other Double-Lined EBs with a very late M dwarf component}
\tablehead{
\colhead{Name} &
\colhead{Period [days]} &
\colhead{$M_{1}$ [$M_{\odot}$]} &
\colhead{$M_{2}$ [$M_{\odot}$]} &
\colhead{$R_{1}$ [$R_{\odot}$]} &
\colhead{$R_{2}$ [$R_{\odot}$]}
}
\startdata
CM Dra & $1.27$ & $0.2310 \pm 0.0009$ & $0.2141 \pm 0.0010$ & $0.2534 \pm 0.0019$ & $0.2396 \pm 0.0015$ \\
SDSS-MEB-1 & $0.407$ & $0.272 \pm 0.020$ & $0.240 \pm 0.022$ & $0.268 \pm 0.010$ & $0.248 \pm 0.009$ \\
2MASSJ04463285+190432 & $0.619$ & $0.47 \pm 0.05$ & $0.19 \pm 0.02$ & $0.57 \pm 0.02$ & $0.21 \pm 0.01$ \\
GJ~3236 & $0.771$ & $0.376 \pm 0.016$ & $0.281 \pm 0.015$ & $0.3795 \pm 0.0084$ & $0.300 \pm 0.015$ \\
\enddata
\tablerefs{CM Dra: \citet{Morales.09}; \citet{Lacy.77}; \citet{Metcalfe.96}; SDSS-MEB-1: \citet{Blake.08}; 2MASSJ04463285+190432: \citet{Hebb.06}; GJ~3236: the parameters listed for this system are determined by giving equal weight to the three models in \citet{Irwin.09b}}
\label{tab:otherEBS}
\ifthenelse{\boolean{emulateapj}}{\end{deluxetable*}}{\end{deluxetable}}

\subsection{Rotational Variables}\label{sec:rot}

\begin{figure}[!ht]
\ifthenelse{\boolean{emulateapj}}{\epsscale{1.2}}{\epsscale{0.9}}
\plotone{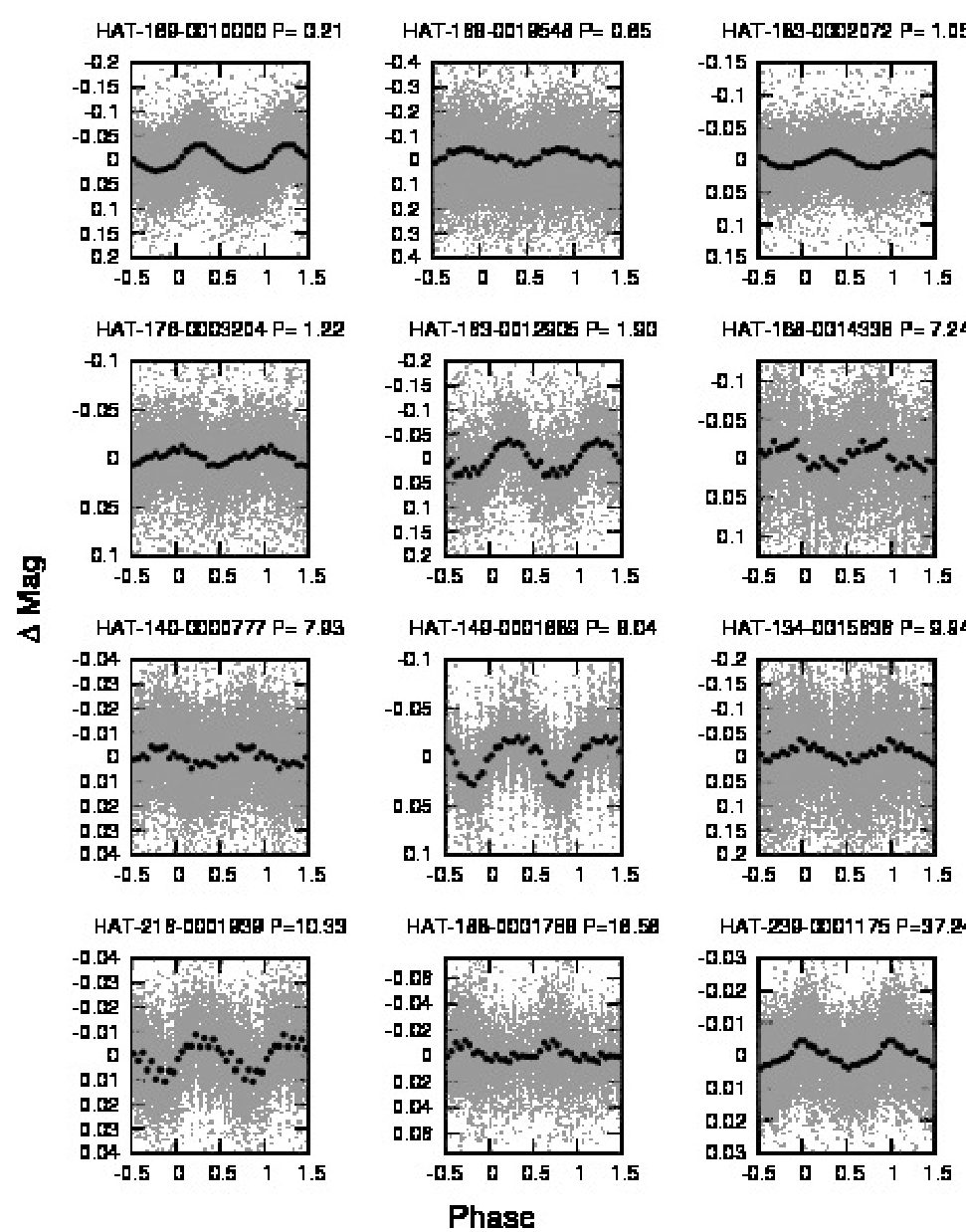}
\caption{Phased EPD light curves for 12 randomly selected rotational
  variables found in the survey. The grey-scale points show all points
  in the light curve, the dark points show the binned light curve. The
  period listed is in days. The zero-point phase is arbitrary. For
  this figure we use the period found by applying AoVHarm to the TFA
  light curves.}
\label{fig:exampleROTlcs}
\end{figure}

\begin{figure}[!ht]
\ifthenelse{\boolean{emulateapj}}{\epsscale{1.2}}{\epsscale{0.45}}
\plotone{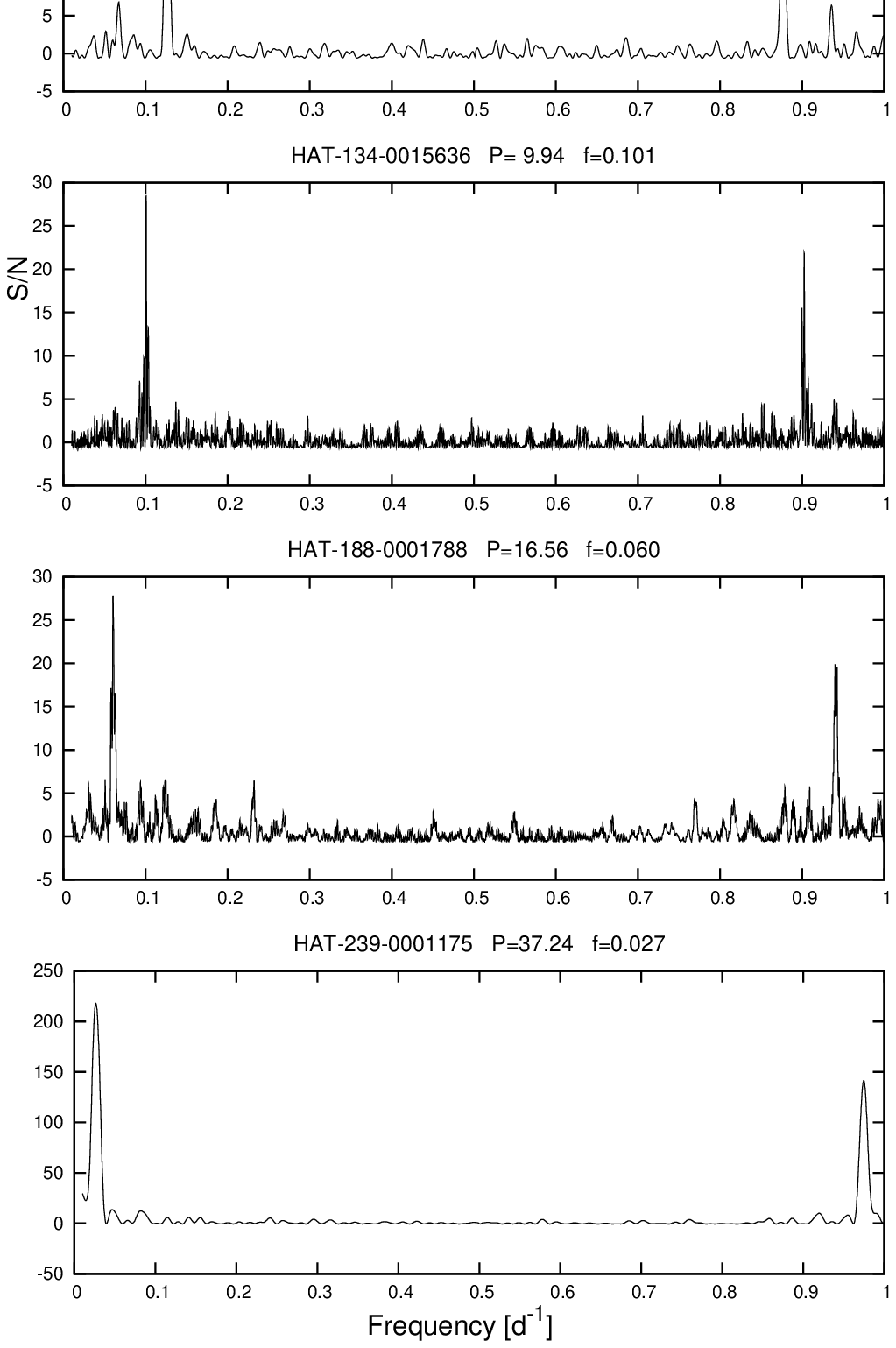}
\caption{AoVHarm periodograms for 5 of the example light curves shown
  in fig.~\ref{fig:exampleROTlcs}. The periodograms have been
  normalized to have zero median and unit standard deviation after
  iterative $5\sigma$ clipping. The peak period/frequency for each
  periodogram is listed in units of days and cycles per day
  respectively. The secondary peak in each case is the sidereal day
  alias of the primary peak. In each case we assume that the highest
  S/N peak corresponds to the correct period. While it is possible,
  though unlikely, that the secondary peak may actually be the true
  period and not the alias, there is no way in general to determine
  this; some ambiguity is inevitable.}
\label{fig:exampleROTAoVHarmspec}
\end{figure}

Figure~\ref{fig:exampleROTlcs} shows example phased light curves for
12 randomly selected rotational variables found in our survey, while
figure~\ref{fig:exampleROTAoVHarmspec} shows the AoVHarm periodograms
for five of these light curves. For the following analysis we only
consider stars in our catalog of periodic noneclipsing variables that
are identified as reliable detections for either the EPD or TFA light
curves and that are not flagged as probable blends. If the variable is
identified as a reliable detection for both the EPD and TFA light
curves, we adopt the period found on the TFA light curve.

\subsubsection{Variability Fraction-Color Relation}

\begin{figure}[!ht]
\ifthenelse{\boolean{emulateapj}}{\epsscale{1.2}}{\epsscale{1.0}}
\plotone{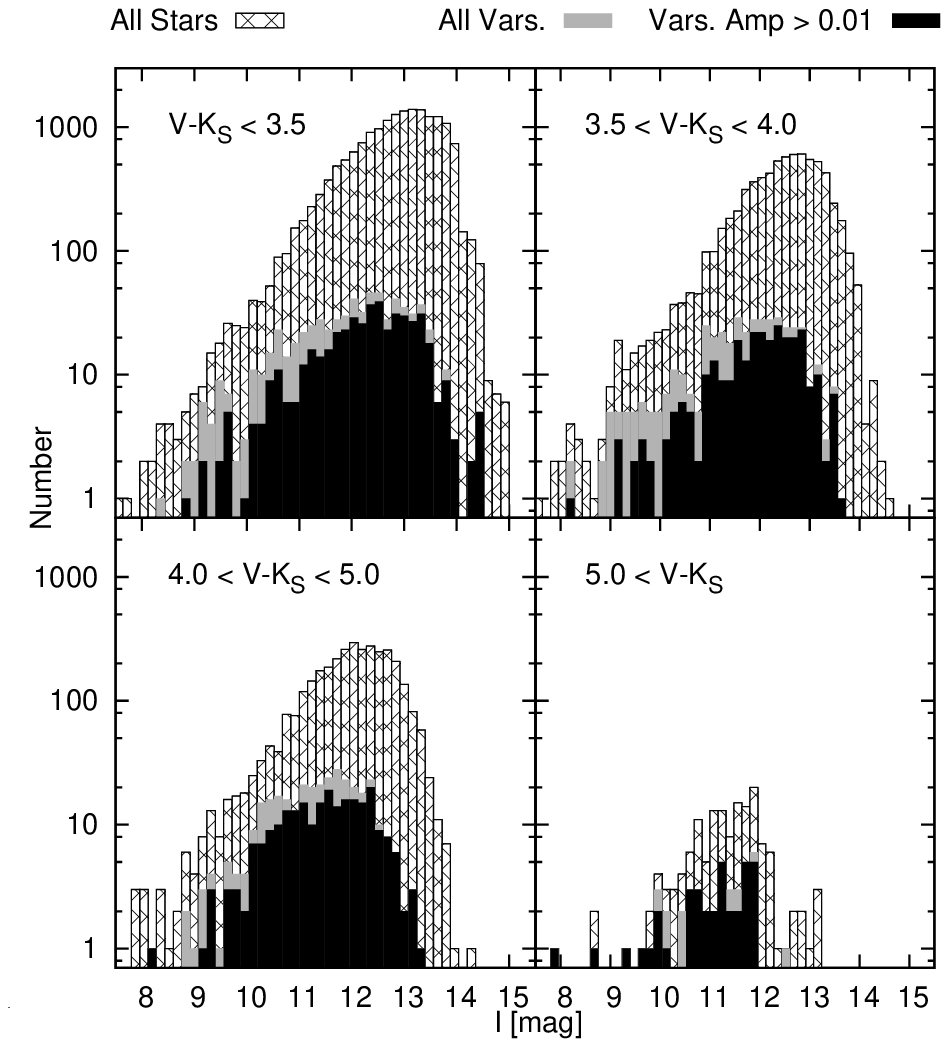}
\caption{The distribution of $I_{C}$ magnitudes for all stars, all
  periodic variable stars, and all periodic variable stars with
  peak-to-peak amplitudes greater than $0.01$~mag is shown for
  different $V-K_{S}$ bins.}
\label{fig:MagDistbyColor}
\end{figure}

In figure~\ref{fig:VarColorHist} we compare the distribution of
$V-K_{S}$ colors for periodic variables with peak-to-peak amplitudes
greater than 0.01~mag to the distribution for all stars in the
sample. We also show the fraction of stars that are variable at this
level as a function of $V-K_{S}$. To estimate the fraction of stars in
our sample that are blends we show the results separately for the full
sample and for a sample restricted to stars that do not have a
neighbor from the 2MASS catalog within $30\arcsec$ and are classified
either as unblended or as unlikely blends. In
figure~\ref{fig:MagDistbyColor} we compare the distribution of $I_{C}$
magnitudes for periodic variables to the distribution for all stars in
the sample.

The plotted relation for the fraction of stars that are variable as a
function of color in figure~\ref{fig:VarColorHist} has been corrected
for completeness by conducting sinusoid injection/recovery simulations
to estimate our detection efficiency. In conducting these tests we
divide the sample into 90 period/amplitude/color bins. We use color
bins of $2.0 < V-K_{S} < 3.5$, $3.5 < V-K_{S} < 4.0$, $4.0 < V-K_{S} <
4.5$, $4.5 < V-K_{S} < 5.0$ and $5.0 < V-K_{S} < 6.0$, three period
bins of 0.1-1~day, 1-10~days, and 10-100~days, and 5 amplitude bins
logarithmically spanning 0.01 to 1.0 mag. We have also considered
using magnitude bins rather than color bins for determining the
completeness correction, and have found that the resulting differences
in the relation shown in figure~\ref{fig:VarColorHist} are
negliglible. The resolution of the grid was set by the available
computational resources. For each bin we randomly select 1000 stars
with the appropriate color (for color bins with fewer than 1000 stars
we select with replacement). For each selected star we then choose a
random period and amplitude drawn from uniform-log distributions over
the bin and inject a sine curve with that period/amplitude and a
random phase into the light curve of the star. If both EPD and TFA
light curves are available for the star we inject the same signal into
both light curves. We do not reduce the amplitude for the injection
into the TFA light curve, this may cause us to slightly overestimate
our detection efficiency for periods longer than $\sim 10~{\rm
  days}$. If the star was identified as a variable or a potential
variable by our survey, we first remove the true variable signal from
the light curve by fitting a harmonic series to the phased light curve
before injecting the simulated signal. We then process the simulated
signals through the AoVHarm algorithm using the same selection
parameters as used for selecting the real variables. We do not apply
by-eye selections on the simulated light curves; we estimate below the
systemic uncertainty that results from this.

To get the completeness corrected variability fraction we weight each
real detected variable by $1/f$ where $f$ is the fraction of simulated
signals that are recovered for the period/amplitude/color bin that the
real variable falls in. We find that we are roughly $\sim 70\%$
complete over our sample of stars for peak-to-peak amplitudes greater
than 0.01 mag and periods between 0.1 and 100 days. For $\sim 97\%$ of
the recovered simulations the recovered frequency is within
$0.001~{\rm day}^{-1}$ of the injected frequency. The recovery rate is
relatively insensitive to the period and color and depends most
significantly on the amplitude. For amplitudes between 0.01 mag and
0.022 mag the recovery rate is $\sim 65\%$, whereas for amplitudes
above 0.05 mag the recovery rate is $\sim 85\%$. Above 0.05 mag the
recovery rate is independent of amplitude.

To estimate the systematic uncertainty due to not applying the by-eye
selection, we note that only $\sim 1\%$ of stars with $P < 10~{\rm
  days}$ that passed the automatic selections for AoVHarm were
classified by eye as unreliable detections. For stars with $P >
10~{\rm days}$, we classified $\sim 30\%$ of the stars passing the
automatic selections as unreliable detections. Assuming that $1\%$ of
the automatically selected simulations with $P < 10~{\rm days}$ would
be rejected by eye, and that $30\%$ of the automatically selected
simulations with $P > 10~{\rm days}$ would be rejected by eye, we put
a lower limit of $\sim 60\%$ on the completeness of our sample. We
conclude, therefore, that the true completeness of our sample for
variable stars with peak-to-peak amplitudes greater than $0.01~{\rm
  mag}$ is between $60 - 70\%$.

As seen in figure~\ref{fig:VarColorHist} the fraction of stars that
are detected as variables increases steeply with decreasing stellar
mass for the sample where blending is not likely to be
significant. While $\la 2\%$ of stars in this sample with $M \ga
0.7~M_{\odot}$ are found to be variable with peak-to-peak amplitude $>
0.01~{\rm mag}$, approximately half of the stars with $M \la
0.2~M_{\odot}$ are detected as variables at this level. We find that
an exponential relation of the form
\begin{equation}\label{eq:fracvar}
{\rm Var.~Frac.} = (\VARFRACCOEFFA{}^{+\VARFRACERRApos{}}_{-\VARFRACERRAneg{}}) e^{(\VARFRACCOEFFB{} \pm \VARFRACERRB{})(V-K_{S})}
\end{equation}
fits the observed relation over the color range $2.0 < V-K_{S} <
6.0$. Note that if the completeness correction is not applied, the
leading coefficient is reduced to
$\VARFRACCOEFFAnocor{}^{+\VARFRACERRAposnocor{}}_{-\VARFRACERRAnegnocor{}}$,
but the coefficient in the exponential is $\VARFRACCOEFFBnocor{} \pm
\VARFRACERRBnocor{}$ which is consistent with the coefficient in
eq.~\ref{eq:fracvar} to within the uncertainties. We conclude
therefore that there does not appear to be a signficant color bias in
our selection of variables and that the color dependence given in
eq.~\ref{eq:fracvar} is correct. Including a by-eye rejection rate of
$30\%$ for light curves with $P > 10~{\rm days}$ increases the leading
coefficient by $0.00021$ and does not change the coefficient in the
exponent. We therefore conclude that the statistical uncertainties
given in eq.~\ref{eq:fracvar} are significantly larger than the
systematic uncertainties in the relation.

For the larger sample of stars the variable fraction-color relation
follows eq.~\ref{eq:fracvar} with a constant value of
$\VARFRACCOEFFBLEND{} \pm \VARFRACERRBLEND{}$ added in. In other
words, for a random star in our sample, there is a $\sim 1.6\%$ chance
that the star will be blended with a nearby variable, show variations
in its light curve with an amplitude $> 0.01~{\rm mag}$, and that the
star will not be flagged as a probable blend by our blending
classification scheme. We estimate that $\NUMfracexpectedblend$ of the
variable stars in the larger sample are blends; most of these are the
bluer stars for which the real variability fraction is comparable to
or less than $1.6\%$.

\ifthenelse{\boolean{emulateapj}}{\begin{figure*}[!ht]}{\begin{figure}[!ht]}
\ifthenelse{\boolean{emulateapj}}{\epsscale{1.2}}{\epsscale{0.6}}
\plotone{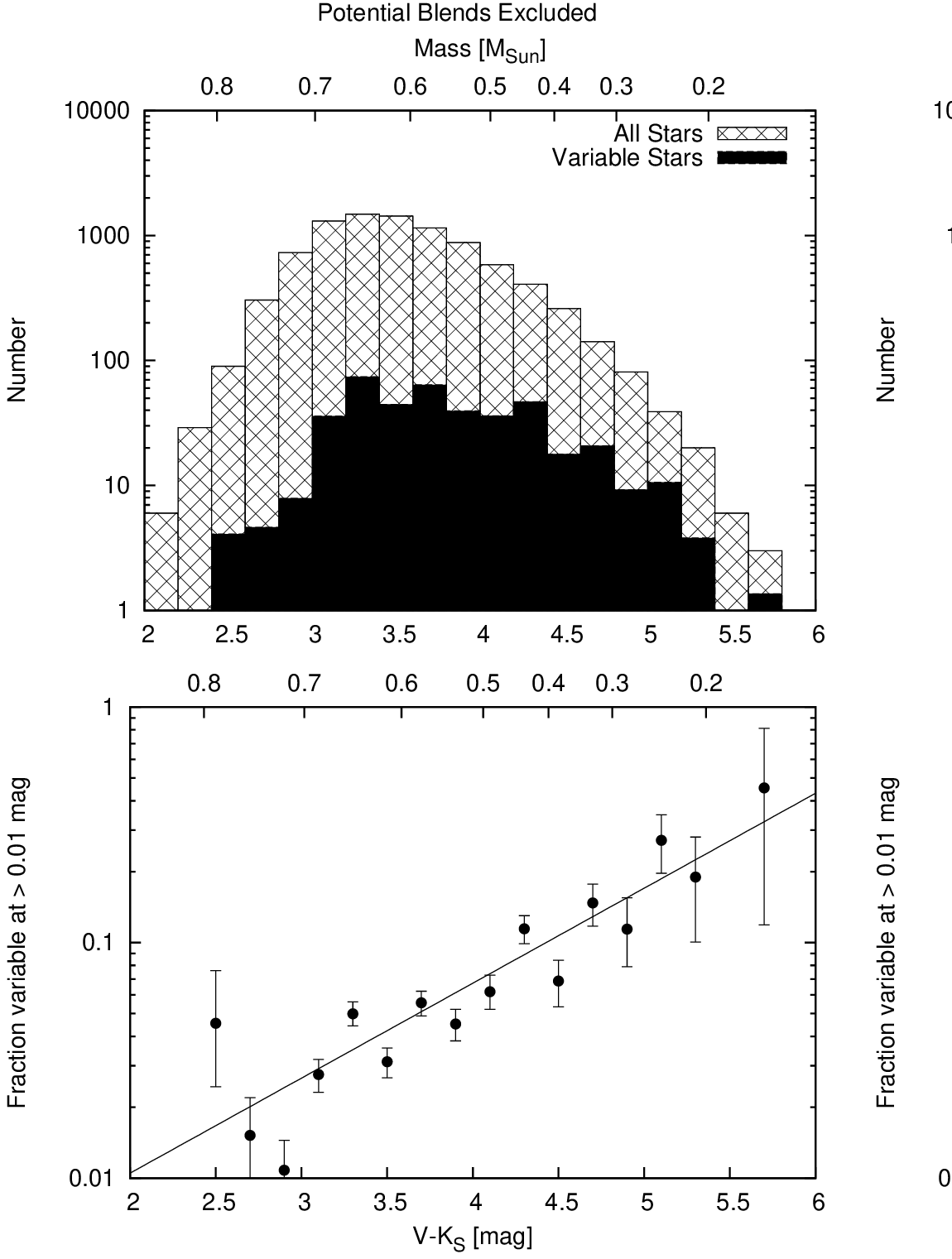}
\caption{{\scriptsize Top: The distribution of $V-K_{S}$ colors for
    variable stars compared to the distribution of $V-K_{S}$ for both
    variable and non-variable stars in the sample. On the right we
    show all \NUMtotstarsinVarColorHistplotRight{} stars with $V-K_{S}
    > 2$ and the \NUMtotvarsinVarColorHistplotRight{} noneclipsing
    periodic variables with $V-K_{S} > 2$ that have a peak-to-peak
    amplitude greater than $0.01~{\rm mag}$, are flagged as being
    secure detections, and are not flagged as probable blends. On the
    left we limit the sample to the
    \NUMtotstarsinVarColorHistplotLeft{} stars and
    \NUMtotvarsinVarColorHistplotLeft{} variables that, in addition to
    the previous restrictions, also do not have any neighbors in the
    2MASS catalog within $30\arcsec$ and are not flagged as potential
    blends. On the top axis we show the corresponding main sequence
    stellar masses determined by combining the empirical $V-K_{S}$
    vs. $M_{K}$ main-sequence for stars in the Solar neighborhood
    given by \citet{Johnson.09} with the mass-$M_{K}$ relation from
    the \citet{Baraffe.98} 4.5~Gyr, solar-metallicity isochrone with
    $L_{\rm mix} = 1.0$. We used the relations from
    \citet{Carpenter.01} to convert the CIT $K$ magnitudes from the
    isochrones into the 2MASS system. The distribution for variable
    stars is biased toward redder $V-K_{S}$ colors relative to the
    distribution for all stars. The decrease in the total number of
    stars in the sample red-ward of $V-K_{S} \sim 3.5$ is due to the
    $V$-band magnitude limit of the PPMX survey. Bottom: The
    completeness corrected fraction of stars that are variable with
    peak-to-peak $R$ or $I_{C}$ amplitude $> 0.01$~mag as a function
    of $V-K_{S}$; we use the same samples of stars on the left and
    right as for the top panels. On the left panel the fraction
    increases exponentially with color (solid line (left), sloped
    dotted line (right), eq.~\ref{eq:fracvar}) such that $\la 2\%$ of
    stars with $M \ga 0.7~M_{\odot}$ are variable with peak-to-peak
    amplitudes $> 0.01~{\rm mag}$ while approximately $50\%$ of stars
    with $M \la 0.2~M_{\odot}$ are variable at this level. On the
    right panel we also include a constant frequency of
    $\VARFRACCOEFFBLEND{} \pm \VARFRACERRBLEND{}$ (constant dotted
    line) that is independent of color, and corresponds to the blend
    frequency. The solid line at right shows the sum of the constant
    term and the exponential term. We estimate that
    $\NUMfracexpectedblend$ of the potential variables included in the
    upper right panel are blends with real variable stars.}}
\label{fig:VarColorHist}
\ifthenelse{\boolean{emulateapj}}{\end{figure*}}{\end{figure}}

\subsubsection{Period-Amplitude Relation}\label{sec:periodamp}

For FGK stars there is a well-known anti-correlation between stellar
activity measured from emission in the H and K line cores, or from
H$\alpha$ emission, and the Rossby number \citep[$R_{O}$, the ratio of
  the rotation period to the characteristic time scale of convection,
  see][]{Noyes.84}, which saturates for short periods. Similar
anti-correlations with saturation have been seen between $R_{O}$ and
the X-ray to bolometric luminosity ratio \citep[e.g.][]{Pizzolato.03},
and between $R_{O}$ and the amplitude of photometric variability
\citep[e.g.][]{Messina.01, Hartman.09}. Main sequence stars with $M
\la 0.35 M_{\odot}$ are fully convective \citep{Chabrier.97}, so one
might expect that the rotation-activity relation breaks down, or
significantly changes, at this mass. Despite this expectation, several
studies have indicated that the rotation-activity relation (measured
using $v \sin i$ and H$\alpha$ respectively) may continue even for
late M-dwarfs \citep{Delfosse.98, Mohanty.03, Reiners.07}. Recently,
however, \citet{West.09} have found that rotation and activity may not
always be linked for these stars.

\begin{figure}[!ht]
\ifthenelse{\boolean{emulateapj}}{\epsscale{1.2}}{\epsscale{0.6}}
\plotone{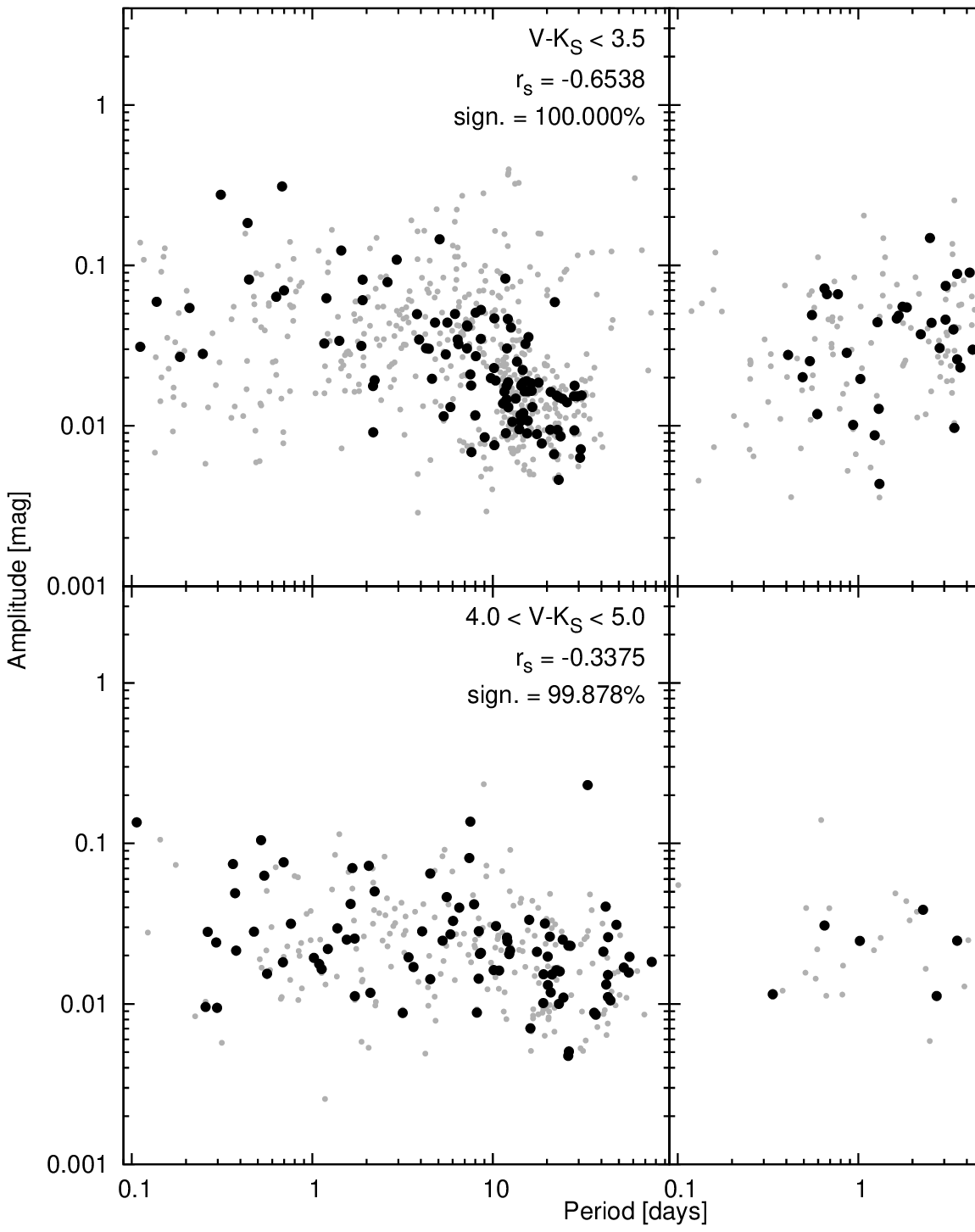}
\caption{Rotation period vs. peak-to-peak photometric amplitude for
  \NUMrobustvarswithEPDamp{} stars. Dark filled circles show variables
  that do not have a 2MASS neighbor within $30\arcsec$ and are flagged
  as unblended or as unlikely blends, while the grey-scale filled
  circles show all other variables that are not flagged as probable
  blends. We divide the sample
  into 4 bins by $V-K_{S}$ color, corresponding roughly to $M \ga 0.6
  M_{\odot}$, $0.5 M_{\odot} \la M \la 0.6 M_{\odot}$, $0.25 M_{\odot}
  \la M \la 0.5 M_{\odot}$, and $M \la 0.25 M_{\odot}$, from blue to
  red. We also list the Spearman rank-order correlation coefficient
  and the statistical significance of the correlation for each sample
  (note that negative values of $r_{S}$ imply that the period and
  amplitude are anti-correlated). The statistics are given for the
  stars shown with the dark filled points. For stars with $V-K_{S} <
  5.0$ the period and peak-to-peak amplitude are anti-correlated at
  high significance (for stars with $4.0 < V-K_{S} < 5.0$ the
  significance is higher than $1 - 10^{-5}$ when the grey-scale points
  are included in the calculation). The relation appears to be
  saturated for periods $\la 5~{\rm days}$, with the saturation period
  increasing for decreasing stellar mass. For stars with $V-K_{S} >
  5.0$ ($M \la 0.25 M_{\odot}$), the period and amplitude are not
  significantly correlated for $P \la 30~{\rm days}$.}
\label{fig:PeriodvsAmp}
\end{figure}

In figure~\ref{fig:PeriodvsAmp} we plot the rotation period against
the peak-to-peak amplitude for stars in several color bins. The
peak-to-peak amplitude is calculated for the EPD light curves as
described in section~\ref{sec:blend}; for stars observed in multiple
fields we take the amplitude of the combined light curve. Stars
without an available EPD light curve are not included in the plot. A
total of \NUMrobustvarswithEPDamp{} stars are included in the plot. We
mark separately the \NUMrobustnonpotblendnoneighborswithEPD{}
variables that are not flagged as potential blends and do not have a
2MASS neighbor within $30\arcsec$. For stars with $V-K_{S} < 5.0$
(corresponding roughly to $M \ga 0.25 M_{\odot}$) the photometric
amplitude and the period are anti-correlated at high
significance. There appears to be a cut-off period, such that the
period and amplitude are uncorrelated for stars with periods shorter
than the cut-off, and are anti-correlated for stars with periods
longer than the cut-off. \citet{Hartman.09} find a saturation
threshold of $R_{O} = 0.31$, which for a $0.6~M_{\odot}$ star
corresponds to a period of $\sim 8~{\rm days}$ \citep[assuming
  $(B-V)_{0} = 1.32$ for stars of this mass, and using the empirical
  relation between the convective time scale and $(B-V)_{0}$
  from][]{Noyes.84}. This is consistent with what we find for the
bluest stars in our sample. For stars with $V-K_{S} > 5.0$ ($M \la
0.25~M_{\odot}$), the period and amplitude are not significantly
correlated, at least for periods $\la 30~{\rm days}$. This result
suggests that the distribution of starspots on late M dwarfs is
uncorrelated with rotation period over a large period range, and is
perhaps at odds with H$\alpha$/$v \sin i$ studies which indicate a
drop in activity for very late M-dwarf stars with $v \sin i \la
10~{\rm km/s}$ \citep[e.g.][]{Mohanty.03}.

\subsubsection{Period-Color/Period-Mass Relation}

\begin{figure}[!ht]
\ifthenelse{\boolean{emulateapj}}{\epsscale{1.2}}{\epsscale{0.55}}
\plotone{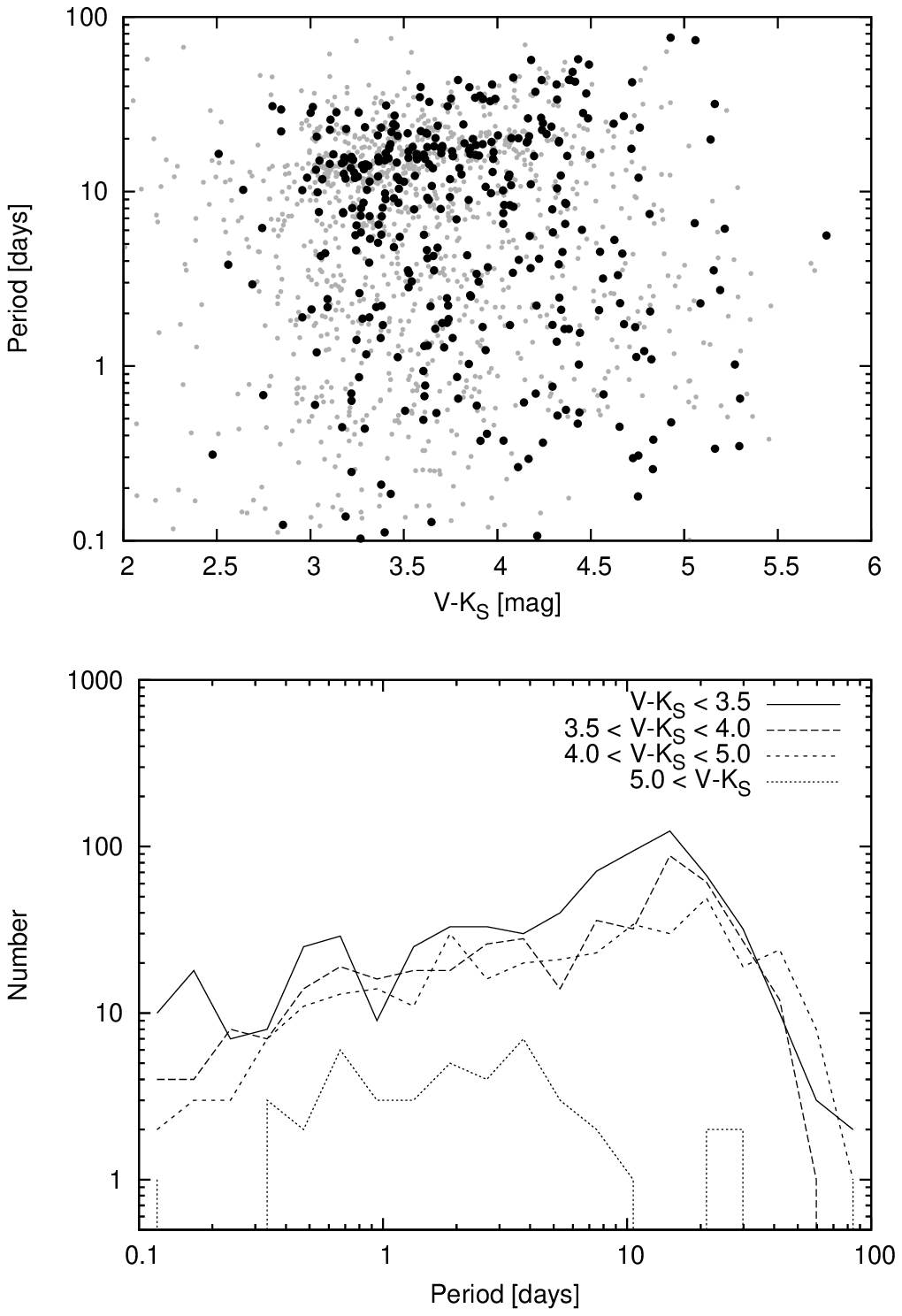}
\caption{Top: Period vs. $V-K_{S}$ for
  \NUMrobustvarsnonblendnoperioderr{} stars in our catalog of periodic
  variables that are flagged as being secure detections for at least
  one detection method, are not flagged as having incorrect period
  determinations for that method, and are not flagged as probable
  blends. There appears to be a
  paucity of stars with $P > 10~{\rm days}$ and $V-K_{S} > 4.5$ or $V
  - K_{S} < 3$. Bottom: The distributions of periods are shown for
  four color bins. For the three bluest color bins there appears to be
  a correlation between period and color, such that the mode period is
  longer for redder stars. For stars with $V-K_{S} \ga 4.5$, on the
  other hand, the period distribution appears to be biased to shorter
  periods. Using a K-S test, we find that the probability that the
  stars in any two of the different color bins are drawn from the same
  distribution is less than $0.01\%$ for all combinations except the
  for the combination of the two intermediate color bins. For that
  combination the probability is $\sim 12\%$.}
\label{fig:PeriodDist}
\end{figure}

In figure~\ref{fig:PeriodDist} we show the relation between period and
color. For stars with $V-K_{S} \la 4.5$ the measured distribution of
$\log P$ is peaked at $\sim 20~{\rm days}$ with a broad tail toward
shorter periods and a more rapid drop-off for longer periods. Note
that the cut-off for longer periods may be due to the correlation
between period and amplitude for these stars; stars with periods
longer than $\sim 20~{\rm days}$ may be harder to detect and not
intrinsically rare. The peak of the distribution appears to be
correlated with color such that redder stars are found more commonly
at longer periods than bluer stars. For stars with $V-K_{S} \ga 4.5$
the distribution changes significantly such that the $\log P$
distribution appears to be more or less flat between $\sim 0.3$ and
$\sim 10~{\rm days}$, while red stars with $P \ga 10~{\rm days}$ are
uncommon.

\begin{figure}[]
\ifthenelse{\boolean{emulateapj}}{\epsscale{1.2}}{\epsscale{0.45}}
\plotone{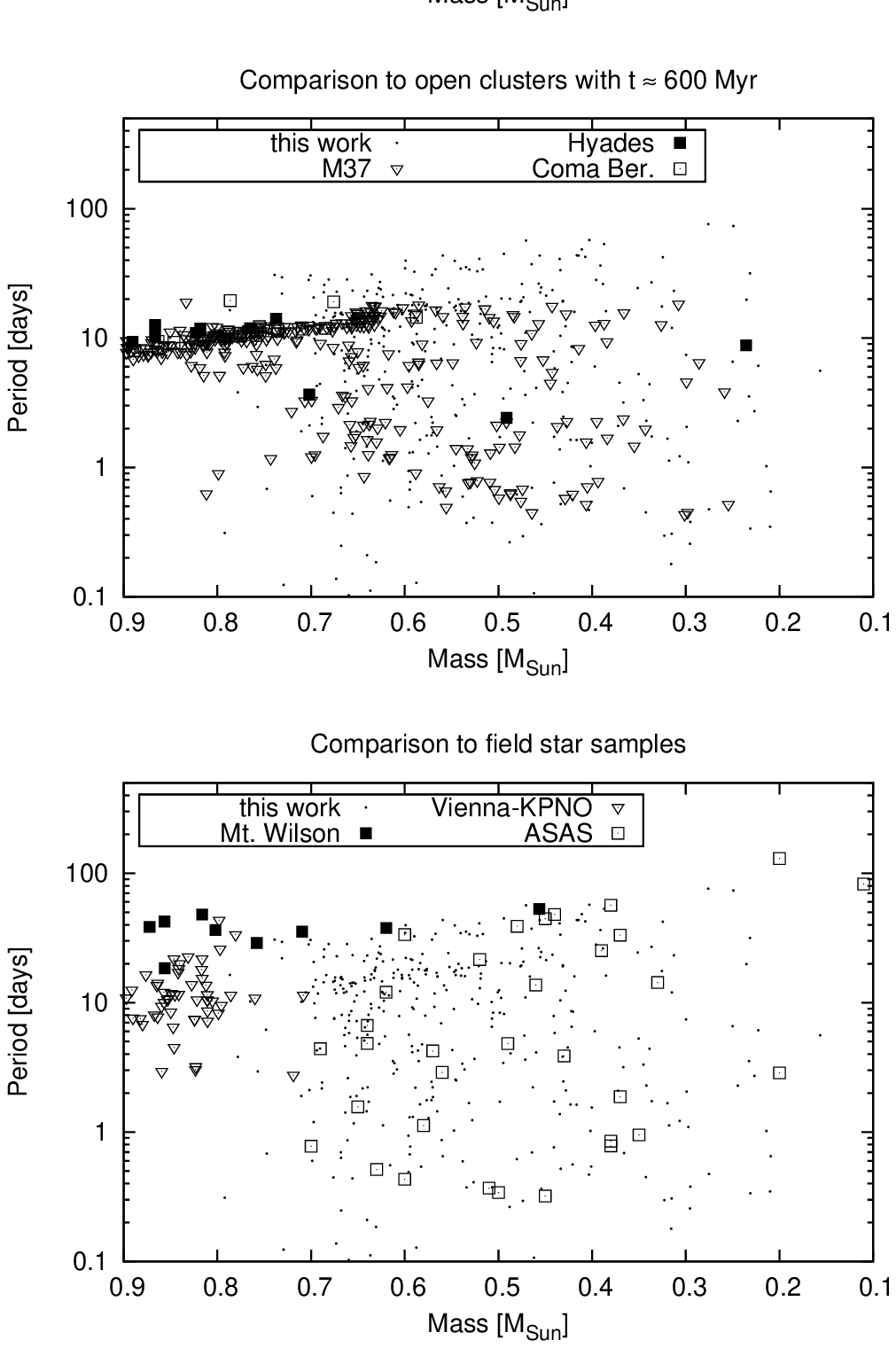}
\caption{{\scriptsize A comparison of the mass-period distribution for stars in our
  survey to the results from other surveys. We only include stars from
  our survey that do not have a 2MASS neighbor within $30\arcsec$ and
  are classified either as an unlikely blend or as unblended. See the
  text for a description of the other data sources. For clarity we
  make the comparison separately for field stars, open clusters with
  $100~{\rm Myr} < t < 500~{\rm Myr}$ and for three open clusters with
  $t \sim 600~{\rm Myr}$. The rotation periods of stars in our sample
  at a given mass appear to be longer, on average, than the periods of
  stars in the open clusters; this is true across all mass ranges
  covered by our survey. The rotation periods from the Vienna-KPNO
  survey appear to be comparable, at a given mass, to the periods of
  stars in our sample, while the periods from the Mt.~Wilson survey
  appear to be generally longer than the periods from our survey. It
  is likely that our survey, the Vienna-KPNO survey and the ASAS
  survey are biased toward high-activity, shorter period stars than
  the Mt.~Wilson survey is.}}
\label{fig:PeriodMassComp}
\end{figure}

In Figure~\ref{fig:PeriodMassComp} we compare the mass-period
distribution for stars in our survey to the results from other surveys
of field stars and open clusters. We choose to use mass for the
comparison rather than observed colors because a consistent set of
colors is not available for all surveys. The masses for stars in our
survey are estimated from their $V-K_{S}$ colors (see
Fig.~\ref{fig:VarColorHist}). 

We take data from the Mount Wilson,
Vienna-KPNO \citep{Strassmeier.00}, and ASAS \citep{Kiraga.07} samples
of field stars with rotation periods. For the Mount Wilson sample we
use the compilation by \citet{Barnes.07}, the original data comes from
\citet{Baliunas.96} and from \citet{Noyes.84}. For the Vienna-KPNO
sample we only consider stars which are listed as luminosity class
V. We estimate the masses for stars in the Mount Wilson and
Vienna-KPNO samples using the same $V-K_{S}$ to mass conversion that
we use for our own sample. The $V$ and $K_{S}$ magnitudes for these
field stars are taken from SIMBAD where available. For the ASAS sample
we use the masses determined from the absolute $M_{V}$ magnitudes of
the stars by \citet{Kiraga.07}. 

We also compare our sample to four open clusters with ages between
$100 - 200$~Myr and to three open clusters with ages of $\sim
600$~Myr. These include: M35 \citep[$\sim 180~{\rm
    Myr}$;][]{Meibom.09}, three clusters observed by the MONITOR
project, including M50 \citep[$\sim 130~{\rm Myr}$;][]{Irwin.09a},
NGC~2516 \citep[$\sim 150~{\rm Myr}$;][]{Irwin.07}, and M34
\citep[$\sim 200~{\rm Myr}$;][]{Irwin.06}, M37 \citep[$\sim 550~{\rm
    Myr}$;][]{Hartman.09}, Coma Berenices \citep[$\sim 600~{\rm
    Myr}$;][]{CollierCameron.09}, and the Hyades \citep[$\sim 625~{\rm
    Myr}$;][]{Radick.87,Radick.95,Prosser.95}. For the MONITOR
clusters we use the mass estimates given in their papers, these are
based on the $I_{C}$-mass relation from the appropriate
\citet{Baraffe.98} isochrone for the age/metallicity of each
cluster. For M35, M37 and the Hyades we use the mass estimates derived
from the $V,I_{C}$-mass relations from the appropriate YREC
\citep{An.07} isochrones for each cluster. For Coma Berenices we use
the mass-$M_{K}$ relation discussed in Fig.~\ref{fig:VarColorHist} and
assume a distance of 89.9~pc for all stars \citep{vanLeeuwen.99}. For
M35 we only include 214 stars from the \citet{Meibom.09} catalog that
lie near the main sequence in $V$, $B$ and $I_{C}$, and we exclude any
stars which have a proper motion membership probability less than
$80\%$, or an RV membership probability less than $80\%$ as determined
by \citet{Meibom.09}. We expect that the stars in our sample, and in
the other field star samples, have a range of ages, but on average
will be older than the stars in the open clusters.

The sample of stars with rotation periods presented here is
substantially richer than is available for other surveys of field
stars. This is especially the case for later spectral types. The
Mt.~Wilson and Vienna-KPNO surveys primarily targeted G and early K
stars, so there is not much overlap in stellar mass between those
samples and our sample. The ASAS survey, on the other hand, covers a
similar mass range as our survey, but for a much smaller sample of
stars (periods were found for 31 out of 180 X-ray active stars). The
few Mt.~Wilson stars with estimated masses $\la 0.8~M_{\odot}$ do show
an anti-correlation between mass and period, and have periods that are
longer than the majority of stars in our sample. The Vienna-KPNO
stars, on the other hand, have periods that cluster around $\sim
10~{\rm days}$, which is closer to the mode of the period distribution
for stars of comparable mass in our sample. Since the Vienna-KPNO
stars were selected as showing spectroscopic evidence for
chromospheric activity before their periods were measured
photometrically, this sample is presumably biased to shorter period
active stars compared to the Mt.~Wilson sample. Given the correlation
between photometric amplitude and rotation period, we would also
expect our sample to be biased toward shorter period stars relative to
the Mt.~Wilson sample.  The ASAS stars with $M \ga 0.5~M_{\odot}$
generally have shorter periods than the stars in our survey; this is
most likely due to the restriction of the ASAS survey to X-ray active
stars.

When compared to the open cluster samples we
see clear evidence for evolution in the rotation periods of low-mass
stars. Stars in the younger clusters have shorter periods at a given
mass, on average, than stars in our sample. The discrepancy becomes
more apparent for stars with $M \la 0.5~M_{\odot}$ for which the
period and mass appear to be positively correlated in the young
cluster samples while they are anti-correlated in our sample. Looking
at the $\sim 600~{\rm Myr}$ clusters, again the periods of stars in
our sample are longer at a given mass, on average, than the periods of
the cluster stars, however in this case the mode of the period
distribution for the cluster stars appears to be closer to the mode of
the period distribution for our sample than it is for the younger
clusters. The lowest mass stars in the older clusters also do not show
as significant a correlation between mass and period as do the lowest
mass stars in the younger clusters. For stars with $M \la
0.3~M_{\odot}$ the available field star and older open cluster samples
are too sparse to draw any conclusions from when comparing to our
sample. For the younger clusters, we note that the distribution of
periods for the lowest mass stars is even more strongly peaked toward
short periods than it is in our sample. This suggests that these stars
do lose angular momentum over time, despite not having a tachocline. A
more detailed comparison of these data to models of stellar angular
momentum evolution is beyond the scope of this paper.

\subsubsection{Period-X-ray Relation}

\begin{figure}[!ht]
\ifthenelse{\boolean{emulateapj}}{\epsscale{1.2}}{\epsscale{1.0}}
\plotone{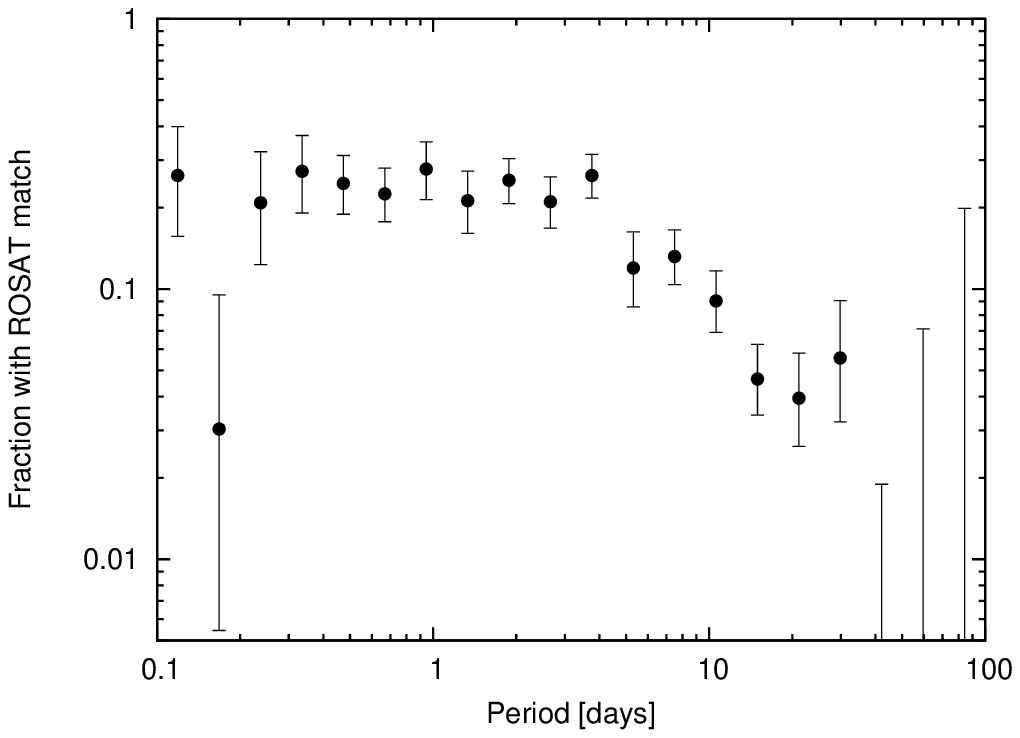}
\caption{The fraction of non-EB periodic variable stars that match to
  a ROSAT source as a function of rotation period. The errorbars for
  the 3 longest period bins show 1$\sigma$ upper-limits. For stars
  with rotation periods less than $\sim$4 days the fraction that
  matches to an X-ray source is constant at $\sim 22\%$, for longer
  periods the fraction decreases as $\sim P^{-0.7}$.}
\label{fig:XrayFractionvsPeriod}
\end{figure}

\begin{figure}[!ht]
\ifthenelse{\boolean{emulateapj}}{\epsscale{1.2}}{\epsscale{1.0}}
\plotone{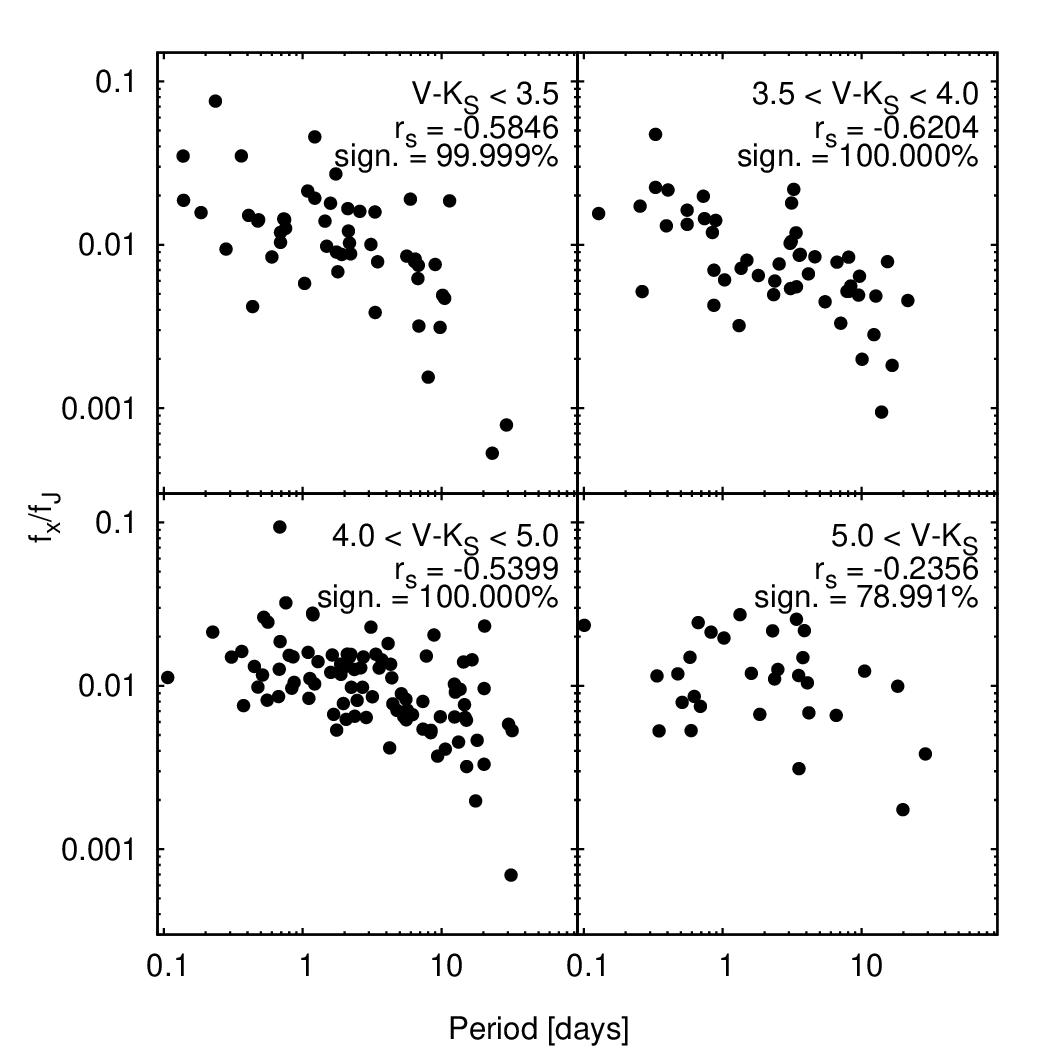}
\caption{The ratio of 0.1-2.4 keV X-ray flux to $J$-band near infrared
  flux vs. the rotation period for non-EB periodic variable stars that
  match to a ROSAT source. We divide the sample into the same 4 color
  bins used in fig.~\ref{fig:PeriodvsAmp}. We also list the Spearman
  rank-order correlation coefficient and the statistical significance
  of the correlation for each sample. The X-ray flux is
  anti-correlated with the rotation period at high significance for
  stars with $V - K_{S} < 5.0$. For stars with $V - K_{S} > 5.0$ ($M
  \la 0.25 M_{\odot}$) there is still a hint of an anti-correlation,
  though it is of low statistical significance.}
\label{fig:logfxfjvsperiod}
\end{figure}

Figure~\ref{fig:XrayFractionvsPeriod} shows the fraction of variables
that match to an X-ray source as a function of period. This fraction
is constant at $\sim 22\%$ for periods less than $\sim$4 days, for
longer periods the fraction that matches to an X-ray source decreases
as $\sim P^{-0.7}$. Following \citet{Agueros.09} we calculate the
ratio of X-ray to J-band flux via
\begin{equation}
\log_{10} (f_{X} / f_{J}) = \log_{10}f_{X} + 0.4J + 6.30
\end{equation}
where 1 count s$^{-1}$ in the 0.1-2.4 keV energy range is assumed to
correspond on average to $f_{X} = 10^{-11}$ erg cm$^{-2}$ s$^{-1}$. In
figure~\ref{fig:logfxfjvsperiod} we plot the flux ratio as a function
of rotation period for samples of variables separated by their $V -
K_{S}$ color. The X-ray flux is anti-correlated with the rotation
period for stars with $M \ga 0.25 M_{\odot}$, for stars with $M \la
0.25 M_{\odot}$ there is still a hint of an anti-correlation, though
it is of low statistical significance (the false alarm probability is
$\sim 20\%$). This result is similar to what we found for the
photometric amplitude-period relation.

\subsection{Flares}\label{sec:flares}

\begin{figure}[!ht]
\ifthenelse{\boolean{emulateapj}}{\epsscale{1.2}}{\epsscale{1.0}}
\plotone{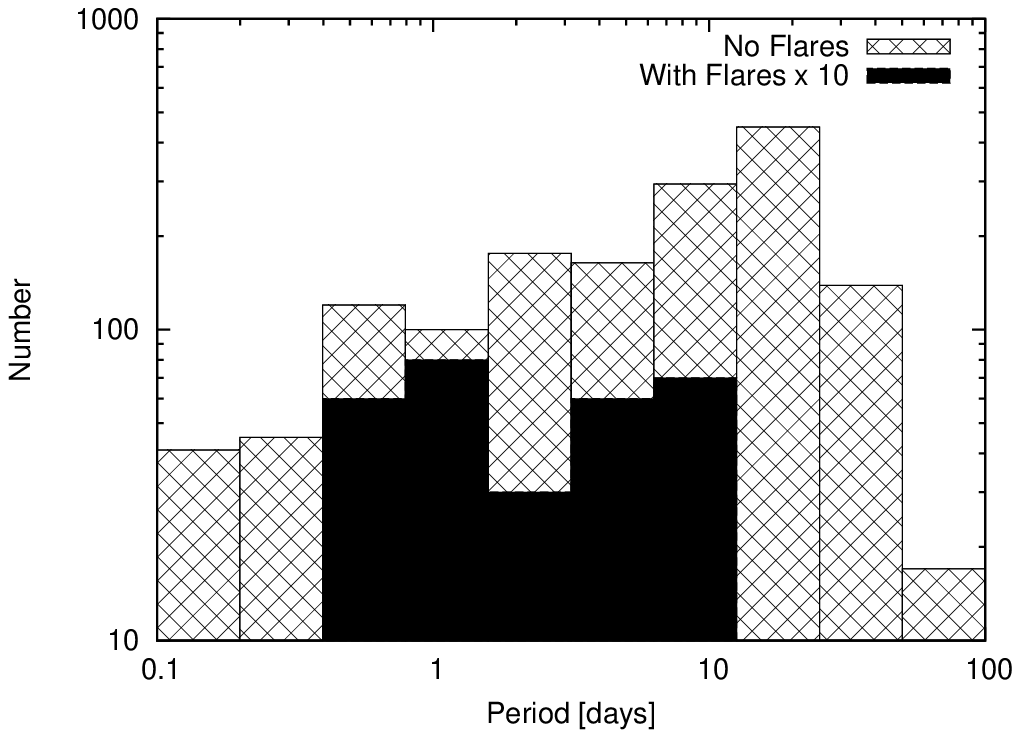}
\caption{Comparison between the period distributions for stars
  with detected large-amplitude long-duration flares, and stars
  without such a flare detected. Note the absence of long-period stars
  with flares. For clarity we have multiplied the flare star period
  distribution by a factor of 10.}
\label{fig:PeriodvsFlareOccurrence}
\end{figure}

In figure~\ref{fig:PeriodvsFlareOccurrence} we compare the distribution of
periods for stars with detected large-amplitude long-duration flares
to the distribution for stars without such flare detections. A total
of \NUMflareswithrobustperorEBnonpererrnonblend{} of the
\NUMflarestars{} flare stars have a robust period determination and
are not flagged as a probable blend. The period detection frequency of
$\sim 50\%$ for flare stars is significantly higher than that for all
other stars ($\la 5\%$). This result is expected if stellar flaring is
associated with the large starspots that give rise to continuous
photometric variations. As seen in figure~\ref{fig:PeriodvsFlareOccurrence},
the distribution of periods for flare stars is concentrated toward
shorter periods than the distribution for non-flare stars. The longest
period found for a flare star is 18.2 days whereas 31\% of the
non-flare stars with period determinations have periods greater than
18.2 days. Conducting a K-S test we find that the probability that the
two samples are drawn from the same distribution is less than
$10^{-4}$.

\begin{figure}[!ht]
\ifthenelse{\boolean{emulateapj}}{\epsscale{1.2}}{\epsscale{0.7}}
\plotone{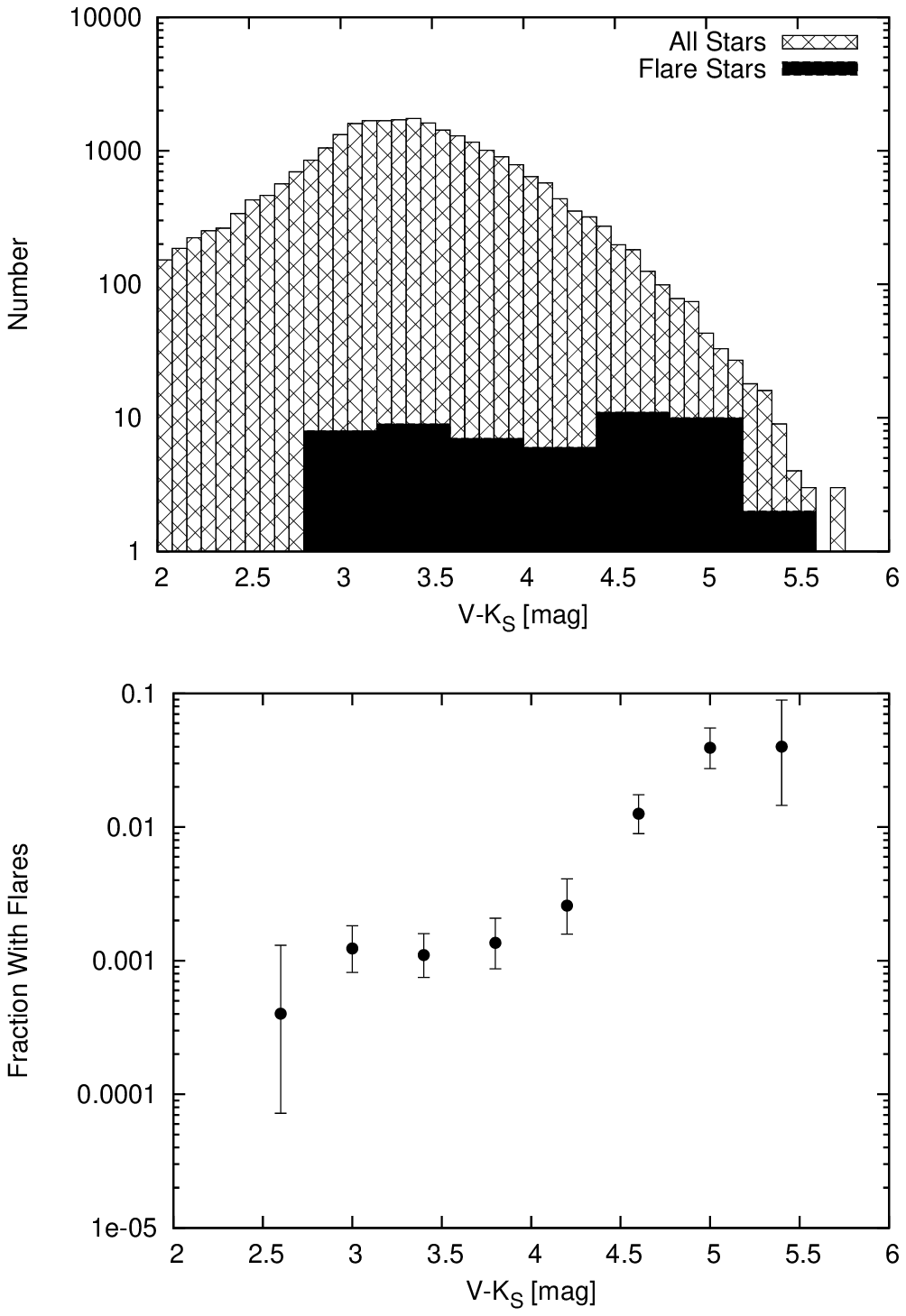}
\caption{Top: The distribution of $V-K_{S}$ colors for the \NUMflarestars{} stars
  with high amplitude flares detected compared to the distribution of
  $V-K_{S}$ colors for all \NUMtotsample{} stars in the sample. The
  distribution for flares stars is biased toward redder $V-K_{S}$
  colors relative to the distribution for all stars. Note that we use
  a 5 times higher binning resolution for the full sample. Bottom:
  The fraction of stars with a high amplitude flare detection is
  plotted against $V-K_{S}$. For stars with $V-K_{S} < 4.0$ less than
  $\sim 0.1\%$ of stars had a flare detected, for stars with $V-K_{S}
  > 4.5$ the fraction is $\ga 1\%$.}
\label{fig:FlareColor}
\end{figure}

Figure~\ref{fig:FlareColor} compares the distribution of $V-K_{S}$ for the flare stars to the distribution for the full sample. From this figure it is clear that flare detections are more common for redder stars.

\subsection{Multi-periodic Variables}\label{sec:multiperiod}

In Section~\ref{sec:gksearch} we described a search for multiperiodic
variability that was applied to a subset of the K/M dwarf light
curves. A number of the stars appear to vary with multiple periods. We
highlight here two of the more significant cases, HAT-094-0007885 and
HAT-123-0003441. These two stars have $V-K_{S} = 3.8$ and $V-K_{S} =
3.9$ respectively, which correspond to stellar masses of $M \sim
0.5~M_{\odot}$. Figure~\ref{fig:powerspectrum_multiperiod} shows the
DFT power spectra for each of these variables after successive
prewhitening cycles. HAT-094-0007885 has two significant frequencies:
$f_{0}=4.5240512~{\rm d}^{-1}$ and $f_{1}=0.2318221~{\rm d}^{-1}$. The
amplitude and/or frequency of both components change during the total
time span of 162 days. This varying phase/amplitude leads to remnant
power near the peak frequencies (or at their aliases). Note that
$(f_{0} - 1)/2 + f_{1} = 1.99215~{\rm d}^{-1}$, so it is possible that
$f_{1}$ is an alias the primary signal $f_{0}$. The nearest 2MASS
source to this star is more than 30\arcsec\ away, and all sources
within 1\arcmin\ are $\ga 2.5$ mag fainter. This star has a blending
flag of 0, indicating that the variation is most likely not due to
blending with a neighbor. HAT-123-0003441 has two significant
frequencies: $f_{0}=2.80539~{\rm d}^{-1}$ and $f_{1}=2.33122~{\rm
  d}^{-1}$. Note that when the frequency search is applied to the
light curves without applying TFA, the two identified frequencies are
$f_{0}=1.80367~{\rm d}^{-1}$ and $f_{1} = 1.32918~{\rm d}^{-1}$, which
are the $1~{\rm d}^{-1}$ aliases of the TFA frequencies. The frequency
corresponding to the period identified with the AoVHarm algorithm
applied to the TFA light curve for this star is $f = 1.80415~{\rm
  d}^{-1}$. This star has 6 faint neighbors within $60\arcsec$, we
assign a blending flag of 1 to this star, indicating that at least the
$f_{0}$ component signal is unlikely to be due to blending with a
nearby variable.

It is unclear what the physical origin of multiple frequencies would
be for either of these stars if they are single $M \sim 0.5~M_{\odot}$
dwarfs. If the variations are not due to nearby stars that are unresolved
in the HATNet images and are aligned by chance with the low-mass
stars, it may be that these are multiple star systems with the
components rotating at different periods.

\ifthenelse{\boolean{emulateapj}}{\begin{figure*}[!ht]}{\begin{figure}[!ht]}
{
\centering
\includegraphics*[width=80mm,angle=0]{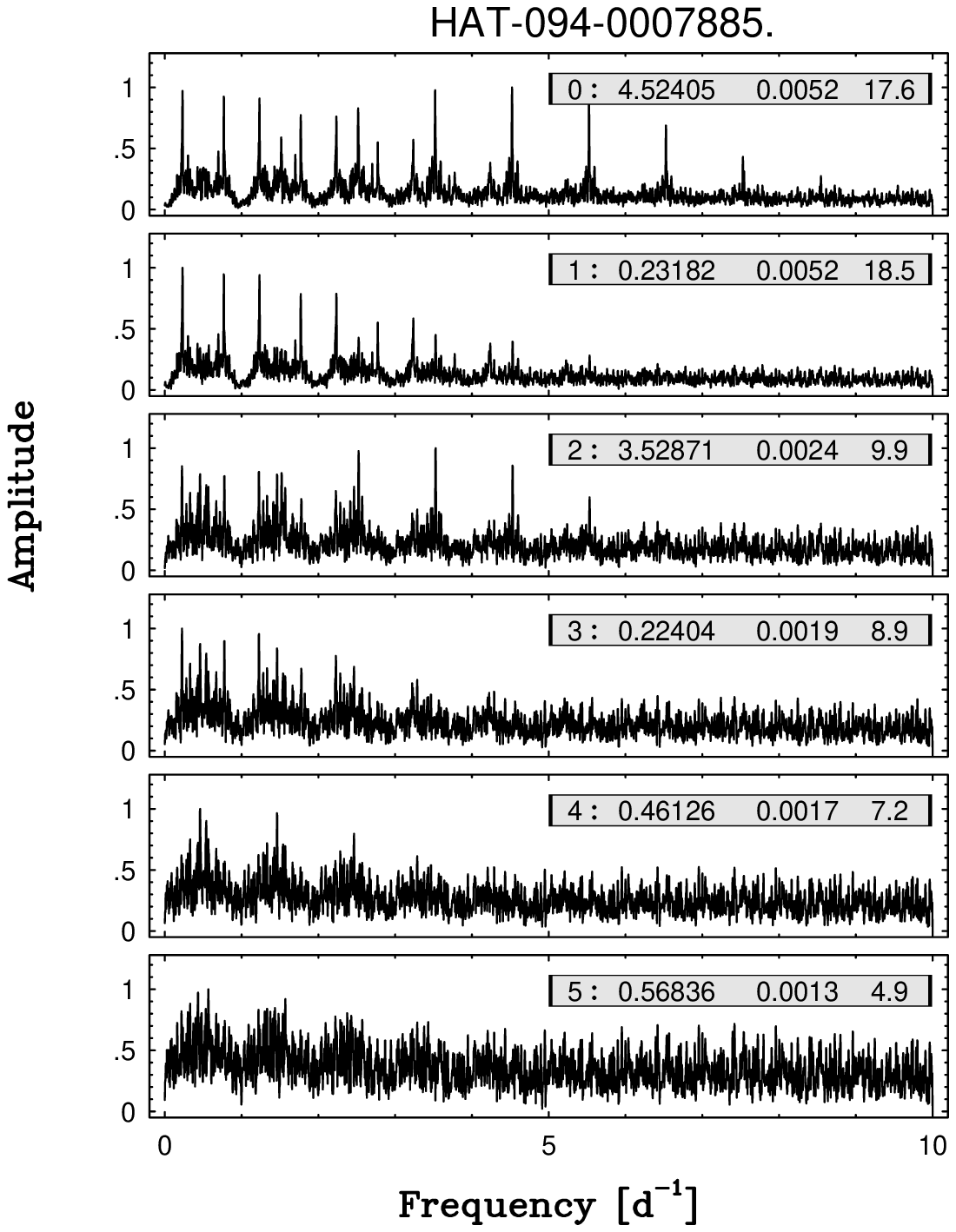}
\hfil
\includegraphics*[width=80mm,angle=0]{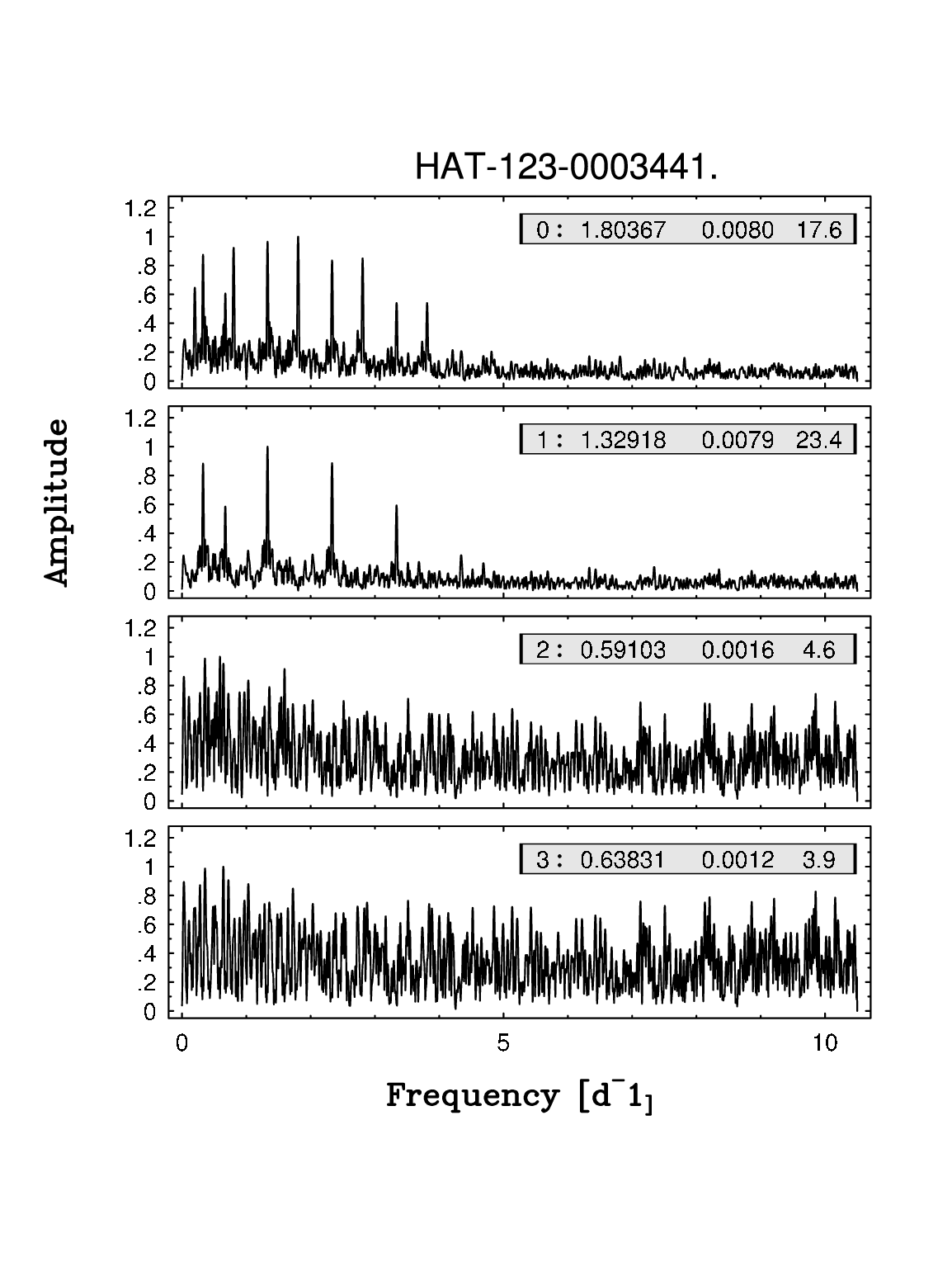}
}
\caption{DFT power spectra for two multi-periodic variables. The power
  spectra are displayed from top to bottom after successively
  prewhitening the light curve against identified frequencies. The
  numbers in each panel give the prewhitening cycle, the peak
  frequency in ${\rm d}^{-1}$, the amplitude of the peak in mag, and
  the S/N of the peak. For HAT-123-0003441 we show the DFT power
  spectra for the light curve without TFA applied.}
\label{fig:powerspectrum_multiperiod}
\ifthenelse{\boolean{emulateapj}}{\end{figure*}}{\end{figure}}

\section{Conclusion}\label{sec:conclusion}

In this paper we have presented the results of a variability survey
conducted with HATNet of field K and M dwarfs selected by color and
proper motion. We used a variety of variability selection techniques
to identify periodic and quasi-periodic variables, and have also
conducted a search for large amplitude, long-duration flare events.
Out of a total sample of \NUMtotsample{} stars we selected
\NUMtotpotentialvar{} that show potential variability, including
\NUMEB{} that show eclipses in their light curves, and
\NUMflarestars{} that show flares. We inspected all automatically
selected light curves by eye, and flagged \NUMrobustplusflares{} stars
(including those with flares) as being secure variability
detections. Because the HATNet images have poor spatial resolution,
variability blending is a significant problem. We therefore
implemented an automated routine to classify selected non-flare
variables as probable blends, potential blends, unlikely blends, or as
unblended. Altogether we found \NUMrobustplusflaresnonblend{}
variables that are classified as secure detections and are not
classified as probable blends. We estimate, however, that 26\% of
these stars may still be blends with fainter variables. The
\NUMrobustplusflaresnonblend{} variables includes \NUMEBnonblend{}
stars that show eclipses in their light curves. We identified
\NUMflareevents{} flare events in \NUMflarestars{} stars,
\NUMflareswithperorEB{} of these stars are also selected as potential
periodic or quasi-periodic variables (all
\NUMflareswithrobustperorEB{} are considered reliable detections, and
\NUMflareswithrobustperorEBnonpererrnonblend{} are not flagged as
probable blends). We matched the sample of potential variables to
other catalogs, and found that \NUMcross{} lie within $2\arcmin$ of a
previously identified variable, while \NUMnotcross{} do not. Including
only flare stars and variables that are classified as secure
detections and are not classified as probable blends,
\NUMrobustplusflaresnonblendknownvar{} (including
\NUMEBnonblendknown{} EBs) lie within $2\arcmin$ of a previously
identified variable, so that \NUMrobustplusflaresnonblendnotknownvar{}
are new identifications.

One of the eclipsing binaries that we identified is the previously
known SB2 system 1RXS~J154727.5+450803. By combining the published RV
curves for the component stars with the HATNet $I$~band light curve,
we obtained initial estimates for the masses and radii of the
component stars (Tab.~\ref{tab:RXJ1547param}). The system is one of
only a handful of known double-lined eclipsing binaries with component
masses less than $0.3~M_{\odot}$. While we caution that the errors on
the component radii are likely to be underestimated due to systematic
errors that have not been considered in this preliminary analysis, it
is interesting that the radii do appear to be larger than predicted if
the system is older than $\sim 200~{\rm Myr}$. With a magnitude $V
\sim 13.4$, this system is only slightly fainter than the well-studied
binary CM Dra ($V \sim 12.90$) which has been the anchor of the
empirical mass-radius relation for very late M
dwarfs. 1RXS~J154727.5+450803 is thus a promising target for more
detailed follow-up to obtain high precision measurements of the
fundamental parameters of the component stars. With additional
follow-up, the large sample of \NUMEBnonblend{} probable late-type eclipsing
binaries presented in this paper should prove fruitful for further
investigations of the fundamental parameters of low-mass stars.

The majority of the variable stars that we have identified are likely
to be BY~Dra type variables, with the measured period corresponding to
the rotation period of the star. This is the largest sample of
rotation periods presented to date for late-type field stars. We
discussed a number of broad trends seen in the data, including an
anti-correlation between the rotation period and the photometric
amplitude of variability, an exponential relation between $V-K_{S}$
color and the fraction of stars that are variable, a positive
correlation between period and the $V-K_{S}$ color for stars with $V -
K_{S} \la 4.5$, a relative absence of stars with $P \ga 10.0~{\rm
  days}$ and $V - K_{S} \ga 4.5$, and an anti-correlation between the
rotation period and the ratio of X-ray to $J$~band flux. The
correlations between period and activity indicators including the
amplitude of photometric variability and the X-ray emission are
consistent with the well-known rotation-age-activity-mass relations
for F, G, and K dwarfs. The data presented here may help in further
refining these relations. Our data hints at a change in the
rotation-activity connection for the least massive stars in the sample
($M \la 0.25~M_{\odot}$). The anti-correlation between period and
amplitude appears to break down for these stars, and similarly the
period-X-ray anti-correlation is less significant for these stars than
for more massive stars. This is potentially at odds with previous
studies which used H$\alpha$ to trace activity and $v \sin i$ to infer
rotation period, and found that the period-activity anti-correlation
extends to very late-type M dwarfs. Comparing our sample to other
field and open cluster samples, we find that the rotation periods of
stars in our sample are generally longer than the periods found in
open clusters with $t \la 620~{\rm Myr}$, which implies that K and M
dwarf stars continue to lose angular momentum past the age of the
Hyades. This appears to be true as well for stars with $M \la
0.25~M_{\odot}$, though these stars generally have shorter periods
than more massive stars in our sample.

We have also conducted a search for flare events in our light
curves, identifying \NUMflareevents{} events in \NUMflarestars{}
stars. Due to the difficulty of distinguishing between a flare and bad
photometry in an automated way, there are likely to be many flare
events in the light curves that we do not identify. We therefore do
not attempt to draw conclusions about the total occurrence rate of
flaring \citep[for a recent determination of this frequency using data
  from the Sloan Digital Sky Survey, see][]{Kowalski.09}. We find that
the distribution of $V-K_{S}$ colors for flare stars is biased toward
red colors, implying that the flare frequency increases with
decreasing stellar mass, which has been known for a long time
(\citealp{Ambartsumyan.70}, see also \citealp{Kowalski.09}). We find
that roughly half the flare stars are detected as periodic variables,
which is a significantly higher fraction than for the full sample of
stars. This is in line with the expectation that stellar flaring is
associated with the presence of significant starspots, and is
consistent with the finding by \citet{Kowalski.09} that the flaring
frequency of active M dwarfs showing H$\alpha$ emission is $\sim 30$
times higher than the flaring frequency of inactive M dwarfs. We also
find that the distribution of periods for flare stars is biased toward
shorter periods, again as expected from the rotation-activity
connection.

Finally we identified several potentially multi-periodic variables
from a search conducted on a subset of the light curves, and we
highlighted two examples. We speculate that the variation in these
cases is either due to blending with nearby variable stars, or else
these may be multi-star systems where each detected period corresponds
to the rotation period of a different component star. As far as we are
aware, there is no known pulsation mechanism in K and M dwarf stars
that would give rise to millimagnitude level photometric variations
with different periods.

\acknowledgements

We would like to thank Lev Or-Tal for calculating the upper limit to
the eccentricity for the eclipsing binary
1RXS~J154727.5+450803. HATNet operations have been funded by NASA
grants NNG04GN74G, NNX08AF23G and SAO IR\&D
grants. G.\'{A}.B. acknowledges support from the Postdoctoral
Fellowship of the NSF Astronomy and Astrophysics Program
(AST-0702843). G.K.~thanks the Hungarian Scientific Research
Foundation (OTKA) for support through grant K-81373. T.M. acknowledges
support from the ISRAEL SCIENCE FOUNDATION (grant No. 655/07). The
Digitized Sky Surveys were produced at the Space Telescope Science
Institute under U.S. Government grant NAG W-2166. The images of these
surveys are based on photographic data obtained using the Oschin
Schmidt Telescope on Palomar Mountain and the UK Schmidt
Telescope. The plates were processed into the present compressed
digital form with the permission of these institutions. This research
has made use of data obtained from or software provided by the US
National Virtual Observatory, which is sponsored by the National
Science Foundation. This research has made use of the SIMBAD database,
operated at CDS, Strasbourg, France.

\begin{appendix}

\section{Catalog of Variable Stars}\label{sec:cat}

We present the data as 3 separate catalogs of candidate variable
stars, including: continuous periodic variables selected with the
AoVHarm technique and not flagged as an EB during the visual
inspection of either the EPD or TFA light curves, eclipsing binaries,
and flare events.
All three catalogs will be made available in electronic form
from the CDS archive server\footnote{http://cdsarc.u-strasbg.fr} as
well as from the HATNet website\footnote{http://hatnet.hu/}. Light
curves for the variable stars will be made available through the
Harvard University Time Series
Center\footnote{http://timemachine.iic.harvard.edu/} and through
NStED\footnote{http://nsted.ipac.caltech.edu}. There are a total of
\NUMinvars{} stars that are in the first catalog, \NUMEB{} candidate
EB stars in the second, and \NUMflareevents{} flare events from
\NUMflarestars{} stars in the third. Note that some of the EBs and
flare stars are included in the primary catalog as well. Altogether
\NUMtotpotentialvar{} stars are included in at least one of the
catalogs.

In the primary catalog of periodic variables we include the internal
HAT-ID of each source, its J2000 position from 2MASS, the periods,
S/N, peak-to-peak amplitudes and quality flags for the AoVHarm
algorithm applied to the TFA and EPD light curves, and the blending
flag described in \S~\ref{sec:blend}.  For convenience we also include
the PPMX proper motion, the 2MASS photometry, the estimated $V$
magnitude for each source, and the number of 2MASS neighbors within
$30\arcsec$ of each source. For both periods we list the peak-to-peak
amplitude in the $I_{C}$ and/or $R$ bands, determined for both the EPD
and TFA light curves phased at that period. See \S~\ref{sec:blend} for
a description of how the amplitude is measured. Note that the TFA
light curves used for the amplitude determination have not been
processed with signal-reconstruction mode TFA, as a result these
amplitudes are generally lower than the real amplitude of the star,
and are lower than the amplitudes measured on the EPD light
curves. Tables~\ref{tab:variables1}-\ref{tab:variables3} at the end of
the paper show the first ten entries in this catalog. For display
purposes we split the catalog into 3 separate tables, in the full
electronic version of the catalog the tables are joined.

In the catalog of candidate eclipsing binary stars we include the
HAT-ID, position, proper motion, 2MASS photometry, and $V$ magnitude,
together with the period, period uncertainty, epochs of primary and
secondary eclipse (if the secondary is detected) and their respective
uncertainties, the durations of the primary and secondary eclipses,
the number of 2MASS neighbors within $30\arcsec$ of the source, and
the blending
flag. Tables~\ref{tab:EBvariables1}-\ref{tab:EBvariables2} at the end
of the paper show the first ten entries in this catalog.

In the catalog of flare events we include the HAT-ID, position, proper
motion, 2MASS photometry, and $V$ magnitude, together with the time of
the flare peak, the peak intensity relative to the non-flaring
intensity ($A$ in Eq.~\ref{eqn:flare}), and the decay time of the
flare ($\tau$ in
Eq.~\ref{eqn:flare}). Tables~\ref{tab:flarevariables1}-\ref{tab:flarevariables2}
at the end of the paper show the first ten entries in this catalog.

\clearpage

\ifthenelse{\boolean{emulateapj}}{\begin{deluxetable*}{lrrrrrrrrrrrr}}{\begin{deluxetable}{lrrrrrrrrrrrr}}
\ifthenelse{\boolean{emulateapj}}{}{\rotate}
\tabletypesize{\footnotesize}
\tablewidth{0pc}
\tablecaption{Catalog of candidate periodic variable stars. I: Astrometric and Photometric Data}
\tablehead{
\colhead{ID} &
\colhead{RA\tablenotemark{a}} &
\colhead{DE\tablenotemark{a}} &
\colhead{pmRA\tablenotemark{b}} &
\colhead{pmDE\tablenotemark{b}} &
\colhead{Jmag\tablenotemark{c}} &
\colhead{Hmag\tablenotemark{c}} &
\colhead{Kmag\tablenotemark{c}} &
\colhead{Vmag} &
\colhead{Vref\tablenotemark{d}} &
\colhead{Nnbr\tablenotemark{e}} &
\colhead{Bl1\tablenotemark{f}} &
\colhead{Bl2\tablenotemark{f}}\\
&
\multicolumn{1}{c}{h:m:s} &
\multicolumn{1}{c}{deg:m:s} &
\multicolumn{1}{c}{mas/yr} &
\multicolumn{1}{c}{mas/yr} &
\multicolumn{1}{c}{mag} &
\multicolumn{1}{c}{mag} &
\multicolumn{1}{c}{mag} &
\multicolumn{1}{c}{mag} &
&
&
&
}
\startdata
HAT-086-0001701 & 00:13:38.42 & +52:45:05.0 &    35.13 &   -36.72 &   9.519 &   8.763 &   8.658 &  12.136 & 1 &  2 & 1 & 9 \\
HAT-086-0004017 & 23:54:23.85 & +54:41:26.9 &    35.30 &    35.51 &   9.565 &   8.804 &   8.643 &  12.235 & 1 &  2 & 2 & 9 \\
HAT-086-0005153 & 23:51:18.20 & +54:35:16.2 &    76.13 &    -7.77 &   9.906 &   9.239 &   9.060 &  12.822 & 1 &  4 & 2 & 9 \\
HAT-086-0005542 & 00:16:56.01 & +50:37:49.0 &   146.68 &   -27.87 &   9.980 &   9.349 &   9.137 &  12.767 & 1 &  3 & 2 & 9 \\
HAT-086-0005602 & 00:16:01.61 & +48:55:56.9 &   -83.81 &    51.47 &  10.008 &   9.409 &   9.186 &  12.870 & 1 &  0 & 3 & 9 \\
HAT-086-0006520 & 23:56:16.37 & +52:42:34.8 &    19.41 &   -45.99 &  10.260 &   9.590 &   9.478 &  12.227 & 1 &  6 & 2 & 9 \\
HAT-086-0008187 & 23:45:32.05 & +55:40:58.0 &   -41.95 &    -6.38 &  10.462 &   9.840 &   9.657 &  12.791 & 1 &  6 & 2 & 9 \\
HAT-086-0009482 & 00:04:57.67 & +49:58:49.1 &    93.81 &     2.31 &  10.594 &   9.933 &   9.757 &  13.413 & 1 &  0 & 1 & 9 \\
HAT-086-0011792 & 23:46:30.54 & +54:48:56.1 &    23.13 &   -21.70 &  10.861 &  10.228 &  10.042 &  13.191 & 1 &  4 & 2 & 9 \\
HAT-086-0013793 & 23:43:29.95 & +51:50:15.8 &    40.15 &    14.37 &  11.126 &  10.514 &  10.366 &  13.280 & 1 &  3 & 2 & 9 \\
\multicolumn{13}{c}{$\ldots$} \\
\enddata
\tablenotetext{a}{J2000, from 2MASS}
\tablenotetext{b}{from PPMX}
\tablenotetext{c}{from 2MASS}
\tablenotetext{d}{Source for Vmag (0: PPMX (Tycho-2), 1: Transformed from USNO-B1.0, 2: Transformed from 2MASS)}
\tablenotetext{e}{Number of other 2MASS sources within $30\arcsec$}
\tablenotetext{f}{Blending Flag, Bl1 is determined from 2MASS, Bl2 is determined from SDSS. (0: unblended, 1: unlikely blend, 2: possible blend, 3: probable blend, 9: No match to SDSS)}
\label{tab:variables1}
\ifthenelse{\boolean{emulateapj}}{\end{deluxetable*}}{\end{deluxetable}}

\ifthenelse{\boolean{emulateapj}}{\begin{deluxetable*}{lrrrrrrr}}{\begin{deluxetable}{lrrrrrrr}}
\ifthenelse{\boolean{emulateapj}}{}{\rotate}
\tabletypesize{\footnotesize}
\tablewidth{0pc}
\tablecaption{Catalog of candidate periodic variable stars. II: AoVHarm EPD}
\tablehead{
\colhead{ID} &
\colhead{P$_{e}$\tablenotemark{a}} &
\colhead{SN$_{e}$\tablenotemark{a}} &
\colhead{Amp$_{eeI}$\tablenotemark{b}} &
\colhead{Amp$_{eeR}$\tablenotemark{b}} &
\colhead{Amp$_{etI}$\tablenotemark{b}} &
\colhead{Amp$_{etR}$\tablenotemark{b}} &
\colhead{Q$_{e}$\tablenotemark{c}} \\
&
\multicolumn{1}{c}{d} &
&
\multicolumn{1}{c}{mag} &
\multicolumn{1}{c}{mag} &
\multicolumn{1}{c}{mag} &
\multicolumn{1}{c}{mag} &
}
\startdata
HAT-086-0001701 & -99.999999 & -999.99 & -9.999 & -9.999 & -9.999 & -9.999 & 2 \\ 
HAT-086-0004017 &  40.868328 &   95.12 & -9.999 &  0.014 & -9.999 &  0.011 & 1 \\ 
HAT-086-0005153 &  16.368924 &   49.27 & -9.999 &  0.010 & -9.999 &  0.003 & 0 \\ 
HAT-086-0005542 &  20.363047 &   64.94 & -9.999 &  0.010 & -9.999 &  0.009 & 0 \\ 
HAT-086-0005602 & -99.999999 & -999.99 & -9.999 & -9.999 & -9.999 & -9.999 & 2 \\ 
HAT-086-0006520 &  11.462731 &   64.66 & -9.999 &  0.024 & -9.999 &  0.005 & 0 \\ 
HAT-086-0008187 & -99.999999 & -999.99 & -9.999 & -9.999 & -9.999 & -9.999 & 2 \\ 
HAT-086-0009482 &  13.561425 &   42.66 & -9.999 &  0.018 & -9.999 &  0.009 & 0 \\ 
HAT-086-0011792 &  79.872558 &  106.96 & -9.999 &  0.022 & -9.999 &  0.002 & 1 \\ 
HAT-086-0013793 &  11.341684 &   38.88 & -9.999 &  0.012 & -9.999 &  0.005 & 0 \\ 
\multicolumn{8}{c}{$\ldots$} \\
\enddata
\tablenotetext{a}{The subscript denotes the light-curve type (e: EPD, t: TFA).}
\tablenotetext{b}{The peak-to-peak amplitude. The first letters in the subscript denotes the period used, the second letter in the subscript denotes the light-curve type used to calculate the amplitude (e: EPD, t: TFA), the last letter denotes the filter (I or R).}
\tablenotetext{c}{Quality flag (0: robust detection, 1: questionable detection, 2: not detected by the given method for the given light curve type).}
\label{tab:variables2}
\ifthenelse{\boolean{emulateapj}}{\end{deluxetable*}}{\end{deluxetable}}
\ifthenelse{\boolean{emulateapj}}{\begin{deluxetable*}{lrrrrrrr}}{\begin{deluxetable}{lrrrrrrr}}
\ifthenelse{\boolean{emulateapj}}{}{\rotate}
\tabletypesize{\footnotesize}
\tablewidth{0pc}
\tablecaption{Catalog of candidate periodic variable stars. III: AoVHarm TFA}
\tablehead{
\colhead{ID} &
\colhead{P$_{t}$\tablenotemark{a}} &
\colhead{SN$_{t}$\tablenotemark{a}} &
\colhead{Amp$_{teI}$\tablenotemark{b}} &
\colhead{Amp$_{teR}$\tablenotemark{b}} &
\colhead{Amp$_{tt}$\tablenotemark{b}} &
\colhead{Amp$_{tR}$\tablenotemark{b}} &
\colhead{Q$_{t}$\tablenotemark{c}} \\
&
\multicolumn{1}{c}{d} &
&
\multicolumn{1}{c}{mag} &
\multicolumn{1}{c}{mag} &
\multicolumn{1}{c}{mag} &
\multicolumn{1}{c}{mag} &
}
\startdata
HAT-086-0001701 &  17.887814 &   41.17 & -9.999 &  0.010 & -9.999 &  0.005 & 0 \\
HAT-086-0004017 & -99.999999 & -999.99 & -9.999 & -9.999 & -9.999 & -9.999 & 2 \\
HAT-086-0005153 & -99.999999 & -999.99 & -9.999 & -9.999 & -9.999 & -9.999 & 2 \\
HAT-086-0005542 &  20.411745 &   29.90 & -9.999 &  0.010 & -9.999 &  0.009 & 0 \\
HAT-086-0005602 &  21.071980 &   47.14 &  0.019 &  0.016 &  0.010 &  0.009 & 1 \\
HAT-086-0006520 &  11.503656 &   20.59 & -9.999 &  0.024 & -9.999 &  0.007 & 0 \\
HAT-086-0008187 &   5.358132 &   22.44 & -9.999 &  0.029 & -9.999 &  0.018 & 0 \\
HAT-086-0009482 & -99.999999 & -999.99 & -9.999 & -9.999 & -9.999 & -9.999 & 2 \\
HAT-086-0011792 & -99.999999 & -999.99 & -9.999 & -9.999 & -9.999 & -9.999 & 2 \\
HAT-086-0013793 & -99.999999 & -999.99 & -9.999 & -9.999 & -9.999 & -9.999 & 2 \\
\multicolumn{8}{c}{$\ldots$} \\
\enddata
\tablenotetext{a}{The subscript denotes the light-curve type (e: EPD, t: TFA).}
\tablenotetext{b}{The peak-to-peak amplitude. The first letters in the subscript denotes the period used, the second letter in the subscript denotes the light-curve type used to calculate the amplitude (e: EPD, t: TFA), the last letter denotes the filter (I or R).}
\tablenotetext{c}{Quality flag (0: robust detection, 1: questionable detection, 2: not detected by the given method for the given light curve type).}
\label{tab:variables3}
\ifthenelse{\boolean{emulateapj}}{\end{deluxetable*}}{\end{deluxetable}}

\ifthenelse{\boolean{emulateapj}}{\begin{deluxetable*}{lrrrrrrrrrrrr}}{\begin{deluxetable}{lrrrrrrrrrrrr}}
\ifthenelse{\boolean{emulateapj}}{}{\rotate}
\tabletypesize{\footnotesize}
\tablewidth{0pc}
\tablecaption{Catalog of candidate eclipsing binaries. I: Astrometric and Photometric Data}
\tablehead{
\colhead{ID} &
\colhead{RA\tablenotemark{a}} &
\colhead{DE\tablenotemark{a}} &
\colhead{pmRA\tablenotemark{b}} &
\colhead{pmDE\tablenotemark{b}} &
\colhead{Jmag\tablenotemark{c}} &
\colhead{Hmag\tablenotemark{c}} &
\colhead{Kmag\tablenotemark{c}} &
\colhead{Vmag} &
\colhead{Vref\tablenotemark{d}} &
\colhead{Nnbr\tablenotemark{e}} &
\colhead{Bl1\tablenotemark{f}} &
\colhead{Bl2\tablenotemark{f}} \\
&
\multicolumn{1}{c}{h:m:s} &
\multicolumn{1}{c}{deg:m:s} &
\multicolumn{1}{c}{mas/yr} &
\multicolumn{1}{c}{mas/yr} &
\multicolumn{1}{c}{mag} &
\multicolumn{1}{c}{mag} &
\multicolumn{1}{c}{mag} &
\multicolumn{1}{c}{mag} &
&
&
&
}
\startdata
HAT-086-0022206 & 23:50:17.08 & +51:11:28.9 &    37.45 &   -20.48 &  11.655 &  11.009 &  10.858 &  13.656 & 1 &  5 & 1 & 9 \\
HAT-087-0016301 & 00:34:50.23 & +53:42:12.1 &   -37.41 &    12.76 &  11.310 &  10.613 &  10.433 &  13.904 & 2 &  5 & 2 & 9 \\
HAT-088-0017192 & 01:38:56.50 & +48:52:40.2 &   -11.17 &   -29.81 &  11.469 &  10.807 &  10.608 &  13.920 & 1 &  2 & 0 & 1 \\
HAT-113-0005699 & 18:09:47.63 & +49:02:55.0 &   -53.66 &   -21.85 &  11.362 &  10.710 &  10.560 &  14.126 & 1 &  1 & 0 & 9 \\
HAT-127-0008153 & 03:04:05.20 & +42:03:10.5 &    53.86 &   -54.80 &  10.650 &   9.983 &   9.730 &  14.159 & 1 &  2 & 2 & 9 \\
HAT-127-0009895 & 03:02:22.68 & +43:40:48.0 &   -76.01 &   -30.99 &  11.018 &  10.375 &  10.198 &  13.890 & 1 &  5 & 3 & 9 \\
HAT-129-0025342 & 04:13:06.81 & +41:21:01.4 &    42.67 &   -26.33 &  11.938 &  11.372 &  11.160 &  14.311 & 1 &  1 & 1 & 9 \\
HAT-131-0026711 & 05:16:36.90 & +48:35:44.3 &   -14.58 &   -32.73 &  11.802 &  11.138 &  10.982 &  14.275 & 1 &  5 & 1 & 9 \\
HAT-132-0012475 & 05:43:20.18 & +47:17:00.1 &     7.93 &   -29.15 &  11.345 &  10.704 &  10.569 &  13.765 & 1 &  1 & 2 & 9 \\
HAT-133-0001901 & 06:31:57.29 & +41:33:57.3 &   -56.65 &   -45.10 &   9.581 &   8.930 &   8.787 &  11.638 & 1 &  2 & 1 & 9 \\
\multicolumn{13}{c}{$\ldots$} \\
\enddata
\tablenotetext{a}{J2000, from 2MASS}
\tablenotetext{b}{from PPMX}
\tablenotetext{c}{from 2MASS}
\tablenotetext{d}{Source for Vmag (0: PPMX (Tycho-2), 1: Transformed from USNO-B1.0, 2: Transformed from 2MASS)}
\tablenotetext{e}{Number of other 2MASS sources within $30\arcsec$}
\tablenotetext{f}{Blending Flag, Bl1 is determined from 2MASS, Bl2 is determined from SDSS (0: unblended, 1: unlikely blend, 2: possible blend, 3: probable blend, 9: no match to SDSS)}
\label{tab:EBvariables1}
\ifthenelse{\boolean{emulateapj}}{\end{deluxetable*}}{\end{deluxetable}}

\ifthenelse{\boolean{emulateapj}}{\begin{deluxetable*}{lrrrrrrrr}}{\begin{deluxetable}{lrrrrrrrr}}
\ifthenelse{\boolean{emulateapj}}{}{\rotate}
\tabletypesize{\footnotesize}
\tablewidth{0pc}
\tablecaption{Catalog of candidate eclipsing binaries. II: Ephemerides}
\tablehead{
\colhead{ID} &
\colhead{Per} &
\colhead{ePer} &
\colhead{HJD0p\tablenotemark{a}} &
\colhead{eHJD0p\tablenotemark{b}} &
\colhead{DURp\tablenotemark{c}} &
\colhead{HJD0s\tablenotemark{a}} &
\colhead{eHJD0s\tablenotemark{b}} &
\colhead{DURs\tablenotemark{c}} \\
&
\multicolumn{1}{c}{d} &
\multicolumn{1}{c}{d} &
\multicolumn{1}{c}{d} &
\multicolumn{1}{c}{d} &
\multicolumn{1}{c}{h} &
\multicolumn{1}{c}{d} &
\multicolumn{1}{c}{d} &
\multicolumn{1}{c}{h}
}
\startdata
HAT-086-0022206 &  0.4290339 &  0.0000062 & 2454431.11547 & 0.00042 & 1.53 & 2454436.90565 & 0.00117 & 1.52 \\
HAT-087-0016301 &  3.1325961 &  0.0003763 & 2454427.60691 & 0.00504 & 5.24 & 2454416.64515 & 0.00442 & 4.63 \\
HAT-088-0017192 &  1.4030920 &  0.0000252 & 2453693.67973 & 0.00066 & 2.15 & 2453680.35434 & 0.00071 & 2.00 \\
HAT-113-0005699 &  0.2278737 &  0.0000013 & 2453901.88979 & 0.00028 & 1.24 & 2453905.42128 & 0.00039 & 1.14 \\
HAT-127-0008153 &  2.5681924 &  0.0000869 & 2452979.53695 & 0.00140 & 1.98 & 2452991.10638 & 0.00215 & 1.53 \\
HAT-127-0009895 &  0.5916726 &  0.0000239 & 2452985.85734 & 0.00187 & 4.16 & 2452990.88691 & 0.05334 & 4.55 \\
HAT-129-0025342 &  4.6109145 &  0.0002340 & 2453726.54374 & 0.00129 & 2.85 & 2453724.34418 & 0.00235 & 3.34 \\
HAT-131-0026711 &  0.6639531 &  0.0000048 & 2454497.17095 & 0.00026 & 1.43 & 2454498.83198 & 0.00170 & 1.42 \\
HAT-132-0012475 &  0.8663223 &  0.0000035 & 2454128.11503 & 0.00145 & 3.12 & 9999999.99999 & 9.99999 & 9.99 \\
HAT-133-0001901 &  0.2929328 &  0.0000001 & 2453390.29599 & 0.00016 & 0.91 & 2453376.38208 & 0.00022 & 0.88 \\
\multicolumn{9}{c}{$\ldots$} \\
\enddata
\tablenotetext{a}{Epoch of primary (p) or secondary (s) eclipse. The value is 9999999.99999 if a secondary eclipse is not detected.}
\tablenotetext{b}{Uncertainty on the epoch of eclipse.}
\tablenotetext{c}{Eclipse duration.}
\label{tab:EBvariables2}
\ifthenelse{\boolean{emulateapj}}{\end{deluxetable*}}{\end{deluxetable}}

\ifthenelse{\boolean{emulateapj}}{\begin{deluxetable*}{lrrrrrrrrr}}{\begin{deluxetable}{lrrrrrrrrr}}
\ifthenelse{\boolean{emulateapj}}{}{\rotate}
\tabletypesize{\footnotesize}
\tablewidth{0pc}
\tablecaption{Catalog of flare events. I: Astrometric and Photometric Data\tablenotemark{a}}
\tablehead{
\colhead{ID} &
\colhead{RA\tablenotemark{b}} &
\colhead{DE\tablenotemark{b}} &
\colhead{pmRA\tablenotemark{c}} &
\colhead{pmDE\tablenotemark{c}} &
\colhead{Jmag\tablenotemark{d}} &
\colhead{Hmag\tablenotemark{d}} &
\colhead{Kmag\tablenotemark{d}} &
\colhead{Vmag} &
\colhead{Vref\tablenotemark{e}} \\
&
\multicolumn{1}{c}{h:m:s} &
\multicolumn{1}{c}{deg:m:s} &
\multicolumn{1}{c}{mas/yr} &
\multicolumn{1}{c}{mas/yr} &
\multicolumn{1}{c}{mag} &
\multicolumn{1}{c}{mag} &
\multicolumn{1}{c}{mag} &
\multicolumn{1}{c}{mag} &
}
\startdata
HAT-087-0002617 & 00:21:57.81 & +49:12:38.0 &   205.00 &   -32.56 &   9.139 &   8.454 &   8.205 &  12.474 & 1 \\ 
HAT-094-0007885 & 05:35:22.84 & +53:16:02.7 &     4.39 &   -35.34 &  10.797 &  10.139 &   9.958 &  13.797 & 1 \\ 
HAT-101-0000851 & 09:59:39.19 & +48:47:54.6 &   -45.67 &     8.38 &   9.686 &   9.002 &   8.817 &  12.137 & 1 \\ 
HAT-103-0001602 & 11:03:17.04 & +49:48:25.5 &   -63.32 &    -7.74 &  10.898 &  10.286 &  10.114 &  13.544 & 1 \\ 
HAT-119-0031226 & 22:02:08.40 & +49:07:17.3 &   -68.41 &     0.39 &  11.102 &  10.434 &  10.228 &  12.777 & 1 \\ 
HAT-123-0002064 & 00:53:06.48 & +48:29:38.5 &   220.82 &  -162.32 &   9.427 &   8.773 &   8.580 &  12.899 & 0 \\ 
HAT-123-0002064 & 00:53:06.48 & +48:29:38.5 &   220.82 &  -162.32 &   9.427 &   8.773 &   8.580 &  12.899 & 0 \\ 
HAT-123-0003289 & 00:41:41.41 & +44:10:53.1 &   -46.23 &   -15.14 &  10.012 &   9.399 &   9.188 &  14.251 & 1 \\ 
HAT-126-0002065 & 02:40:52.51 & +44:52:36.5 &   215.74 &     3.85 &   9.277 &   8.696 &   8.461 &  13.197 & 0 \\ 
HAT-128-0007422 & 03:42:45.17 & +43:27:40.1 &    30.96 &   -64.10 &  10.435 &   9.830 &   9.568 &  14.611 & 1 \\ 
\multicolumn{10}{c}{$\ldots$} \\
\enddata
\tablenotetext{a}{Stars may be listed more than once if they have multiple flare events}
\tablenotetext{b}{J2000, from 2MASS}
\tablenotetext{c}{from PPMX}
\tablenotetext{d}{from 2MASS}
\tablenotetext{e}{Source for Vmag (0: PPMX (Tycho-2), 1: Transformed from USNO-B1.0, 2: Transformed from 2MASS)}
\label{tab:flarevariables1}
\ifthenelse{\boolean{emulateapj}}{\end{deluxetable*}}{\end{deluxetable}}
\ifthenelse{\boolean{emulateapj}}{\begin{deluxetable*}{lrrrr}}{\begin{deluxetable}{lrrrr}}
\ifthenelse{\boolean{emulateapj}}{}{\rotate}
\tabletypesize{\footnotesize}
\tablewidth{0pc}
\tablecaption{Catalog of flare events. II: Flare data\tablenotemark{a}}
\tablehead{
\colhead{ID} &
\colhead{HJD\tablenotemark{b}} &
\colhead{Amp\tablenotemark{c}} &
\colhead{$\tau$\tablenotemark{d}} &
\colhead{Nobs\tablenotemark{e}}\\
&
\multicolumn{1}{c}{d} &
&
\multicolumn{1}{c}{d} &
}
\startdata
HAT-087-0002617 & 2454437.61429 &  0.130 & 0.0047 & 26 \\
HAT-094-0007885 & 2454463.88452 &  0.219 & 0.0092 &  7 \\
HAT-101-0000851 & 2454162.63179 &  0.200 & 0.0176 & 26 \\
HAT-103-0001602 & 2454563.96193 &  0.279 & 0.0212 & 29 \\
HAT-119-0031226 & 2453360.74326 &  0.552 & 0.0346 & 31 \\
HAT-123-0002064 & 2454337.05672 &  0.118 & 0.0053 & 11 \\
HAT-123-0002064 & 2454391.64587 &  0.482 & 0.0065 & 15 \\
HAT-123-0003289 & 2454338.90758 &  0.236 & 0.0087 & 23 \\
HAT-126-0002065 & 2452999.82533 &  0.277 & 0.0056 &  9 \\
HAT-128-0007422 & 2453387.68872 &  0.346 & 0.0216 & 13 \\
\multicolumn{5}{c}{$\ldots$} \\
\enddata
\tablenotetext{a}{Stars may be listed more than once if they have multiple flare events}
\tablenotetext{b}{Time of flare peak}
\tablenotetext{c}{Peak intensity relative to the non-flaring intensity.}
\tablenotetext{d}{1/e decay time of the flare.}
\tablenotetext{e}{Number of observations included in fitting the flare.}
\label{tab:flarevariables2}
\ifthenelse{\boolean{emulateapj}}{\end{deluxetable*}}{\end{deluxetable}}

\ifthenelse{\boolean{emulateapj}}{\begin{deluxetable*}{llrrr}}{\begin{deluxetable}{llrrr}}
\ifthenelse{\boolean{emulateapj}}{}{\rotate}
\tabletypesize{\footnotesize}
\tablewidth{0pc}
\tablecaption{Cross-matches to previously identified variable stars}
\tablehead{
\colhead{ID} &
\colhead{Alternate ID} &
\colhead{Mflag\tablenotemark{a}} &
\colhead{BlFlag1\tablenotemark{b}} &
\colhead{BlFlag2\tablenotemark{b}}
}
\startdata
HAT-086-0005602 &                  V0544 Cas & 1 & 3 & 9 \\ 
HAT-086-0022206 &                  V1001 Cas & 0 & 1 & 9 \\ 
HAT-091-0010976 &                  V0524 Per & 0 & 2 & 9 \\ 
HAT-091-0016338 &                  V0485 Per & 0 & 1 & 9 \\ 
HAT-112-0001188 & 1SWASP J173659.28+485946.1 & 0 & 0 & 9 \\ 
HAT-128-0006973 &                  NSV 01380 & 1 & 3 & 9 \\ 
HAT-128-0022221 &                     EZ Per & 1 & 3 & 9 \\ 
HAT-141-0000663 &                     BT UMa & 0 & 0 & 0 \\ 
HAT-144-0000282 & 1SWASP J132712.10+455826.4 & 0 & 0 & 1 \\ 
HAT-144-0000526 &                     DE CVn & 0 & 0 & 1 \\ 
\multicolumn{5}{c}{$\ldots$} \\
\enddata
\tablenotetext{a}{0: the match is correct, 1: the two stars are not the same, but may be blended in the HAT images.}
\tablenotetext{b}{Automatically generated blending flag. BlFlag1 is determined from 2MASS, BlFlag2 is determined from SDSS. 0: unblended, 1: unlikely blend, 2: potential blend, 3: probable blend, 4: source is a flare star, no blending info is generated, 9: no SDSS match.}
\label{tab:cross}
\ifthenelse{\boolean{emulateapj}}{\end{deluxetable*}}{\end{deluxetable}}

\ifthenelse{\boolean{emulateapj}}{\begin{deluxetable*}{llr}}{\begin{deluxetable}{llr}}
\ifthenelse{\boolean{emulateapj}}{}{\rotate}
\tabletypesize{\footnotesize}
\tablewidth{0pc}
\tablecaption{Cross-matches to ROSAT X-ray sources}
\tablehead{
\colhead{ID} &
\colhead{RosatID\tablenotemark{a}} &
\colhead{lgfxfj\tablenotemark{b}}
}
\startdata
HAT-087-0001873 & 1RXS J002854.1+502226 & -1.7955 \\ 
HAT-087-0002547 & 1RXS J003633.7+553728 & -2.2528 \\ 
HAT-087-0002617 & 1RXS J002158.0+491245 & -2.1767 \\ 
HAT-093-0001111 & 1RXS J044853.2+552725 & -2.3278 \\ 
HAT-094-0004894 & 1RXS J050718.8+530713 & -2.3805 \\ 
HAT-094-0005389 & 1RXS J053001.7+484932 & -2.0147 \\ 
HAT-094-0006428 & 1RXS J050809.3+532437 & -1.7462 \\ 
HAT-094-0009460 & 1RXS J050012.4+532033 & -1.6652 \\ 
HAT-094-0384094 & 1RXS J053625.1+533618 & -2.2575 \\ 
HAT-103-0001602 & 1RXS J110318.0+494822 & -1.8031 \\ 
\multicolumn{3}{c}{$\ldots$} \\
\enddata
\tablenotetext{a}{ID of matching ROSAT source}
\tablenotetext{b}{Estimated $\log(f_x/f_J)$ (ratio of X-ray flux to J-band flux).}
\label{tab:xray}
\ifthenelse{\boolean{emulateapj}}{\end{deluxetable*}}{\end{deluxetable}}

\end{appendix}

\end{document}